\documentclass[10pt,aps,a4paper,prd,floatfix,onecolumn,notitlepage,nofootinbib]{revtex4-2}

\usepackage[left=2cm,right=2cm,top=2cm,bottom=2cm]{geometry}

\usepackage[utf8]{inputenc}
\usepackage[british,UKenglish,USenglish,american]{babel}
\usepackage{amsmath}

\usepackage{graphicx}
\usepackage{subcaption}
\usepackage{booktabs}

\usepackage{setspace}%\onehalfspacing

\usepackage{slashed}

\usepackage{ytableau}

\usepackage{mathrsfs}
\usepackage{amssymb}
\usepackage{bbm}

\usepackage{nicefrac}

\usepackage{microtype}

\usepackage{parskip}

\usepackage{physics}

\usepackage{xcolor}

\newcommand\numberthis{\addtocounter{equation}{1}\tag{\theequation}}

\DeclareMathOperator{\diag}{diag}

\newcommand{\p}{\prime}

\renewcommand{\bar}{\overline}

\usepackage{bm}\renewcommand{\vec}{\bm}

\newcommand{\e}{\mathrm{e}}

\newcommand{\Plrep}[2]{\ell_{#1,#2}}
\newcommand{\Plbarrep}[2]{\bar{\ell}_{#1,#2}}

\newcommand{\Pl}[1]{\ell_{#1}}

\newcommand{\SU}[1]{\mathrm{SU}(#1)}
\newcommand{\U}[1]{\mathrm{U}(#1)}
\newcommand{\Z}[1]{\mathrm{Z}(#1)}

\newcommand{\Rep}{\mathcal{R}}
\newcommand{\Fund}{\mathrm{F}}
\newcommand{\Adj}{\mathrm{A}}

\begin{document}
\raggedbottom

\title{
Building models of quarks and gluons with an arbitrary number of colors using Cartan-Polyakov loops
}

\author{Renan Câmara Pereira}
\email{renan.pereira@student.uc.pt}
\affiliation{Centro de Física da Universidade de Coimbra (CFisUC), 
Department of Physics, University of Coimbra, P-3004 - 516  Coimbra, Portugal}

\author{Pedro Costa}
\email{pcosta@uc.pt}
\affiliation{Centro de Física da Universidade de Coimbra (CFisUC), 
Department of Physics, University of Coimbra, P-3004 - 516  Coimbra, Portugal}

\begin{abstract}
In this work we introduce the concept of Cartan-Polyakov loops, a special subset of Polyakov loops in the fundamental and antifundamental representation of the $\SU{N_c}$ group, $\Plrep{\Fund}{k}$ and $\Plbarrep{\Fund}{k}$ respectively, with charges $k = 1, \ldots, (N_c-1)/2$. It constitutes a sufficient set of independent degrees of freedom and it is used to parametrize the thermal Wilson line. Polyakov loops not contained in this set are classified as non-Cartan-Polyakov loops. Using properties of the characteristic polynomial of the thermal Wilson line, we write a non-Cartan-Polyakov loop charge decomposition formula. This formalism allows one to readily build effective models of quarks and gluons with an arbitrary number of colors. We apply it to the Polyakov$-$Nambu$-$Jona-Lasinio model and to an effective glue model, in the mean field approximation, showing how to directly extend these models to higher values of $N_c$.
\end{abstract}

\maketitle

\section{Introduction}

Gauge theories with the symmetry group $\SU{N}$, are fundamental building blocks to our current understanding of the laws of physics at the microscopic level. The theory of strong interactions, Quantum Chromodynamics (QCD), is a perfect example of a gauge theory based on the local invariance under transformations of the $\SU{N_c}$ symmetry group, with $N_c$ being the number of colors. This theory has several important properties like scaling, asymptotic freedom, chiral symmetry breaking and confinement \cite{Skands:2012ts}.

The interplay of the properties of QCD alongside its non-perturbative nature at low energies, leads the way to an extremely rich phase structure for strongly interacting matter, also known as the QCD phase diagram \cite{Halasz:1998qr}. Originally, only confined and deconfined phases were displayed in the diagram however, since then, other features have been discovered or proposed, such as, the existence of the quark-gluon plasma at high temperatures, the existence of the conjectured critical endpoint at finite temperature and baryon density, the possible diquark condensation in high density baryon matter or the confined chirally symmetric phase \cite{Hansen:2019lnf}, to name a few.

However, in order to study QCD in these energy regimes we are limited in options due to the non-perturbative behavior of the theory. From a theoretical point of view, one can use lattice QCD calculations, Dyson$-$Schwinger equations and effective models. Lattice methods are the main tool to understand strongly interacting matter from first principles, at least for finite temperature and zero chemical potential. At finite density, lattice QCD suffers from the infamous sign problem which renders the importance sampling, necessary in Monte Carlo simulations, ineffective \cite{Schmidt:2017bjt}.

Deviating from the real world of three colors and studying systems with increasing $N_c$, or considering the large $N_c$ limit (also known as ’t Hooft limit, originally introduced in Ref. \cite{tHooft:1973alw}), has been demonstrated to be a really good tool to better understand the behavior of the QCD properties \cite{Witten:1979kh,Jenkins:1998wy,Bonanno:2011yr,Lucini:2013qja,Richardson:2021xiu,Lucha:2021mwx,Flores-Mendieta:2021wzh}, allowing for the exploration of the non-perturbative behavior of QCD in a systematic way \cite{Witten:1979kh,Jenkins:1998wy,Lucini:2013qja}. The large $N_c$ limit is taken by considering a large number of colors, while fixing $\lambda=g^2 N_c$, with $g$ the strong coupling and $\lambda$ the so-called 't Hooft coupling \cite{tHooft:1973alw}.

Indeed, the large $N_c$ limit of the theory should still exhibit confinement and chiral symmetry breaking since the theory remains asymptotically free, with a growing coupling at low energies \cite{Jenkins:1998wy,Hernandez:2020tbc}. Not only that, some lattice studies, the phenomenological OZI (Okubo$-$Zweig$-$Iizuka) rule and the reproduction of ratios of low-lying meson masses \cite{Karthik:2022fdb}, also seem to point in the direction of a large $N_c$ limit of QCD resembling the real world case of $N_c=3$ \cite{Teper:1997tq,Teper:1998te,DiVecchia:1999yr,DeGrand:2016pur,GarciaPerez:2020gnf,Hernandez:2020tbc}. Although QCD in the large $N_c$ limit is a simpler theory, even in such approximation it has not yet to been solved \cite{Polchinski:1991tw,Makeenko:2009dw}. Non-abelian gauge theories, with the gauge group $\SU{N}$, in the large $N$ limit are also very important in the study of string theories, in particular, in studies involving the conjectured AdS/CFT correspondence, i.e., the duality between gauge theories and theories of gravity \cite{Maldacena:1997re,Aharony:1999ti,Lucini:2013qja}. Indeed, one can also use this correspondence to study the QCD phase diagram \cite{Yang:2020hun}.

In the spirit of the large $N_c$ limit, Pisarski et al. proposed the existence of a new phase of matter in the phase diagram of strongly interacting matter, the so-called quarkyonic phase, expected when the number of colors, $N_c$, is very large \cite{McLerran:2007qj}. It was conjectured that, in this limit, the QCD phase diagram can be separated in three distinct phases, two at lower temperatures and one at high temperatures. The latter is the usual quark-gluon plasma, constituting the deconfined phase of matter. At low temperatures there would be two distinct phases: a mesonic phase, described by confined matter with zero baryon number density and, at higher chemical potentials, the so-called quarkyonic phase, described by confined matter with finite baryon number density \cite{McLerran:2007qj,McLerran:2008ua}. The conjectured quarkyonic phase of matter was discussed and studied in Refs. \cite{McLerran:2008ua,Torrieri:2010gz,Buisseret:2011ms,Yang:2020hun,Kovacs:2022zcl}, being also explored in the context of neutron stars in Refs. \cite{Giacosa:2017mis,Margueron:2021dtx}.

From the phenomenological point of view, several effective models of QCD have been built in order to study the properties of the conjectured quarkyonic phase. Some of these models were based on the Polyakov$-$Nambu$-$Jona-Lasinio (PNJL) model, in which quark-quark interactions are point-like, with a static gluon field in the temporal direction acting as a background field. These models possess all the global symmetries of QCD, including the confinement/deconfinement transition in the sense of the breaking of the $\Z{N_c}$ symmetry and which can be measured by the expectation value of the Polyakov loop. This quantity is an exact order parameter in the pure glue theory but only approximate when quarks are introduced in the Lagrangian. The Polyakov loop is extremely used as an order parameter to study the deconfinement transition in both ab-initio calculations like lattice QCD, in Dyson-Schwinger equations, and in phenomenological models like the Polyakov-quark-meson model or the PNJL model. 

%The NJL model, for instance, is widely used as an effective model of QCD to study the properties of the phase diagram of strongly interacting matter, meson behaviour at finite temperature and density 

In this work we show how to build models of quarks and gluons with an arbitrary number of colors, in a systematic way. More precisely, we will build a version of the Polyakov$-$Nambu$-$Jona-Lasinio model, in the mean field approximation and consider a glue effective potential, both for any number of colors. Being able to study such models systematically can be quite useful in order to better understand the deconfinement phase transition, the conjectured quarkyonic phase and the properties of quarks and gluons.

\section{Center symmetry and the Polyakov loop}

In the imaginary time formalism, often used to study quantum field theory at finite temperature, boson and fermion fields must respect periodic and antiperiodic boundary conditions, respectively. Using this prescription, it can be shown that, at the Lagrangian level, the boson fields of the Yang-Mills theory have symmetric boundary conditions with respect to transformations of the discrete group $\Z{N_c}$, the center of the $\SU{N_c}$. The introduction of fermion fields, on the other hand, explicitly breaks such invariance. Hence, Yang-Mills theory at finite temperature have boundary conditions that are invariant under $\Z{N_c}$ transformations while QCD does not share the same property, due to the presence of the quark fields. However, even in the pure glue theory, such symmetry can be spontaneously broken. 

The Polyakov loop, $\Phi$, is a gauge invariant quantity, that can be used as an exact order parameter for the spontaneous symmetry breaking of the $\Z{N_c}$ symmetry in the Yang-Mills theory \cite{Pisarski:2002ji}: if the Polyakov loop is zero, the $\Z{N_c}$ symmetry is preserved while, if it is non-zero, the symmetry is broken.

One can check the gauge invariance of the Polyakov loop by considering the gauge transformation of the thermal Wilson loop, $L \qty( \vec{x} ) $, from which one can define the Polyakov loop. The Wilson line can be defined along a path in spacetime, from an initial position ($x_i$) to a final position ($x_f$), as follows:
\begin{align}
W ( x_i,x_f ) = 
\mathcal{P} \exp 
\qty[ ig \int_{x_i}^{x_f} \dd{z^\mu} A_\mu (z) ] .
\end{align}
Here, $\mathcal{P}$ is the path-ordering operator, $g$ is the strong interaction coupling and $A_\mu$ is the gauge field\footnote{
Under a gauge transformation, the gauge field $A_\mu (x)$ transforms as:
\begin{align}
A_\mu (x) 
\xrightarrow[]{\text{g.t.}}
A_\mu^\p (x) = 
\Omega (x) \qty( A_\mu + \frac{1}{ig} \partial_\mu ) \Omega^\dagger (x).
\end{align}}. The Wilson line of the gauge transformed field is given by:
\begin{align*}
W(x_i,x_f)
\xrightarrow[]{\text{g.t.}}
W^\p(x_i,x_f) 
= 
\Omega (x_f) W(x_i,x_f) \Omega^\dagger (x_i) .
\end{align*}
By considering a closed path ($x_i=x_f=x$), one gets the so-called Wilson loop, which is a gauge invariant quantity due to the cyclic property of the trace:
\begin{align*}
\tr \qty[ W(x_i,x_f) ]
\xrightarrow[]{\text{g.t.}}
\tr
\qty[ W^\p(x_i,x_f) ]
& = 
\tr \qty[
\Omega (x_f) W(x_i,x_f) \Omega^\dagger (x_i) 
]
\\
& = 
\tr \qty[ W(x_i,x_f) ] . 
\numberthis
\end{align*}
Considering the Wilson loop wrapped in imaginary time (i.e., at finite temperature with periodic boundary conditions) one defines the thermal Wilson line,
\begin{align}
L \qty( \vec{x} ) 
=
\mathcal{P} \exp 
\qty[ i \int_0^\beta \dd{\tau} A_4 (\vec{x},\tau) ] .
\label{def_thermalWilsonLine}
\end{align}
Here, $A_4=i g {A}_\mu^a T_a \delta_0^\mu$, is the gluon field in the temporal direction ($T_a$ are the $\SU{N_c}$ generators in the fundamental representation, ${A}_\mu^a$ are the gluon fields in the adjoint representation and the index $a$ runs from $0$ to $N_c^2-1$). This quantity is also gauge invariant. Finally, one can define the Polyakov loop on a spatial coordinate $\vec{x}$, in the fundamental representation of the $\SU{N_c}$ group, $\Phi(\vec{x})$, as the trace of the thermal Wilson loop with the field $A_4$ written in the fundamental representation \cite{Pisarski:2002ji,Fukushima:2003fw}:
\begin{align}
\Phi (\vec{x})= 
\frac{1}{N_c} 
\tr\qty[
L_{\Fund} \qty( \vec{x} ) ] .
\label{polyakov.loop.definiton}
\end{align}
Here, the trace is made over color space and the index $\Fund$ indicates that the matrix $A_4$ is written in the fundamental representation of the $\SU{N_c}$ group. The Polyakov loop transforms akin to the fermion fields and only the identity element of the $\Z{N_c}$ group maintains the symmetry indeed, under a global $\Z{N_c}$ transformation, it transforms as a field with charge one \cite{Pisarski:2002ji,Fukushima:2017csk}.

The physical interpretation behind the two previously discussed phases, corresponding to zero and non-zero Polyakov loop, can be understood by evaluating the Helmholtz free energy of a single quark in a background of gluons.  To this end, one can solve the static Dirac equation coupled to a static background field and use it to calculate the Helmholtz free energy of a single quark in a background of gluons \cite{McLerran:1981pb,Pisarski:2002ji,QuarkGluonPlasma}. It yields that, if the Polyakov loop is finite, a single quark can be created in the gluonic background with a finite amount of energy while, if the Polyakov loop is zero, an infinite amount of energy is required \cite{Pisarski:2002ji,QuarkGluonPlasma}. Hence, a finite Polyakov loop would correspond to a deconfined phase of matter while, a zero Polyakov loop to a confined phase.

As already pointed out, the Polyakov loop is often incorporated in phenomenological models of QCD, like the PNJL or quark-meson models, in order to study the properties of the phase diagram of strongly interacting matter and explore the relation between the restoration of chiral symmetry and the deconfinement transition. In the following sections we will show how to build both a PNJL model and an effective glue potential, with an arbitrary number of colors, which can be used to explore the properties of strongly interacting matter.

%In particular, a PNJL model defined for an arbitrary number of colors could be used to study the phase diagram of QCD and determine the behavior of the critical endpoint for increasing $N_c$ \cite{Kovacs:2022zcl}.

\section{Cartan-Polyakov loops}
\label{cartanPolyakovLoops}

In this section, we will introduce the concept of Cartan-Polyakov loops which will enable one to readily derive quark and gluon models with an arbitrary number of colors. 

Consider the thermal Wilson line defined in Eq. (\ref{def_thermalWilsonLine}), $L$. Allying the unity determinant property of this quantity (it is an element of the $\SU{N_c}$ group) with the fact that, at every spacetime point, this matrix can be diagonalized with a gauge transformation \cite{Pisarski:2002ji}, one can conclude that this quantity has a total of $N_c-1$ degrees of freedom, the same number as the rank of the $\SU{N_c}$ group \cite{Zee:2016fuk}. Knowing the value of $N_c-1$ traces of different powers of this matrix, allows one to build a system of $N_c-1$ equations to the same number of unknown matrix entries. Hence, the traces can be used to parametrize the thermal Wilson line \cite{Hidaka:2009ma}.

As introduced earlier, the Wilson line in the fundamental representation transforms as a field with charge $+1$, under a $\Z{N_c}$ symmetry transformation. The antifundamental representation transforms as a field with charge $-1$. As pointed out in Ref. \cite{Dumitru:2003hp}, the charges are defined modulo $N_c$, meaning that the antifundamental charge, $-1$, is equivalent to the charge $N_c-1$.

One can build Polyakov loops in higher dimensional representations, using combinations of fundamental and antifundamental Wilson lines. Examples include Wilson lines in the adjoint representation of $\SU{N_c}$, two-index symmetric, antisymmetric and other representations, with higher index counts. These different possible representations transform under $\Z{N_c}$ as fields with different charges. For more information on different representations of the Wilson line, see \cite{Dumitru:2003hp,Gupta:2007ax,Abuki:2009dt}. In particular, in Ref. \cite{Dumitru:2003hp}, it was shown how the classification procedure can be made in the large $N_c$ limit.

In $\SU{3}$, the fundamental and antifundamental Polyakov loops constitute a sufficient set from which one can build any other higher representation Polyakov loop. For larger values of $N_c$ this is not the case \cite{Ayyar:2018ppa} and a higher number of degrees of freedom must be considered. In this work we will use Polyakov loops with different charges, in the fundamental representation, in order to easily build effective models containing both gluon and quark degrees of freedom, for an arbitrary $N_c$. More specifically, we will show how to build the Polyakov$-$Nambu$-$Jona-Lasinio model, in the mean field approximation and an effective gluon potential, for increasing values of $N_c$. Such constructions can be very helpful in better understanding the thermodynamics of QCD at large $N_c$, as well as, studying the properties of the conjectured quarkyonic phase in the QCD phase diagram  or determining the behavior of the critical endpoint for increasing $N_c$ \cite{Kovacs:2022zcl}.

We define the Polyakov loop with charge $k$, as the trace of the $k-$power of the thermal Wilson line, in a particular representation, $\Rep$, of the $\SU{N_c}$ group. Polyakov loops with negative charge, are defined analogously, as the trace of the $k-$power of the adjoint thermal Wilson line \cite{Fukushima:2017csk}. Formally, we write:
\begin{align}
\Plrep{\Rep}{k} \qty( \vec{x} ) 
= 
\frac{1}{N_\Rep} \tr \big[ L_\Rep (\vec{x})^k \big] ,
\label{polyakov_loop_charge_q_def}
\\
\Plbarrep{\Rep}{k} \qty( \vec{x} ) 
= 
\frac{1}{N_\Rep} \tr \big[ L_\Rep^\dagger (\vec{x})^k \big] .
\label{polyakov_loop_ADJ_charge_q_def}
\end{align}
Here, $N_\Rep$ is the dimension of the representation. In the literature, the Polyakov loop variable in the fundamental representation is denoted by $\Phi$ and $\bar{\Phi}$, see Eq.(\ref{polyakov.loop.definiton}). Thus, in our notation, the following equivalence holds: $\Phi=\Plrep{\Fund}{1}$ and $\bar{\Phi}=\Plbarrep{\Fund}{1}$. The zero charge Polyakov loop is trivially defined by the normalized trace of the $N_\Rep \times N_\Rep$ identity matrix, $\Plrep{\Rep}{0}=\Plbarrep{\Rep}{0}=1$. In other words, the Polyakov loop with charge $k$ is the normalized trace of a thermal Wilson line wrapped $k$ times around the compact imaginary time dimension \cite{Polchinski:1991tw}:
\begin{align}
\Plrep{\Rep}{k} \qty( \vec{x} ) 
=
\frac{1}{N_\Rep} \tr
\mathcal{P} \exp 
\qty[ i \int_0^{k\beta} \dd{\tau} A_4 (\vec{x},\tau) ] .
\end{align}

Interestingly, for a given number of colors, these objects are not independent of each other. Indeed, a given Polyakov loop with charge $k$ can be written as a function of other Polyakov loops, with different charges. This is connected to a previously mentioned property of the thermal Wilson line: it has a total of $N_c-1$ independent degrees of freedom.  As discussed earlier, one can parametrize the thermal Wilson line by calculating the trace of $N_c-1$ powers of this matrix. Since a Polyakov loop with charge $k$ is simply the trace of the $k-$power of the Wilson thermal line, there is a total of $N_c-1$ independent charges. This immediately poses the question about which charges should be chosen as degrees of freedom to parametrize a phenomenological model.

This motivates the introduction of a special subset of Polyakov loops, which we denote by Cartan-Polyakov loops. For a specific number of colors, $N_c$, we denote the subset of $N_c-1$ Polyakov loops, with charges chosen to be independent of each other, by Cartan-Polyakov loops and the remaining ones, by non-Cartan-Polyakov loops. In this work we define the Cartan-Polyakov loop set to be constituted by Polyakov loops in the fundamental and antifundamental representations of the $\SU{N_c}$ group, $\Plrep{\Fund}{k}$ and $\Plbarrep{\Fund}{k}$, with charges $k=1,\ldots,(N_c-1)/2$. Hence, the Cartan-Polyakov loop set, $\mathcal{C}_{N_c}$, is given by:
\begin{align}
\mathcal{C}_{N_c}
=
\qty{
\Plrep{\Fund}{1},\ldots,\Plrep{\Fund}{(N_c-1)/2},
\Plbarrep{\Fund}{1},\ldots,\Plbarrep{\Fund}{(N_c-1)/2}
} .
\label{CartaPolyakovLoopSet}
\end{align} 
Although this choice is made \textit{a priori} and may seem arbitrary, it is guided by simplicity: although other choices exist\footnote{For instance, one could choose $\Plrep{\Fund}{k}$ and $\Plbarrep{\Fund}{k}$ with $k=(N_c-1)/2 ,\ldots,(N_c-1)$ or, only consider the fundamental representation, $\Plrep{\Fund}{k}$ with $k=1,\ldots,(N_c-1)$.}, this particular one produces straightforward one-to-one mappings from the Cartan loops to non-Cartan loops while, for other choices, this may not be the case. For $N_c=3$, for example, the Cartan loops, are given by the set $\mathcal{C}_{3}=\qty{ \Plrep{\Fund}{1}, \Plbarrep{\Fund}{1} }$. For the cases in which $N_c$ is an even number, we opted to remove the element $\Plbarrep{\Fund}{(N_c-1)/2}$ from the set defined in Eq. (\ref{CartaPolyakovLoopSet}). The reason for this is based on the fact that, for an even number of colors, there is freedom to choose between $\Plrep{\Fund}{(N_c-1)/2}$ and $\Plbarrep{\Fund}{(N_c-1)/2}$. In other words, picking one of them to include in the Cartan set allows one to write the other in terms of the other loops in the Cartan set. In this work, for simplicity, we mainly focus on results for odd $N_c$: for such case there is a symmetry between the number of $\Plrep{\Fund}{k}$ and $\Plbarrep{\Fund}{k}$ contained in the Cartan set.

In the following, we will write recurrence relations, which will allow us to, within a fixed representation of the $\SU{N_c}$ group, decompose a non-Cartan-Polyakov loop with charge $k$, in terms of Polyakov loops with smaller charges, $k-1,k-2,\ldots$ The idea is to use such recurrence relation in order to build effective models of quarks and gluons whose only degrees of freedom are the Cartan-Polyakov loops. For example, when building a model with $N_c=5$, the Cartan loops, are given by the set $\mathcal{C}_{5}=\qty{ \Plrep{\Fund}{1}, \Plrep{\Fund}{2}, \Plbarrep{\Fund}{1},\Plbarrep{\Fund}{2} }$. However, in the construction process, non-Cartan-Polyakov loops such as, $\Plrep{\Fund}{3}$, $\Plrep{\Fund}{4}$, $\Plbarrep{\Fund}{3}$, $\Plbarrep{\Fund}{4}$, will arise. Using the recurrence relation that we will derive in the next section, enables one to write these non-Cartan-Polyakov loops as functions of loops with smaller charge, more specifically the ones present in the set $\mathcal{C}_{5}$.

We note that it may not be always possible to write effective models in terms of Cartan-Polyakov loops. Indeed in some models arises the necessity to involve elements of the thermal Wilson line matrix \cite{Abuki:2009dt}. The mapping between the matrix elements of $L_\Rep$ and the Cartan-Polyakov loops is not straightforward.

\newpage
\subsection{Decomposition of non-Cartan-Polyakov loops}
\label{decompositionNonCartanPolyakovLoops}

In order to derive the non-Cartan-Polyakov loop charge decomposition formula, we will make use of an identity between the coefficients of the characteristic polynomial of an unitary matrix. Consider the characteristic polynomial of the thermal Wilson line in a particular representation, $L_\Rep$. It can be written as (see Appendix \ref{characteristic_polynomial_appendix}):
\begin{align*}
p \qty( L_\Rep,\lambda ) 
& = 
\det \qty[ L_\Rep - \lambda I ] 
\\
& =
\qty(-1)^{N_\Rep}
\qty(
\lambda^{N_\Rep}
+ 
\sum_{i=1}^{{N_\Rep}-1}
c_i ( L_\Rep ) \lambda^{{N_\Rep}-i} 
+
\qty(-1)^{N_\Rep}  \det \qty[ L_\Rep ]
) .
\numberthis
\label{characteristic_polynomial_LRep}
\end{align*}
Here, $c_i ( L_\Rep )$ are coefficients calculated using (see Eq. (\ref{c_k_definition})):
\begin{align}
c_k(L_\Rep)
& = 
\frac{ (-1)^k }{ k! }
\det \qty[ J_{\Rep,k} ] .
\label{ckLRep_def}
\end{align}
The matrix $J_{\Rep,k}$ will be essential in the derivation of the reducing formula. It is a $k \times k$, square matrix defined by (see Eq. (\ref{c_k_definition})): 
\begin{align*}
J_{\Rep,k}
& = 
\mqty(
\tr L_\Rep               & 1       & 0       & 0       & 0      & 0   \\ 
\tr L_\Rep^2               & \tr L_\Rep   & 2       & 0       & 0      & 0   \\ 
\tr L_\Rep^3              & \tr L_\Rep^2   & \tr L_\Rep    & 3       & 0      & 0   \\ 
\vdots                & \vdots  & \vdots  & \vdots  & \ddots & 0   \\ 
\tr L_\Rep^{k-1} & \tr L_\Rep^{k-2} & \tr L_\Rep^{k-3} & \tr L_\Rep^{k-4} & \dots  & k-1 \\ 
\tr L_\Rep^{k}     & \tr L_\Rep^{k-1} & \tr L_\Rep^{k-2} & \tr L_\Rep^{k-3} & \dots  & \tr L_\Rep 
) 
\\
& = 
N_\Rep
\mqty(
\Plrep{\Rep}{1}              & \frac{1}{N_\Rep}                        & 0                        & 0                        & 0      & 0          \\ 
\Plrep{\Rep}{2}     &  \Plrep{\Rep}{1}              & \frac{2}{N_\Rep}                        & 0                        & 0      & 0          \\ 
\Plrep{\Rep}{3}     & \Plrep{\Rep}{2}               & \Plrep{\Rep}{1}              & \frac{3}{N_\Rep}                        & 0      & 0          \\ 
\vdots                   & \vdots                   & \vdots                   & \vdots                   & \ddots & 0          \\ 
\Plrep{\Rep}{k-1} & \Plrep{\Rep}{k-2} & \Plrep{\Rep}{k-3} & \Plrep{\Rep}{k-4} & \dots  & \frac{ k-1 }{N_\Rep}        \\ 
\Plrep{\Rep}{k}     & \Plrep{\Rep}{k-1} & \Plrep{\Rep}{k-2} & \Plrep{\Rep}{k-3} & \dots  & \Plrep{\Rep}{1}
)  .
\numberthis
\label{JRepk_matrix_def}
\end{align*}
$\Plrep{\Rep}{k}$ was defined in Eq. (\ref{polyakov_loop_charge_q_def}). One can also write it in terms of matrix elements, $(J_{\Rep,k})_{i,j}$ as,
\begin{align}
(J_{\Rep,k})_{i,j} 
=
N_\Rep \Plrep{\Rep}{i-j+1} H\qty(i-j) 
+ 
i \delta_{ i,j-1 } .
\end{align}
Here, $H\qty(n)$ is the discrete version of the Heaviside step function, $H\qty(n)=0$ if $n<0$ or $H\qty(n)=1$ if $n\geq 0$. 

We will also need the matrix $\bar{J}_{\Rep,k}$, which has an identical structure to the one defined above but with the replacement, $\Plrep{\Rep}{k} \to \Plbarrep{\Rep}{k} $. Its definition comes from the characteristic polynomial of the adjoint thermal Wilson line, $p ( L_\Rep^\dagger ,\lambda )$, with coefficients given by, $c_k(L_\Rep^\dagger) = \nicefrac{ (-1)^k }{ k! } \det \qty[ \bar{J}_{\Rep,k} ]$.

Excluding the upper triangular section of the matrix $J_{\Rep,k}$, all its entries are Polyakov loops in the representation $\Rep$, with different charges, ranging from $1$ to $k$. Also, the higher charge Polyakov loop, $\Plrep{\Rep}{k}$, is only present in the entry $(k,1)$. All the other entries in the matrix are numbers or Polyakov loops of charge smaller than $k$. The idea is to isolate this quantity from the matrix $J_{\Rep,k}$ and derive a closed expression for $\Plrep{\Rep}{k}$ in terms of Polyakov loops of smaller charge i.e., $\Plrep{\Rep}{k-1}, \Plrep{\Rep}{k-2}, \ldots, \Plrep{\Rep}{1}$ and $\Plbarrep{\Rep}{k-1}, \Plbarrep{\Rep}{k-2}, \ldots, \Plbarrep{\Rep}{1}$.

To this end, consider the Laplace expansion of the determinant on the first column of the matrix $J_{\Rep,k}$. It can be written as:
\begin{align*}
\det \qty[J_{\Rep,k}]
& = 
\sum_{i=1}^k  
\qty(-1)^{i+1} N_\Rep \Plrep{\Rep}{i} \det \qty[J_{\Rep,k}^{(i,1)}] 
\\
& =  
\qty(-1)^{k+1} N_\Rep \Plrep{\Rep}{k} \det \qty[J_{\Rep,k}^{(k,1)}] 
+
N_\Rep
\sum_{i=1}^{k-1}  
\qty(-1)^{i+1} \Plrep{\Rep}{i} \det \qty[J_{\Rep,k}^{(i,1)}] . 
\numberthis
\label{detJ_Repk}
\end{align*}
Here, $J_{\Rep,k}^{(i,j)}$ is the square $(k-1)\times (k-1)$ matrix, obtained by removing the $i-$th line and $j-$th column of matrix $J_{\Rep,k}$. The determinant in the first term, $\det [J_{\Rep,k}^{(k,1)}]$, can be readily calculated by the product of the diagonal entries since the matrix $J_{\Rep,k}^{(k,1)}$ is of the lower triangular type. It yields:
\begin{align}
\det \qty[J_{\Rep,k}^{(k,1)}] 
= 
1 \times 2 \times 3 \times \ldots \times (k-1) 
=
\frac{ k! }{ k } .
\label{detJRepk_(k,1)}
\end{align}
The determinant of the matrix $J_{\Rep,k}$ can then be written as:
\begin{align}
\det \qty[J_{\Rep,k}]
& = 
\qty(-1)^{k+1} N_\Rep \frac{ k! }{ k } \Plrep{\Rep}{k} 
+
N_\Rep
\sum_{i=1}^{k-1}  
\qty(-1)^{i+1} \Plrep{\Rep}{i} \det \qty[J_{\Rep,k}^{(i,1)}] .
\label{detJ_RepkSimplified}
\end{align}
Using the coefficients of the characteristic polynomial of the thermal Wilson line, defined in Eq. (\ref{ckLRep_def}), we can write the determinant in the left-hand side of the above equation as:
\begin{align}
(-1)^{-k} k! c_k(L_\Rep)
& = 
\qty(-1)^{k+1} N_\Rep \frac{ k! }{ k } \Plrep{\Rep}{k} 
+
N_\Rep
\sum_{i=1}^{k-1}  
\qty(-1)^{i+1} \Plrep{\Rep}{i} \det \qty[J_{\Rep,k}^{(i,1)}] .
\label{detJ_RepkSimplified2}
\end{align}
At this point we still cannot solve the equation for $\Plrep{\Rep}{k}$ since the coefficients, $c_k(L_\Rep)$, still strongly depend on $\Plrep{\Rep}{k}$. So, we make use of a special property of these coefficients. If an $N-$dimensional square matrix $M$ is unitary, then the following holds:
\begin{align}
c_j (M) & = \qty(-1)^N \det \qty[ M ] c_{N-j} (M^\dagger),
\label{relation_cj_and_cjdagger_in_paper}
\end{align}
with $j=0,1,2,3,\dots,N-1,N$. Here, $c_j (M)$ and $c_{j} (M^\dagger)$ are the $j-$coefficients of the characteristic polynomial of the matrices $M$ and $M^\dagger$, respectively. This property is shown in Appendix \ref{characteristic_polynomial_appendix}. For our purposes, it will allow us to write $c_k(L_\Rep)$ in terms of $c_{N_\Rep-k}(L_\Rep^\dagger)$. This might not sound as much progress however, the highest charge Polyakov loop present in the coefficient $c_{N_\Rep-k}(L_\Rep^\dagger)$ is $\Plbarrep{\Rep}{N_\Rep-k}$. As long as $N_\Rep-k>0$, all the Polyakov loops present in $c_{N_\Rep-k}(L_\Rep^\dagger)$ will have a smaller charge than $\Plrep{\Rep}{k}$. Hence, applying the identity written in Eq. (\ref{relation_cj_and_cjdagger_in_paper}) to the thermal Wilson line, $L_\Rep$, we can write Eq. (\ref{detJ_RepkSimplified2}) as:
\begin{align}
(-1)^{N_\Rep-k} k! 
c_{N_\Rep-k} (L_\Rep^\dagger)
& = 
\qty(-1)^{k+1} N_\Rep \frac{ k! }{ k } \Plrep{\Rep}{k} 
+
N_\Rep
\sum_{i=1}^{k-1}  
\qty(-1)^{i+1} \Plrep{\Rep}{i} \det \qty[J_{\Rep,k}^{(i,1)}] .
\end{align} 
Additionally, we have used the fact that the thermal Wilson line is an special unitary matrix, $\det \qty[ L_\Rep ]=1$. Finally, we can solve this equation for $\Plrep{\Rep}{k}$ to yield the generating formula to decompose a non-Cartan-Polyakov loop into combinations of loops with smaller charges:
\begin{align*}
\Plrep{\Rep}{k} 
& = 
(-1)^{k+1}
\frac{ k }{ N_\Rep \qty(N_\Rep-k)! }
\det \qty[ \bar{J}_{\Rep,N_\Rep-k} ]
-
\frac{ k }{ k! }
\sum_{i=1}^{k-1}  
\qty(-1)^{i-k} 
\Plrep{\Rep}{i} \det \qty[J_{\Rep,k}^{(i,1)}] .
\numberthis
\label{lk_decomposing_formula_JRk}
\end{align*}
In the above, we used Eq. (\ref{ckLRep_def}) to write, $c_{N_\Rep-k}(L_\Rep^\dagger)
= 
\nicefrac{ (-1)^{N_\Rep-k} }{ \qty(N_\Rep-k)! }
\det \qty[ \bar{J}_{\Rep,N_\Rep-k} ] $ and applied the identity, $(-1)^{2N_\Rep-3k-1} = (-1)^{k+1} $. The most important feature of this equation is that the right-hand side does not depend on the non-Cartan-Polyakov loop of charge $k$. Indeed, both $\det \qty[ \bar{J}_{\Rep,N_\Rep-k} ]$ and $\det \qty[J_{\Rep,k}^{(i,1)}]$, depend only on Polyakov loops with charge smaller than $k$.

One can also derive a very similar formula using the characteristic polynomial of $-L_\Rep$, instead of $L_\Rep$ i.e., $p \qty( -L_\Rep,\lambda ) $. This polynomial can be written as:
\begin{align*}
p \qty( -L_\Rep,\lambda ) 
& = 
\det \qty[ -L_\Rep - \lambda I ] 
\\
& =
\qty(-1)^{N_\Rep}
\qty(
\lambda^{N_\Rep}
+ 
\sum_{i=1}^{{N_\Rep}-1}
c_i ( -L_\Rep ) \lambda^{{N_\Rep}-i} 
+
\qty(-1)^{N_\Rep}  \det \qty[ -L_\Rep ]
) .
\numberthis
\label{characteristic_polynomial_minusLRep}
\end{align*}
Similar to the previous characteristic polynomial, the coefficients are given by:
\begin{align}
c_k(-L_\Rep)
& = 
\frac{ (-1)^k }{ k! }
\det \qty[ R_{\Rep,k} ] .
\label{ckMinusLRep_def}
\end{align}
The  $k \times k$ matrix $R_{\Rep,k}$ used above, is defined by:
\begin{align*}
R_{\Rep,k}
& =
N_\Rep  
\mqty(
-\Plrep{\Rep}{1}              & \frac{1}{N_\Rep}                        & 0                        & 0                        & 0      & 0          \\ 
\Plrep{\Rep}{2}     & -\Plrep{\Rep}{1}              & \frac{2}{N_\Rep}                        & 0                        & 0      & 0          \\ 
- \Plrep{\Rep}{3}     & \Plrep{\Rep}{2}               & -\Plrep{\Rep}{1}              & \frac{3}{N_\Rep}                       & 0      & 0          \\ 
\vdots                   & \vdots                   & \vdots                   & \vdots                   & \ddots & 0          \\ 
(-1)^{k-1} \Plrep{\Rep}{k-1} & (-1)^{k-2} \Plrep{\Rep}{k-2} & (-1)^{k-3} \Plrep{\Rep}{k-3} & (-1)^{k-4} \Plrep{\Rep}{k-4} & \dots  & \frac{ k-1 }{N_\Rep}        \\ 
(-1)^{k} \Plrep{\Rep}{k}     & (-1)^{k-1} \Plrep{\Rep}{k-1} & (-1)^{k-2} \Plrep{\Rep}{k-2} & (-1)^{k-3} \Plrep{\Rep}{k-3} & \dots  & - \Plrep{\Rep}{1}
)  .
\numberthis
\label{RRepk_matrix_def}
\end{align*}
The matrix elements are given by:
\begin{align}
(R_{\Rep,k})_{i,j} 
=
(-1)^{i-j+1} N_\Rep \Plrep{\Rep}{i-j+1} H\qty(i-j) 
+ 
i \delta_{ i,j-1 } .
\end{align}
Following the same steps as before, one can arrive at the following decomposition formula:
\begin{align*}
\Plrep{\Rep}{k} 
& = 
(-1)^{N_\Rep+1} 
\frac{ k }{ N_\Rep (N_\Rep-k)! } \det \qty[ \bar{R}_{\Rep,N_\Rep-k} ] 
-
\frac{ k }{ k! }
\sum_{i=1}^{k-1} \Plrep{\Rep}{i} \det \qty[R_{\Rep,k}^{(i,1)}] .
\numberthis
\label{lk_decomposing_formula_RRk}
\end{align*}
Once again, the matrix $\bar{R}_{\Rep,k}$, has the same structure as the matrix defined in Eq. (\ref{RRepk_matrix_def}) with the replacement: $\Plrep{\Rep}{k} \to  \Plbarrep{\Rep}{k} $.

We highlight that both formulas, given in Eqs. (\ref{lk_decomposing_formula_JRk}) and (\ref{lk_decomposing_formula_RRk}), are only valid for $k<N_\Rep$. In order to decompose non-Cartan loops with charge $k\geq N_\Rep$, one can use the Cayley-Hamilton theorem. Using this theorem, we can write:
\begin{align}
L_\Rep^{N_\Rep}
+ 
\sum_{i=1}^{{N_\Rep}-1}
c_i ( L_\Rep ) L_\Rep^{{N_\Rep}-i} 
+
\qty(-1)^{N_\Rep}  
= 0 .
\label{cayleyHamilton}
\end{align}
Multiplying by $p-$powers of the thermal Wilson line, dividing by the dimension of the representation and taking the trace, yields the following equality:
\begin{align}
\frac{1}{N_\Rep}
\tr[
L_\Rep^{N_\Rep+p}
]
= 
-
\frac{1}{N_\Rep}
\sum_{i=1}^{{N_\Rep}-1}
c_i ( L_\Rep ) 
\tr[
L_\Rep^{{N_\Rep}+p-i} 
]
+
\qty(-1)^{N_\Rep+1} \frac{1}{N_\Rep} 
\tr[ L_\Rep^{p} 
].
\label{cayleyHamilton2}
\end{align}
Hence, recognizing the different Polyakov loops, for $k \geq N_\Rep+p$, we can write:
\begin{align}
\Plrep{\Rep}{k} 
= 
\qty(-1)^{N_\Rep+1}
\Plrep{\Rep}{k-N_\Rep} 
-
\sum_{i=1}^{{N_\Rep}-1}
c_i ( L_\Rep ) 
\Plrep{\Rep}{k-i} 
.
\label{cayleyHamilton3}
\end{align}
This non-Cartan-Polyakov loop decomposition equation is valid for $k \geq N_\Rep$. One can also use the identity given in Eq. (\ref{relation_cj_and_cjdagger_in_paper}) in order to introduce the adjoint Polyakov loops.

In order to show some examples in which Eqs. (\ref{lk_decomposing_formula_JRk}) and (\ref{cayleyHamilton3}) are useful, lets consider Polyakov loops in the fundamental representation of the $\SU{N_c}$ group, with increasing number of colors. Consider first $N_c=3$. For this particular case, the Cartan loops are given by the set $\mathcal{C}_3=\qty{ \Plrep{\Fund}{1}, \Plbarrep{\Fund}{1} }$. Polyakov loops with higher charge (so-called non-Cartan-Polyakov loops) can be written as functions of the loops in this set. The charge 2 loop, $\Plrep{\Fund}{2} $, can be calculated using Eq. (\ref{lk_decomposing_formula_JRk}). It yields:
\begin{align}
\Plrep{\Fund}{2} 
& = 
3
\Plrep{\Fund}{1}^2
-
2 \Plbarrep{\Fund}{1} .
\label{PlFund2Nc3}
\end{align}
For the higher charges, we use Eq. (\ref{cayleyHamilton3}):
\begin{align*}
\Plrep{\Fund}{3}
& = 
1 
- \frac{9}{2} \Plrep{\Fund}{1}^3
+ \frac{9}{2} \Plrep{\Fund}{1} \Plrep{\Fund}{2}
\\
& =
1 
- 9 \Plrep{\Fund}{1} \Plbarrep{\Fund}{1} 
+ 9 \Plrep{\Fund}{1}^3 ,
\numberthis
\\
\Plrep{\Fund}{4}
& = 
\Plrep{\Fund}{1} 
- \frac{9}{2} \Plrep{\Fund}{1}^2 \Plrep{\Fund}{2}
+ \frac{3}{2} \Plrep{\Fund}{2}^2
+ 3 \Plrep{\Fund}{1} \Plrep{\Fund}{3} 
\\
& =
4 \Plrep{\Fund}{1} 
- 36 \Plrep{\Fund}{1} ^2 \Plbarrep{\Fund}{1} 
+ 6 \Plbarrep{\Fund}{1}^2 
+ 27 \Plrep{\Fund}{1}^4 ,
\numberthis
\\
\Plrep{\Fund}{5}
& = 
\Plrep{\Fund}{2} 
-\frac{9}{2} \Plrep{\Fund}{1}^2 \Plrep{\Fund}{3}
+ \frac{3}{2} \Plrep{\Fund}{2} \Plrep{\Fund}{3} 
+ 3 \Plrep{\Fund}{1} \Plrep{\Fund}{4}
\\
& = 
15 \Plrep{\Fund}{1}^2
+ 81 \Plrep{\Fund}{1}^5
- 135 \Plrep{\Fund}{1}^3 \Plbarrep{\Fund}{1}
+ 45 \Plrep{\Fund}{1} \Plbarrep{\Fund}{1}^2
- 5 \Plbarrep{\Fund}{1} ,
\numberthis
\\
& \;\; \vdots
\end{align*}
Naturally, the non-Cartan-Polyakov loops decomposition formula maintains the symmetry charge of the loop. For each decomposed non-Cartan-Polyakov loop, each term in its decomposition, has the same charge as the original non-Cartan-Polyakov loop. In the case of $\Plrep{\Fund}{2}$, for example, both $\Plrep{\Fund}{1}^2$ and $\Plbarrep{\Fund}{1}$ transform as fields of charge $+2$. One can readily obtain the respective adjoint Polyakov loops by replacing, $\Plrep{\Rep}{k} \to \Plbarrep{\Rep}{k}$ and $\Plbarrep{\Rep}{k} \to \Plrep{\Rep}{k}$.

Increasing the number of colors to $N_c=5$ changes the set that constitute the Cartan loops. In this new case the set is: $\mathcal{C}_5=\qty{ \Plrep{\Fund}{1},\Plrep{\Fund}{2},  \Plbarrep{\Fund}{1} ,  \Plbarrep{\Fund}{2} }$. Once again, using Eq. (\ref{lk_decomposing_formula_JRk}), we can decompose the non-Cartan loops, with charge $k<5$, in terms of the Cartan set as follows:
\begin{align*}
\Plrep{\Fund}{3}
& = 
\frac{15}{2} \Plbarrep{\Fund}{1}^2 
- \frac{3}{2} \Plbarrep{\Fund}{2} 
- \frac{25}{2} \Plrep{\Fund}{1}^3 
+ \frac{15}{2} \Plrep{\Fund}{1} \Plrep{\Fund}{2}
,
\numberthis
\label{elleFund3Nc5}
\\
\Plrep{\Fund}{4}
& =
\frac{125}{6}\Plrep{\Fund}{1}^4 
- 4 \Plbarrep{\Fund}{1}
- 25 \Plrep{\Fund}{1}^2 \Plrep{\Fund}{2} 
+ \frac{20}{3} \Plrep{\Fund}{1} \Plrep{\Fund}{3} 
+ \frac{5}{2} \Plrep{\Fund}{2}^2 
\\
& =
50 \Plrep{\Fund}{1} \Plbarrep{\Fund}{1}^2 
- 10 \Plrep{\Fund}{1} \Plbarrep{\Fund}{2} 
- 4 \Plbarrep{\Fund}{1}
- \frac{125}{2}\Pl{1}^4 
+ 25 \Plrep{\Fund}{1}^2 \Plrep{\Fund}{2} 
+ \frac{5}{2} \Plrep{\Fund}{2}^2  .
\numberthis
\label{elleFund4Nc5}
\end{align*}
Again, the higher charge Polyakov loops can be decomposed using Eq. (\ref{cayleyHamilton3}):
\begin{align*}
\Plrep{\Fund}{5}
& = 
1
- \frac{625}{24} \Plrep{\Fund}{1}^5 
+ \frac{625}{12} \Plrep{\Fund}{1}^3 \Plrep{\Fund}{2}
- \frac{125}{6}  \Plrep{\Fund}{1}^2 \Plrep{\Fund}{3} 
- \frac{125}{8}  \Plrep{\Fund}{1} \Plrep{\Fund}{2}^2 
+ \frac{25}{4}   \Plrep{\Fund}{1} \Plrep{\Fund}{4} 
+ \frac{25}{6}   \Plrep{\Fund}{2} \Plrep{\Fund}{3}
\\
& =
1 +
\frac{625}{4} \Plrep{\Fund}{1}^2 \Plbarrep{\Fund}{1}^2
- \frac{125}{4} \Plrep{\Fund}{1}^2 \Plbarrep{\Fund}{2}
- 25 \Plrep{\Fund}{1} \Plbarrep{\Fund}{1}
+ \frac{125}{4} \Plrep{\Fund}{2} \Plbarrep{\Fund}{1}^2
- \frac{25}{4} \Plrep{\Fund}{2} \Plbarrep{\Fund}{2}
- \frac{625}{4} \Plrep{\Fund}{1}^5
+ \frac{125}{4} \Plrep{\Fund}{1} \Plrep{\Fund}{2}^2  ,
\numberthis
\\
\Plrep{\Fund}{6}
& = 
\Plrep{\Fund}{1}
+ 5 \Plrep{\Fund}{1} \Plrep{\Fund}{5}
- \frac{625}{24} \Plrep{\Fund}{2} \Plrep{\Fund}{1}^4
+ \frac{125}{6} \Plrep{\Fund}{1}^3 \Plrep{\Fund}{3}
+ \frac{125}{4} \Plrep{\Fund}{1}^2 \Plrep{\Fund}{2}^2 
\\
& \quad \quad \quad \quad \quad \quad
- \frac{25}{2} \Plrep{\Fund}{1}^2 \Plrep{\Fund}{4}
- \frac{125}{6} \Plrep{\Fund}{1} \Plrep{\Fund}{2} \Plrep{\Fund}{3} 
+ \frac{5}{3} \Plrep{\Fund}{3}^2
+ \frac{15}{4} \Plrep{\Fund}{2} \Plrep{\Fund}{4}
- \frac{25}{8} \Plrep{\Fund}{2}^3
\\
& = 
6 \Plrep{\Fund}{1}
-75 \Plrep{\Fund}{1}^2 \Plbarrep{\Fund}{1}
+ 375 \Plrep{\Fund}{1} \Plbarrep{\Fund}{1}^2 \Plrep{\Fund}{2}
- 75 \Plrep{\Fund}{1} \Plrep{\Fund}{2} \Plbarrep{\Fund}{2}
+ \frac{375}{4} \Plbarrep{\Fund}{1}^4
\\
& \quad \quad \quad \quad \quad \quad
+ \frac{15}{4} \Plbarrep{\Fund}{2}^2
- 15 \Plbarrep{\Fund}{1} \Plrep{\Fund}{2} 
- \frac{75}{2} \Plbarrep{\Fund}{1}^2 \Plbarrep{\Fund}{2}
- \frac{1875}{4} \Plrep{\Fund}{2} \Plrep{\Fund}{1}^4
+ \frac{375}{2}  \Plrep{\Fund}{1}^2 \Plrep{\Fund}{2}^2
+ \frac{25}{4} \Plrep{\Fund}{2}^3 ,
\numberthis
\\
& \;\; \vdots
\nonumber
\end{align*}

The formalism developed earlier allows one to go on decomposing non-Cartan-Polyakov loops for even higher values of $N_c$. Naturally, as $N_c$ increases, the number of terms in each decomposition gets larger and larger. For $N_c=9$, the set is $\mathcal{C}_9=\qty{ \Plrep{\Fund}{1},\Plrep{\Fund}{2},\Plrep{\Fund}{3},\Plrep{\Fund}{4},  \Plbarrep{\Fund}{1} ,  \Plbarrep{\Fund}{2}, \Plbarrep{\Fund}{3} , \Plbarrep{\Fund}{4} }$, for $N_c=11$, one must also join $\Plrep{\Fund}{5}$ and $\Plbarrep{\Fund}{5}$ and so on. The number of terms that appear in the decomposition of non-Cartan-Polyakov loops into Cartan-Polyakov loops naturally increases with the number of colors. For example, for $N_c=11$, the first non-Cartan loop, $\Plrep{\Fund}{6}$, has 17 terms while, for $N_c=21$, the first non-Cartan is $\Plrep{\Fund}{11}$, which contains 97 different terms.

These relations can be checked by hand if one considers a particular gauge choice. In the so-called Polyakov loop gauge, the $A_4$ field in the fundamental representation is diagonal and static \cite{Megias:2004hj} and it can be expressed by the $N_c$ eigenphases $q_i$, $A_4=\diag\qty(iq_1,iq_2,\ldots,iq_{N_c})$ \cite{Lo:2021qkw}. Since $A_4$ is traceless, the constraint $\sum_{i=1}^{N_c} q_i=0$ applies. Thus, in this particular gauge, Polyakov loops of different charges can be written as:
\begin{align}
\Plrep{\Fund}{k} & = 
\frac{1}{N_c} 
\sum_{j=1}^{N_c} \e^{ i k q_{j} }
=
\frac{1}{N_c} 
\qty( 
\sum_{j=1}^{N_c-1} \e^{ i k q_{j} }
+
\e^{ -\sum_{j=1}^{N_c-1} i k q_{j} } 
) ,
\label{PlrepFundPolyGauge}
\\
\Plbarrep{\Fund}{k} 
& = 
\frac{1}{N_c} 
\sum_{j=1}^{N_c} \e^{ -i k q_{j} }
=
\frac{1}{N_c} 
\qty( 
\sum_{j=1}^{N_c-1} \e^{ -i k q_{j} }
+
\e^{ \sum_{j=1}^{N_c-1} i k q_{j} } 
) .
\label{PlbarrepFundPolyGauge}
\end{align}
Using these gauge fixed Polyakov loops in the fundamental representation, one can check the results coming from the non-Cartan-Polyakov decomposition formulas.

\section{Applications}

In this section we show two examples of how to construct models of quarks and gluons with any number of colors. First, we consider the PNJL model in the mean field approximation. Secondly, we consider a glue effective potential.

\subsection{The mean-field Polyakov$-$Nambu$-$Jona-Lasinio model with an arbitrary $N_c$}

The NJL model is widely used as an effective model of QCD due to its simplicity, symmetry properties and ability to display dynamical chiral symmetry breaking in the vacuum and restoration at high temperatures/densities. Some applications of this model include the study of in-medium meson behavior, transport coefficients of quark matter, the equation of state of hybrid neutron stars and the phase diagram of strongly interacting matter with great emphasis on the location of the conjectured critical endpoint. This model however, lacks the ability to describe the confinement-deconfinement transition. In the Polyakov version of the model, a static gluonic background field is coupled to the quarks sector, allowing for the description of both the chiral and deconfinement transitions \cite{Fukushima:2003fw,Mocsy:2003qw,Ratti:2005jh,Hansen:2006ee}.

The Polyakov loop can be included in the model by minimally coupling a background gluon field in the time direction, $A_0 = -iA_4$ \cite{Ratti:2005jh}, to the Lagrangian density via the covariant derivative. It can be defined as:
\begin{align}
\mathcal{L} 
\qty( \psi, \bar{\psi}, A_4 , T )
& = 
\bar{\psi} \qty(i\slashed{D}-\hat{m}) \psi
+ 
\mathcal{L}_\mathrm{int} \qty( \psi, \bar{\psi} )
-
\mathcal{U}_\mathrm{eff}\qty( A_4 , T ) .
\label{generaPNJL_lagrangian}
\end{align}
Here, $\psi$ is the quark field, $D^\mu = \partial^\mu - i \delta_0^\mu A^0 $ is the covariant derivative and $\hat{m}=\diag  \qty{ m_1, m_2, \ldots , m_{N_f} } $ is the quark current mass matrix (with $N_f$ the number of quark flavors). The different quark-quark interactions are contained in the term $\mathcal{L}_\mathrm{int} \qty( \psi, \bar{\psi} )$ and, for our purposes, its exact definite form is not important. This interaction term not only includes the dynamical chiral symmetry breaking 4-quark scalar-pseudoscalar interaction, $\mathcal{L} \supset (\bar{\psi} \lambda_a \psi)^2 + (\bar{\psi} i \gamma_5 \lambda_a \psi)^2$ ($\lambda_a$ are the generators of the $\U{N_f}$ algebra), but it also includes other multi-quark interactions, like the 't Hooft determinant, eight quark-quark interactions, explicit chiral symmetry breaking interactions, vector interactions, etc \cite{Moreira:2018xsp,CamaraPereira:2020rtu,Pereira:2021xxv,Ferreira:2021osk}. The effective potential, $\mathcal{U}_\mathrm{eff}\qty( A_4 , T )$, incorporates the spontaneous breaking of the $\Z{N_c}$ symmetry at some finite temperature. This potential will be discussed in much detail in later sections where we will define a version of it for any number of colors.

The Euclidean generating functional for the PNJL model, can be written as:
\begin{align}
\mathcal{Z}
\qty[ \eta, \bar{\eta} , j_4 ]
\propto 
\int 
\mathcal{D} A_4
\mathcal{D}\overline{\psi} \mathcal{D} \psi \exp 
\qty[ 
- \mathcal{S}_E \qty[ \bar{\psi},\psi, A_4 ]
+ \overline{\psi} \eta  
+ \bar{\eta} \psi 
+A_4 j_4
] . 
\end{align}
Here, $\mathcal{S}_E$ is the Euclidean action of the PNJL model. The path integral over the temporal gluon field $A_4$ is to be made over all of its $N_c-1$ components and $j_4$ is a source for the temporal gluon field.

To deal with the path integral over the quark fields, we apply the mean field approximation. In this scheme, the multi-quark interactions present in $\mathcal{L}_\mathrm{int}$, can be written as the product of quark bilinear operators with the introduction of mean fields, $\expval{\phi_i}$. After the linearization of the product between these operators, a quadratic Lagrangian in the quark fields is obtained and the quark fields can be integrated out from the generating function \cite{CamaraPereira:2020rtu}. The generating functional becomes:
\begin{align}
\mathcal{Z}
\qty[ \expval{\phi_i},  j_4 ]
\propto 
\int 
\mathcal{D} A_4
\exp 
\qty[
- \mathcal{S}_E \qty[ \expval{\phi_i}, A_4 ]
+ A_4 j_4  
] . 
\label{generatingFunctionalPNJL1}
\end{align}
The only thing left to do is to solve the path integral over the temporal gluon field. Different approaches to deal with the path integral over the $A_4$ field, can be found in the literature, ranging from using the Functional Renormalization Group, a Gaussian model or using the Weiss mean field approximation, see \cite{Abuki:2009dt,Haas:2013qwp,Lo:2018wdo}. One can parametrize the $\SU{N_c}$ group volume in terms of the diagonal elements of the $\SU{N_c}$ algebra, i.e., the Cartan elements of the algebra \cite{Rossnerthesis,Hell:2009by,Hellthesis,Sasaki:2012bi}. Hence, we can separate the integration over $A_4$ in Eq. (\ref{generatingFunctionalPNJL1}) in two contributions, one made over the diagonal generators of the $\SU{N_c}$ algebra and another over the non-diagonal generators. One can integrate over the non-diagonal components to yield the so-called Haar measure of the $\SU{N_c}$ group, $H$ \cite{Hell:2008cc,Hell:2009by,DROUFFE19831}:
\begin{align}
H (A_4^{\diag}) =
\frac{ 1 }{ V_{\SU{N_c}} }
\int  \prod_{i \notin \diag } \dd{A_4^{(i)}} .
\end{align}
Here, $A_4^\mathrm{diag}$ is the set of diagonal components of $A_4$. In the already introduced Polyakov gauge, $A_4$ is diagonal and traceless, meaning it can be parametrized in terms of the Cartan elements, $A_4=\sum_{i \in  \diag} A_4^{(i)} t_i $ with $t_i$ the diagonal generators of the $\SU{N_c}$ algebra i.e., the Cartan subalgebra. For $N_c=3$, for instance, we can write $A_4=A_4^{(3)} \lambda_3 + A_4^{(8)} \lambda_8 $, with $\lambda_3$ and $\lambda_8$ the diagonal Gell-Mann matrices. Thus, in Eq. (\ref{generatingFunctionalPNJL1}) one can integrate out the non-diagonal elements of $A_4$ leaving only the path integral over the diagonal components, which contains all the physics of the model \cite{Hell:2008cc,Hell:2009by}:
\begin{align}
\mathcal{Z}
\qty[ \expval{\phi_i},  j_4 ]
\propto 
\int \prod_{j \in \diag } \mathcal{D}{A_4^{(j)}}
\exp 
\qty[
- \mathcal{S}_E \qty[ \expval{\phi_i}, A_4 ]
+ A_4 j_4  
+ \kappa \ln \qty[ H (A_4^{\diag}) ]
] . 
\label{generatingFunctionalDiagA4Integral}
\end{align}
Here, $\kappa$ is some proportionality constant and $H [A_4^{\diag}]$ is the Haar measure, which results from integrating the non-diagonal components of $A_4$. The Haar measure can be absorbed on the definition of the Polyakov effective potential, $\mathcal{U}_\mathrm{eff}$. As a matter of fact, the Haar measure has been understood to be a crucial part of the potential by favoring the confined phase \cite{Fukushima:2017csk}.

At this point one needs to solve the path integral over the diagonal components of the field $A_4$. The standard approach is to consider that the functional integration is dominated by the stationary point: quantum fluctuations are neglected and only the classical configuration contributes to the path integral \cite{Roessner:2006xn,Pereira:2021xxv}. Thus, the functional integration can be dropped and  the diagonal components of the $A_4$ field can be treated as mean fields. The values of these fields are calculated by requiring the effective action to be stationary, $\dv*{ \mathcal{S}_E }{ A_4^{(i)} } =0$, with $A_4^{(i)}$ the Cartan elements.

Usually, for $N_c=3$, the requirements of stationarity of the effective action with respect to the diagonal components of the $A_4$ field are replaced by stationarity with respect to the values of the charge one Polyakov loop and its adjoint in the fundamental representation, $\Phi$ and $\bar{\Phi}$ \cite{Roessner:2006xn}. Since these quantities are functions of $A_4^{\diag}$, $\Phi=\Phi[A_4^{\diag}]$ and $\bar{\Phi}=\bar{\Phi}[A_4^{\diag}]$, changing variables is allowed as long as the Jacobian of the transformation is non-singular \cite{Haag:1958vt,CHISHOLM1961469,KAMEFUCHI1961529}. For other values of $N_c$, $\Plrep{\Fund}{1}=\Phi$ and $\Plbarrep{\Fund}{1}=\bar{\Phi}$ are not enough degrees of freedom to express the Haar measure. Thus, for other values of $N_c$, we will consider that these change of variables are allowed and replace the variables of the model: instead of using the set of Cartan elements of the $A_4$ field, we will use the Cartan-Polyakov loop set defined in Eq. (\ref{CartaPolyakovLoopSet}).

Within the mean field approximation, the Lagrangian density of the PNJL model can be written as the one of fermionic quasiparticles with an effective mass, $M$, chemical potential, $\mu$, and effective mean field potential, $U$. Using the Matsubara formalism, one can calculate the thermodynamic potential of the model at finite temperature and chemical potential, $\Omega \qty(N_c, T,\mu)$ \cite{Fukushima:2003fw,Ratti:2005jh}. It is given by:
\begin{align*}
\Omega (N_c, T,\mu)
& =
U(\expval{\phi_i}) + 
\mathcal{U}_\mathrm{eff}( A_4^{\diag} , T )
\\
& 
- 2T
\tr
\int_{\mathrm{reg}}
\frac{\dd[3]{p}}{ \qty(2 \pi)^3 }
\qty{
\beta N_c E 
+
\ln 
\det \qty[ 1 + L_\Fund \e^{ -\beta \qty( E - \mu ) } ] 
+
\ln 
\det \qty[ 1 + L_\Fund^\dagger \e^{ -\beta \qty( E + \mu ) } ]
} .
\numberthis
\label{thermoPotPNJLNc3}
\end{align*}
Here, $E = \sqrt{p^2+M^2}$ and $\beta$ is the inverse temperature, $\beta={T}^{-1}$. The potential $U(\expval{\phi_i})$ contains the contribution coming from the mean fields, $\expval{\phi_i}$. The trace must be calculated over flavor indexes while the determinant is calculated over color space. The mean fields are fixed by requiring for thermodynamic consistency, $\dv*{\Omega}{ \expval{\phi_i} } = 0$ \cite{Buballa:2003qv}. The subscript ``reg'' in the integration, serves only as an acknowledgment for the fact that the PNJL model has divergent integrals and some regularization scheme must be employed to render these integrals finite. Some techniques include the 3-momentum or the Pauli-Villars regularization schemes. 

Until now, we have not yet specified the number of colors. In order to obtain the thermodynamic potential for a particular number of colors, one must write the effective Polyakov loop potential and calculate the determinant inside the logarithms, for the chosen $N_c$.

\subsubsection{The fermionic determinant with an arbitrary $N_c$}

Lets start by evaluating the fermionic contribution for a given number of colors. The fermionic determinants are given by:
\begin{align}
\det \qty[ 1 + L_\Fund \e^{ -\beta \qty( E - \mu ) } ] 
=
h_-^{N_c} \det \qty[ h_-^{-1} I + L_\Fund ]   ,
\label{determinantLF}
\\
\det \qty[ 1 + L_\Fund^\dagger \e^{ -\beta \qty( E + \mu ) } ]
=
h_+^{N_c} \det \qty[ h_+^{-1} I + L_\Fund^\dagger ] .
\label{determinantLFdagger}
\end{align}
Here, we have defined the non-zero quantities:
\begin{align}
h_\pm = \e^{ -\beta \qty( E \pm {\mu} ) } .
\label{h_pm_definition}
\end{align}
The goal is to calculate the determinants for an arbitrary $N_c$ using only the corresponding set of Cartan-Polyakov loops. To this end, we will write the determinants in Eqs. (\ref{determinantLF}) and (\ref{determinantLFdagger}) in their respective characteristic polynomials. These polynomials will have $(N_c-1)$ coefficients, $c_k(-M)$, that have to be calculated. Calculating the $k-$th coefficient, implies the calculation of another determinant, as laid out in Eq. (\ref{ckLRep_def}). Considering the characteristic polynomial $p \qty( -L_\Rep, h_-^{-1} ) $ (see Eq. (\ref{characteristic_polynomial_minusLRep})), we can write Eq. (\ref{determinantLF}) as:
\begin{align*}
h_-^{N_c} \det \qty[ L_\Fund + h_-^{-1} I ] 
& =
1
+ 
\sum_{k=1}^{{N_c}-1}
c_k ( -L_\Fund ) h_-^{k} 
+
h_-^{N_c} .
\numberthis
\label{determinantLF2}
\end{align*}
We used the fact the dimension of the fundamental representation is, $N_\Fund=N_c$ and that $\det \qty[ -L_\Fund ] = \qty(-1)^{N_c}$.

In the sum, the difficulty of calculating the determinant increases with the number of colors. For a fixed number of colors, calculating the $k-$th coefficient in the sum, amounts to solving the determinant of the $R_{\Rep,k}$ matrix, see Eq. (\ref{ckMinusLRep_def}). Not only that, the $k-$th coefficient in the sum will necessarily contain Polyakov loops, of charge $\qty[1,\ldots,k]$. This can be seen by observing the structure of the matrix $R_{\Rep,k}$, defined in Eq. (\ref{RRepk_matrix_def}). As introduced in this work, by definition, the Cartan-Polyakov loops have charge in the interval $\qty[1,\ldots,(N_c-1)/2]$. Hence, the coefficients in the sum, with $k=(N_c-1)/2+1,\ldots,N_c-1$, will unavoidably contain non-Cartan-Polyakov loops.  Since the goal is to write the model exclusively as a function of Cartan-Polyakov loops, this feature poses the difficult task of expanding the determinant of Eq. (\ref{determinantLF2}) and then applying the charge decomposition formula, given in Eq. (\ref{lk_decomposing_formula_JRk}), to systematically decompose all non-Cartan-Polyakov loops into Cartan ones. In order to solve this issue, we will use a trick to simplify the practical calculations while, at the same time, automatically re-write the sum as a function of Cartan-Polyakov loops only. For simplicity, in the following discussion, we will focus on the case in which the number of colors is odd. The even number of colors case is a straightforward extension.

Consider the identity given in Eq. (\ref{relation_cj_and_cjdagger_in_paper}), which states that we can relate  the coefficient $c_k(-L_\Fund)$, with the coefficient $c_{N_c-k}(-L_\Fund^\dagger)$. If we employ this identity, we can break the original sum in Eq. (\ref{determinantLF2}) in the middle, in two distinct sums and write,
\begin{align}
\sum_{k=1}^{{N_c}-1}
c_k ( -L_\Fund ) h_-^{k} 
=
\sum_{k=1}^{(N_c-1)/2}
c_k ( -L_\Fund ) h_-^{k} 
+ 
\sum_{k=(N_c+1)/2}^{{N_c}-1}
c_{N_c-k} ( -L_\Fund^\dagger ) h_-^{k} ,
\label{breakingSum}
\end{align}
where we have used $c_j (-L_\Fund) = c_{N-j} (-L_\Fund^\dagger) $. One can check that both the first and second sums, on the right-hand side of this equation, contains coefficients that will only produce Polyakov loops with charges in the interval $\qty[1,\ldots,(N_c-1)/2]$. This range of charges is the same as the one which defines the Cartan-Polyakov loop set. Hence, the determinant can be written only as a function of Cartan-Polyakov loops.

From an analytically point of view, another nice feature of this trick is that by employing it, one only needs to calculate half of the coefficients. At first glance, one has to calculate $N_c-1$ coefficients, starting with the coefficient with $k=1$ up to $k=N_c-1$. However, one can calculate the coefficients, $c_k ( -L_\Fund )$, up to $k=(N_c-1)/2$ and the remaining coefficients, related to $c_k ( -L_\Fund^\dagger )$, can be obtained from the substitution, $\Plrep{\Fund}{k} \to \Plbarrep{\Fund}{k}$.

Thus, for an odd number of colors, we can re-write the sum using Eq. (\ref{breakingSum}), and write the determinant in Eq. (\ref{determinantLF}) as:
\begin{align*}
\det \qty[ 1 + L_\Fund \e^{ -\beta \qty( E - \mu ) } ] 
& =
1
+ 
\sum_{k=1}^{(N_c-1)/2}
c_k ( -L_\Fund ) h_-^{k} 
+ 
\sum_{k=(N_c+1)/2}^{{N_c}-1}
c_{N_c-k} ( -L_\Fund^\dagger ) h_-^{k} 
+
h_-^{N_c} .
\numberthis
\label{determinantLF3}
\end{align*}
For the case in which $N_c$ is an even number, the idea is the same with the small caveat that one of the sums will have an additional term. 

The calculation of the determinant, defined in Eq. (\ref{determinantLFdagger}), follows the same steps. One gets the same result as the one shown in Eq. (\ref{determinantLF3}) with the following replacements: $-\to+$, $L_\Fund \to L_\Fund^\dagger$ and $L_\Fund^\dagger \to L_\Fund$.

The explicit calculation of the coefficients follows from Eq. (\ref{ckMinusLRep_def}). Hence, one can readily calculate the determinants defined in Eqs. (\ref{determinantLF}) and (\ref{determinantLFdagger}), for an arbitrary $N_c$. The problem of writing the fermionic part of the PNJL model for an arbitrary number of colors is reduced to calculating the $c_k(-L_\Fund)$ coefficients up to $(N_c-1)/2$. The first coefficients are:
\begin{align}
c_1(-L_\Fund) & = 
N_c \Plrep{\Fund}{1} ,
\\
c_2(-L_\Fund) & = 
\frac{N_c}{2} 
\qty( 
N_c \Plrep{\Fund}{1}^2 
- 
\Plrep{\Fund}{2} 
) ,
\\ 
c_3(-L_\Fund) & =
\frac{N_c}{6} 
\qty( 
N_c^2 \Plrep{\Fund}{1}^3 
- 3 N_c \Plrep{\Fund}{1} \Plrep{\Fund}{2} 
+ 2 \Plrep{\Fund}{3}
) ,
\\
c_4(-L_\Fund) & =
\frac{N_c}{24} 
\qty( 
N_c^3 \Plrep{\Fund}{1}^4 
- 6 N_c^2 \Plrep{\Fund}{1}^2 \Plrep{\Fund}{2}  
+ 3 N_c \Plrep{\Fund}{2}^2 
+ 8 N_c \Plrep{\Fund}{1} \Plrep{\Fund}{3}
- 6 \Plrep{\Fund}{4}
) ,
\\
& \;\; 
\vdots \nonumber
\end{align}
As an example, for $N_c=9$, the fermionic determinant containing $L_\Fund$, is given by:
\begin{align*}
\det \qty[ 1 + L_\Fund \e^{ -\beta \qty( E - \mu ) } ] 
& =  
1  
\\
& + 
9
\Plrep{\Fund}{1}
\e^{ -\beta \qty( E - \mu ) }
\\
& + 
\frac{9}{2} \qty( 9 \Plrep{\Fund}{1}^2 - \Plrep{\Fund}{2} ) 
\e^{ -2 \beta \qty( E - \mu ) }
\\
& + 
\frac{3}{2} \qty( 81 \Plrep{\Fund}{1}^3 - 27 \Plrep{\Fund}{1} \Plrep{\Fund}{2} + 2 \Plrep{\Fund}{3} ) 
\e^{ -3\beta \qty( E - \mu ) }
\\
& + 
\frac{9}{8} \qty( 243 \Plrep{\Fund}{1}^4 - 162 \Plrep{\Fund}{1}^2 \Plrep{\Fund}{2} + 9 \Plrep{\Fund}{2}^2 + 24 \Plrep{\Fund}{1} \Plrep{\Fund}{3} - 2 \Plrep{\Fund}{4} ) 
\e^{ -4\beta \qty( E - \mu ) }
\\
& + 
\frac{9}{8} \qty( 243 \Plbarrep{\Fund}{1}^4 - 162 \Plbarrep{\Fund}{1}^2 \Plbarrep{\Fund}{2}  + 9 \Plbarrep{\Fund}{2}^2 + 24 \Plbarrep{\Fund}{1} \Plbarrep{\Fund}{3} - 2 \Plbarrep{\Fund}{4} ) 
\e^{ -5\beta \qty( E - \mu ) }
\\
& + 
\frac{3}{2} \qty( 81 \Plbarrep{\Fund}{1}^3 - 27 \Plbarrep{\Fund}{1} \Plbarrep{\Fund}{2} + 2 \Plbarrep{\Fund}{3} ) 
\e^{ -6\beta \qty( E - \mu ) }
\\
& + 
\frac{9}{2} \qty( 9 \Plbarrep{\Fund}{1}^2 - \Plbarrep{\Fund}{2} ) 
\e^{ -7\beta \qty( E - \mu ) }
\\
& + 
9 
\Plbarrep{\Fund}{1}
\e^{ -8\beta \qty( E - \mu ) }
\\
& + 
\e^{ -9\beta \qty( E - \mu ) } .
\numberthis
\end{align*}
The determinant containing $L_\Fund^\dagger$, can be obtained from the above expression with the appropriate substitutions: $\Plrep{\Fund}{k} \to \Plbarrep{\Fund}{k}$, $\Plbarrep{\Fund}{k} \to \Plrep{\Fund}{k}$ and $\mu \to -\mu$.

\subsubsection{ The Polyakov loop effective potential with an arbitrary $N_c$ }
\label{polyakovEffectivePotentialPNJLArbitraryNc}

We now turn our attention to the effective Polyakov loop potential, $\mathcal{U}_\mathrm{eff}$, which can be written using the Ginzburg$-$Landau theory of phase transitions. Within this approach, the potential must respect the symmetry properties of the original theory.

For $N_c=3$, one can build this potential using the charge one Polyakov loop in the fundamental representation (and its conjugate), $\Plrep{\Fund}{1}=\Phi$ and $\Plbarrep{\Fund}{1}=\bar{\Phi}$. Different phenomenological potentials were introduced in the literature including the logarithmic, polynomial and polynomial-logarithmic. They all share a kinetic term, proportional to $\Phi \bar{\Phi}$ and higher order terms that are essential to build a bounded potential, with the correct symmetries (the kinetic term introduces an additional $\U{1}$ symmetry which must be removed by other terms \cite{Rossnerthesis,Hellthesis,Rainerthesis}), which reproduces a first-order phase transition at some finite temperature. 

In the logarithm potential, the non-quadratic term is given by the logarithm of the Haar measure. As already seen, this contribution naturally appears when one integrates out the non-diagonal components of the $A_4$ field \cite{Fukushima:2003fw,Hell:2008cc}. This potential, for $N_c=3$, is widely defined as \cite{Fukushima:2003fw,Roessner:2006xn,Costa:2010zw}:
\begin{align}
\frac{ \mathcal{U}_\mathrm{eff}\qty( \Phi, \bar{\Phi} , T ) }{ T^4 } 
= 
-\frac{1}{2} a\qty( T ) \bar{\Phi} \Phi 
+ 
b\qty( T ) \ln \qty[ 1 - 6 \bar{\Phi} \Phi + 4 \qty( \bar{\Phi}^3 + \Phi^3  ) - 3 \qty( \bar{\Phi} \Phi )^2   ] .
\label{logarithmicPolyakovPotential}
\end{align} 
The quantity inside the logarithm is the Haar measure for $N_c=3$. The temperature dependent parameters \cite{Roessner:2006xn,Costa:2010zw} are:
\begin{align}
a\qty( T ) 
& = 
a_0 + a_1 \qty( \frac{T_0}{T} ) + a_2 \qty( \frac{T_0}{T} )^2 ,
\label{eq:a}
\\
b\qty( T ) 
& =  
b_3 \qty( \frac{T_0}{T} )^3 .
\label{eq:b}
\end{align} 
The parameters $T_0$, $a_0$, $a_1$, $a_2$ and $a_3$ are fixed by reproducing lattice QCD results \cite{Boyd:1996bx,Kaczmarek:2002mc} with $T_0$ fixing the temperature of the phase transition \cite{Ratti:2005jh}. The parameter $a_0$ is fixed by requiring the Steffan-Boltzmann limit when $T \to \infty$ while the parameter $b_3$ is fixed by requiring the second-order phase transition to occur at $T=T_0$ \cite{Ratti:2005jh,Roessner:2006xn}. The gluon contribution to the Steffan-Boltzmann pressure is $\nicefrac{ P_\mathrm{SB} }{ T^4 } 
= \nicefrac{ 2 ( N_c^2 - 1 ) \pi^2 }{ 90 }$. Hence, the $a_0$ parameter can be readily defined at diverging temperatures, $T \to \infty$, $a_0=\nicefrac{ 4 ( N_c^2 - 1 ) \pi^2 }{ 90 }$. Thus, one can conclude that this parameter is $N_c$ dependent. Since $b_3$ can be calculated from the remaining parameters, including $a_0$, it is also $N_c$ dependent. When building an effective Polyakov loop potential for any number of colors, this has to be taken into account.

For $N_c=3$, following lattice QCD calculations at zero chemical potential, the $T_0$ parameter is usually fixed to $270$ MeV \cite{Karsch:2001cy,Borsanyi:2012ve}. Different approaches have been used in the literature to fix the value of $T_0$, specially when studying systems at finite chemical potential. As a matter or fact, in Ref. \cite{Schaefer:2007pw}, the parameter $T_0$ was given an explicit dependence on the number of flavors of quarks considered, due to renormalization group arguments.

The polynomial potential is very similar with the major difference that higher powers of $\Plrep{\Fund}{1}=\Phi$ and $\Plbarrep{\Fund}{1}=\bar{\Phi}$ are incorporated in the potential \cite{Pisarski:2000eq,Ratti:2005jh}. The polynomial-logarithmic potential, as stated in the name, is a mixture between the logarithmic and polynomial potentials \cite{Lo:2013hla}. Another important property of the logarithmic potential is its diverging nature in the limit $\Plrep{\Fund}{1}=\Phi \to 1$ and $\Plbarrep{\Fund}{1}=\bar{\Phi} \to 1$. This ensures that the Polyakov loops do not exceed unity \cite{Roessner:2006xn}.

The question that now arises is the following: how to define an effective Polyakov loop potential for any given value of the number of colors? The studies performed in Refs. \cite{McLerran:2008ua,Buisseret:2011ms}, built effective Polyakov loop potentials for any number of colors based on the polynomial potential in order to study the phase diagram at large $N_c$. However, such potentials were written only as a function of $\Plrep{\Fund}{1}=\Phi$ and $\Plbarrep{\Fund}{1}=\bar{\Phi}$ and Polyakov loops with larger charges were neglected. At large $N_c$, the fundamental loop with charge one may indeed be the most important degree of freedom however, in this work, we want to write a general expression for any $N_c$, without any \textit{a priori} approximations (apart from the mean field one). After building such model, then one may perform a large $N_c$ approximation.

In this work, we propose to simply extend the logarithmic potential to any value of $N_c$. Using the polynomial potential, would mean having to consider different interaction terms between different Cartan-Polyakov loops in such a way to build $\Z{N_c}$ invariant and $\U{1}$ breaking interactions. The Haar measure automatically gives interaction terms which respect the symmetries of the model, yielding an appropriate effective potential at any number of $N_c$. To this end we consider that the potential can be written solely in terms of the previously introduced Cartan-Polyakov loops. Hence, for the kinetic part of the potential we simply consider a sum over quadratic terms of the Cartan-Polyakov loop set:
\begin{align}
\sum_{i=1}^{\frac{N_c-1}{2}}
\Plrep{\Fund}{i}
\Plbarrep{\Fund}{i}.
\label{kineticPartPolyakovLoopEffectivePotential}
\end{align}
This term is invariant under $\Z{N_c}$ and under an unwanted $\U{1}$ symmetry which must be removed by the remaining contributions. The non-quadratic part of the potential will be given by the Haar measure, for a given $N_c$, written exclusively in terms of Cartan-Polyakov loops. In a later discussion, we will cover the details of how one can accomplish this task. Hence, the effective Polyakov loop potential, with an arbitrary $N_c$, can be expressed as:
\begin{align}
\frac{ \mathcal{U}_\mathrm{eff}\qty( \mathcal{C}_{N_c} , T ) }{ T^4 } 
=
-\frac{1}{2} A\qty(N_c, T ) 
\sum_{i=1}^{ \frac{N_c-1}{2} }
\Plrep{\Fund}{i}
\Plbarrep{\Fund}{i}
+ 
B\qty( N_c, T ) 
\ln \qty[ H( \mathcal{C}_{N_c} ) ] .
\label{logarithmicPolyakovPotentialNc}
\end{align}
Here, we define the temperature and color dependent parameters $A\qty(N_c, T ) $ and $B\qty( N_c, T ) $, in the same way as it was done for the temperature parameters $a(T)$ and $b(T)$ in Eqs. (\ref{eq:a}) and (\ref{eq:b}):
\begin{align}
A\qty(N_c, T ) 
& = 
A_0(N_c) + A_1(N_c) \qty[ \frac{T_0}{T} ] + A_2(N_c) \qty[ \frac{T_0}{T} ]^2 ,
\\
B\qty( N_c, T ) 
& =  
B_3(N_c ) \qty[ \frac{T_0}{T} ]^3 .
\end{align} 
As before, the $A_0(N_c)$ parameter have to be fixed in such a way to recover the Steffan-Boltzmann limit for a gas of gluons with $\SU{N_c}$ symmetry. The remaining parameters must ensure that a second-order phase transition occurs at $T_0$ at least for $N_c=3$. At higher values of $N_c$ other behaviors might be expected \cite{Lucini:2013qja}. We did not introduce a color dependence in the $T_0$ parameter. Some lattice results, seem to indicate that $T_0$ very weakly depends on the number of colors \cite{Lucini:2002ku,Lucini:2003jp,Liddle:2008kk,Lucini:2013qja}. Indeed, in such studies, the deconfinement temperature $T_0$, as a function of the number of colors, was found to be,
\begin{align}
T_0 (N_c)
& = 
\sqrt{\sigma}
\qty(
\alpha_1 + \frac{ \alpha_2 }{ N_c^2 }
),
\end{align}
where, $\sigma$ is the $N_c-$independent string tension. In \cite{Lucini:2002ku,Lucini:2003jp}, the values obtained were $\alpha_1=0.582(15)$ and $\alpha_2=0.43(13)$, in Ref. \cite{Liddle:2008kk}, $\alpha_1=0.9026(23)$ and $\alpha_2=0.880(43)$ while, in Ref. \cite{Lucini:2012wq}, $\alpha_1=0.5949(17)$  and  $\alpha_2=0.458(18)$.

The exact functional dependencies, on the number of colors, of the different parameters present in the effective Polyakov loop potential, is beyond the scope of the present work. In order to find such dependencies, one can perform phenomenological studies involving results coming from lattice QCD with $N_c \geq 2$. The current status of the studies seems to agree that the transition of pure Yang-Mills theory is of second order for both $N_c=2$ and $N_c=3$, while it is of first-order for $N_c \geq 4$ \cite{Lucini:2013qja}. The most straightforward  way to give some $N_c$ dependence to the parameters would be to simply consider the Steffan-Boltzmann requirement enforced on $A_0$ and fix the remaining parameters at $N_c=3$.

We now turn our attention to evaluating the argument inside the logarithm of Eq. (\ref{logarithmicPolyakovPotentialNc}), $H( \mathcal{C}_{N_c} )$, for a given $N_c$. Since we wrote the fermionic determinant and the kinetic part of the potential, using Cartan-Polyakov loops as degrees of freedom, we also want to write the non-quadratic part of the potential in terms of such variables. In the following, we show how one can write the argument of the logarithm in Eq. (\ref{logarithmicPolyakovPotentialNc}), the Haar measure, in terms of Cartan-Polyakov loops. The Haar measure is given by \cite{Zhang:2010kn,Fukushima:2017csk,haarMeasureBrauner,Lo:2021qkw,DROUFFE19831}:
\begin{align}
H(q_1,q_2,\ldots,q_{N_c}) 
=
\prod_{1 \le i < j	 }^{N_c}
\abs{ \e^{ i q_i } - \e^{ i q_j } }^2 .
\end{align}
Here, $q_i$ are $N_c$ eigenphases \cite{Lo:2021qkw}. This measure can be written using the Vandermonde determinant. The Vandermonde matrix, $V$, of a set of variables $z_1,\ldots,z_m$, is defined as \cite{haarMeasureBrauner}:
\begin{align}
V (z_1,\ldots,z_m)
=
\mqty
(
1 & z_1 & z_1^2 & \dots & z_1^{n-1}\\
1 & z_2 & z_2^2 & \dots & z_2^{n-1}\\
1 & z_3 & z_3^2 & \dots & z_3^{n-1}\\
\vdots & \vdots & \vdots & \ddots &\vdots \\
1 & z_m & z_m^2 & \dots & z_m^{n-1}
) .
\end{align}
The matrix elements of this quantity can readily be written:
\begin{align}
\qty(V)_{ij}
& =
z_i^{j-1} .
\label{vandermondeMatrixElements}
\end{align}
The determinant of this matrix, known as the Vandermonde determinant, can be calculated to yield:
\begin{align}
\det \qty[V ( z_1, \ldots, z_{N_c} ) ] 
=
\prod_{1 \le i < j \le N_c}
\qty( z_j - z_i ) .
\end{align}
The relation between the Vandermonde determinant and the Haar measure is\footnote{
We used the following property:
\begin{align*}
\abs{ \det \big[ V \big] }^2
=
\det \big[ V^* \big] 
\det \big[ V\big] 
=
\det \big[ V^\dagger \big] 
\det \big[ V \big] 
=
\det \big[ V^\dagger V \big].
\end{align*}
}:
\begin{align*}
H(q_1,q_2,\ldots,q_{N_c}) 
& =
\abs{ \det 
\big[ 
V( \e^{ i q_1 }, \ldots, \e^{ i q_{N_c} } ) 
\big] }^2
\\
& =
\det 
\big[ 
V^\dagger ( \e^{ i q_1 }, \ldots, \e^{ i q_{N_c} } ) 
V( \e^{ i q_1 }, \ldots, \e^{ i q_{N_c} } ) 
\big] 
\\
& =
\det \big[ W(q_1,\ldots,q_{N_c}) \big] .
\numberthis
\end{align*}
Where, $W(q_1,\ldots,q_{N_c}) = V^\dagger ( \e^{ i q_1 }, \ldots, \e^{ i q_{N_c} } ) 
V( \e^{ i q_1 }, \ldots, \e^{ i q_{N_c} } ) $. The matrix elements of $W(q_1,\ldots,q_{N_c})$ are given by:
\begin{align*}
\qty( W )_{ij}
& = 
\sum_{l=1}^{N_c}
\qty(V^\dagger)_{il} \qty(V)_{lj}
\\
& = 
\sum_{l=1}^{N_c}
\qty(V^*)_{li} \qty(V)_{lj}
\\
& = 
\sum_{l=1}^{N_c}
{z^*_l}^{i-1}
z_l^{j-1}
\\
& = 
\sum_{l=1}^{N_c}
z_l^{j-i} .
\numberthis
\end{align*}
Here, $z_l = \e^{ i q_l }$. Recognizing that the diagonal entries are equal to $N_c$ and that the non-diagonal entries of this matrix are the fundamental Polyakov loops with different charges (see Eqs. (\ref{PlrepFundPolyGauge}) and (\ref{PlbarrepFundPolyGauge})), one can write the matrix $W$ as \cite{haarMeasureBrauner}:
\begin{align*}
W
& =
\mqty
(
N_c & \sum_{l=1}^{N_c}z_l & \sum_{l=1}^{N_c}z_l^{2} & \dots & \sum_{l=1}^{N_c} z_l^{N_c-1} \\
\sum_{l=1}^{N_c} \bar{z}_l & N_c & \sum_{l=1}^{N_c} z_l & \dots & \sum_{l=1}^{N_c} z_l^{N_c-2}\\
\sum_{l=1}^{N_c} \bar{z}_l^{2} & \sum_{l=1}^{N_c} \bar{z}_l & N_c & \dots & \sum_{l=1}^{N_c} z_l^{N_c-3}\\
\vdots & \vdots & \vdots & \ddots &\vdots \\
\sum_{l=1}^{N_c} \bar{z}_l^{N_c-1} & \sum_{l=1}^{N_c} \bar{z}_l^{N_c-2} & \sum_{l=1}^{N_c} \bar{z}_l^{N_c-3} & \dots & N_c
) 
\\
& = 
N_c
\mqty
(
1 & \Plrep{\Fund}{1} & \Plrep{\Fund}{2} & \dots & \Plrep{\Fund}{N_c-1} \\
\Plbarrep{\Fund}{1} & 1 & \Plrep{\Fund}{1} & \dots & \Plrep{\Fund}{N_c-2}\\
\Plbarrep{\Fund}{2} & \Plbarrep{\Fund}{1} & 1 & \dots & \Plrep{\Fund}{N_c-3}\\
\vdots & \vdots & \vdots & \ddots &\vdots \\
\Plbarrep{\Fund}{N_c-1} & \Plbarrep{\Fund}{N_c-2} & \Plbarrep{\Fund}{N_c-3} & \dots & 1
) .
\numberthis
\end{align*}
Thus, the Haar measure is exclusively written in terms of Polyakov loops of different charges (and its adjoints), in the fundamental representation. It is given by:
\begin{align}
H(
\Plrep{\Fund}{1},\ldots,\Plrep{\Fund}{N_c-1},
\Plbarrep{\Fund}{1},\ldots,\Plbarrep{\Fund}{N_c-1}
) 
& =
\det \qty[ W ] .
\label{haarMeasureCartanAndNonCartanLoops}
\end{align}

Using the recursion formula in Eq. (\ref{lk_decomposing_formula_JRk}), it is possible to write the Haar measure above using only the set of Cartan-Polyakov loops, $\mathcal{C}_{N_c}$. Hence, $
H(
\Plrep{\Fund}{1},\ldots,\Plrep{\Fund}{N_c-1},
\Plbarrep{\Fund}{1},\ldots,\Plbarrep{\Fund}{N_c-1}
) 
=
H( \mathcal{C}_{N_c} ) $. For $N_c=3$, for example, we can write the Haar measure as:
\begin{align*}
H(
\Plrep{\Fund}{1},
\Plbarrep{\Fund}{1}
) 
& =
27
\mqty
|
1 & \Plrep{\Fund}{1} & \Plrep{\Fund}{2}       \\
\Plbarrep{\Fund}{1} & 1 & \Plrep{\Fund}{1}    \\
\Plbarrep{\Fund}{2} & \Plbarrep{\Fund}{1} & 1 \\
|  .
\numberthis
\end{align*}
Using Eq. (\ref{PlFund2Nc3}), we can write $\Plrep{\Fund}{2}=\Plrep{\Fund}{2}(\Plrep{\Fund}{1},\Plbarrep{\Fund}{1})$, $\Plbarrep{\Fund}{2}=\Plbarrep{\Fund}{2}(\Plrep{\Fund}{1},\Plbarrep{\Fund}{1})$ and, after the evaluation of the determinant, obtain the same expression as the one inside the logarithm in Eq. (\ref{logarithmicPolyakovPotential}). For our purposes, we can drop the overall constant $N_c^{N_c}$ that arise when calculating the determinant of the matrix $W$. Since we are interested in the logarithm of the Haar measure, we can write: $\ln\qty[N_c^{N_c} \qty(\ldots) ]= N_c \ln\qty[N_c ] + \ln\qty[ \ldots  ]$. The term $N_c \ln\qty[N_c ]$ can be absorbed in the overall normalization constant of the generating functional, leaving only the logarithm of the Haar measure, see Eq. (\ref{generatingFunctionalDiagA4Integral}). From this definition one can expect the complexity of the calculation to largely increase for increasing values of $N_c$, leading to polynomials with several interaction terms. For instance, for $N_c=5$, the Haar measure is given by:
\begin{align*}
H
\qty(
\Plrep{\Fund}{1},\Plrep{\Fund}{2},  \Plbarrep{\Fund}{1} ,  \Plbarrep{\Fund}{2}
) 
& =
5^5
\mqty
|
1 & \Plrep{\Fund}{1} & \Plrep{\Fund}{2} & \Plrep{\Fund}{3}( \mathcal{C}_{5} )  & \Plrep{\Fund}{4}( \mathcal{C}_{5} )        
\\
\Plbarrep{\Fund}{1} & 1 & \Plrep{\Fund}{1} & \Plrep{\Fund}{2} & \Plrep{\Fund}{3}( \mathcal{C}_{5} ) 
\\
\Plbarrep{\Fund}{2} & \Plbarrep{\Fund}{1} & 1 & \Plrep{\Fund}{1} & \Plrep{\Fund}{2} 
\\
\Plbarrep{\Fund}{3} ( \mathcal{C}_{5} ) & \Plbarrep{\Fund}{2} & \Plbarrep{\Fund}{1} & 1 & \Plrep{\Fund}{1}  
\\
\Plbarrep{\Fund}{4} ( \mathcal{C}_{5} ) & \Plbarrep{\Fund}{3} ( \mathcal{C}_{5} ) & \Plbarrep{\Fund}{2} & \Plbarrep{\Fund}{1} & 1
| .
\numberthis
\end{align*}
Here, $\Plrep{\Fund}{3}$, $\Plrep{\Fund}{4}$, $\Plbarrep{\Fund}{3}$ and $\Plbarrep{\Fund}{4}$, are non-Cartan-Polyakov loops that must be decomposed in terms of Cartan loops using the previously derived decomposition formulas, in particular, Eqs. (\ref{elleFund3Nc5}) and (\ref{elleFund4Nc5}). Written in terms of the Cartan-Polyakov loop set, the Haar measure for $N_c=5$ has 106 different symmetry conserving interaction terms. For $N_c=7$ and $N_c=9$ this number increases to 3589 and 164856, respectively. The reason for this rapid increase in the number of terms can be traced back to combinatorics: increasing $N_c$ increases the number of elements inside the Cartan set, which in turn, boosts the number of possible $\Z{N_c}$ invariant combinations of Cartan-Polyakov loops.

\subsubsection{The thermodynamic potential with an arbitrary $N_c$}

Using the results obtained so far, we are finally able to write a thermodynamic potential for the PNJL model, in the mean field approximation, for an arbitrary number of colors:
\begin{align*}
\Omega & \qty(N_c,T,\mu)
=
U\qty(N_c, \expval{\phi_i}) 
+ 
\mathcal{U}_\mathrm{eff}\qty( \mathcal{C}_{N_c} , T ) 
- 2 N_c
\tr
\int_{\mathrm{reg}}
\frac{\dd[3]{p}}{ \qty(2 \pi)^3 } E 
\\
&  
- 2 T
\tr
\int_{\mathrm{reg}}
\frac{\dd[3]{p}}{ \qty(2 \pi)^3 }
\ln 
\qty[
1
+ 
\sum_{k=1}^{\frac{N_c-1}{2}}
c_k ( -L_\Fund ) \e^{ -k \beta \qty( E - {\mu} ) }
+ 
\sum_{k=\frac{N_c+1}{2}}^{{N_c}-1}
c_{N_c-k} ( -L_\Fund^\dagger ) \e^{ -k\beta \qty( E - {\mu} ) }
+
\e^{ -N_c \beta \qty( E - {\mu} ) } 
]
\\
&  
- 2T
\tr
\int_{\mathrm{reg}}
\frac{\dd[3]{p}}{ \qty(2 \pi)^3 }
\ln 
\qty[
1
+ 
\sum_{k=1}^{\frac{N_c-1}{2}}
c_k ( -L_\Fund^\dagger ) \e^{ -k \beta \qty( E + {\mu} ) } 
+ 
\sum_{k=\frac{N_c+1}{2}}^{{N_c}-1}
c_{N_c-k} ( -L_\Fund ) \e^{ -k\beta \qty( E + {\mu} ) } 
+
\e^{ - N_c \beta \qty( E + {\mu} ) }
] .
\numberthis
\label{thermodynamicPotentialArbitraryNc}
\end{align*}
As already stated, $\mathcal{C}_{N_c}$, is the set of Cartan-Polyakov loops, the coefficients $c_k ( -L_\Fund )$ and $c_k ( -L_\Fund^\dagger )$  can be calculated with Eq. (\ref{ckMinusLRep_def}) and the effective Polyakov loop potential is given by Eq. (\ref{logarithmicPolyakovPotentialNc}). The mean field potential, $U\qty(N_c, \expval{\phi_i}) $, can also be $N_c$ dependent. For instance, in the NJL model, the coupling of the four scalar-pseudoscalar quark-quark interaction can be considered as being inversely proportional to $N_c$ due to the QCD counting rules \cite{Ripka:1997zb}. As already discussed, the values of the mean fields, $\expval{\phi_i}$, and of the Cartan-Polyakov loops (elements of the $\mathcal{C}_{N_c}$ set) can be obtained by deriving the gap equations of the model: $\dv*{\Omega}{\expval{\phi_i}}=\dv*{\Omega}{\mathcal{C}_{N_c}^i}=0$, where $\mathcal{C}_{N_c}^i$ is the $i-$th element of the Cartan set.

It is important to note that the fermionic part of the thermodynamic potential written above, by construction, only depends on Cartan-Polyakov loops. However, the effective Polyakov loops potential, $\mathcal{U}_\mathrm{eff}\qty( \mathcal{C}_{N_c} , T )$, contains the Haar measure which depends explicitly on non-Cartan-Polyakov loops, in the fundamental representation, with charges up to $N_c-1$ (see Eq. (\ref{haarMeasureCartanAndNonCartanLoops})). Of course, as already discussed, one can make use of the decomposition formula given in Eq. (\ref{lk_decomposing_formula_JRk}) (or Eq. (\ref{lk_decomposing_formula_RRk})) in order to write the Haar measure in terms Cartan-Polyakov loops only. However, when dealing with the mean field approximation, the values of the Cartan-Polyakov loops are fixed by solving the gap equations, $\dv*{\Omega}{\mathcal{C}_{N_c}^i}=0$. The derivative of the Haar measure with respect to an element of the Cartan-Polyakov loop set, $\Plrep{\Fund}{k}$ (the same can be applied to derivatives with respect to $\Plbarrep{\Fund}{k}$), is:
\begin{align}
\dv{H}{ \Plrep{\Fund}{k} }
=
H \, \tr \qty[
W
\dv{W}{ \Plrep{\Fund}{k} }
] .
\end{align}
Here, we used Eq. (\ref{haarMeasureCartanAndNonCartanLoops}) without the $N_c^{N_c}$ coefficient and $\dv*{W}{ \Plrep{\Fund}{k} }$ is a tangent matrix, whose matrix elements are derivatives of Polyakov loops, in the fundamental representation, with charges $1,\ldots,N_c-1$, with respect to $\Plrep{\Fund}{k}$ (or $\Plbarrep{\Fund}{k}$). Explicitly, it is given by:
\begin{align}
\dv{W}{ \Plrep{\Fund}{k} } = 
\mqty
(
0 & \dv{\Plrep{\Fund}{1}}{ \Plrep{\Fund}{k} } & \dv{\Plrep{\Fund}{2}}{ \Plrep{\Fund}{k} } & \dots & \dv{\Plrep{\Fund}{N_c-1}}{ \Plrep{\Fund}{k} } \\
\dv{\Plbarrep{\Fund}{1}}{ \Plrep{\Fund}{k} } & 0 & \dv{\Plrep{\Fund}{1}}{ \Plrep{\Fund}{k} } & \dots & \dv{\Plrep{\Fund}{N_c-2}}{ \Plrep{\Fund}{k} }\\
\dv{\Plbarrep{\Fund}{2}}{ \Plrep{\Fund}{k} } & \dv{\Plbarrep{\Fund}{1}}{ \Plrep{\Fund}{k} } & 0 & \dots & \dv{\Plrep{\Fund}{N_c-3}}{ \Plrep{\Fund}{k} }\\
\vdots & \vdots & \vdots & \ddots &\vdots \\
\dv{\Plbarrep{\Fund}{N_c-1}}{ \Plrep{\Fund}{k} } & \dv{\Plbarrep{\Fund}{N_c-2}}{ \Plrep{\Fund}{k} } & \dv{\Plbarrep{\Fund}{N_c-3}}{ \Plrep{\Fund}{k} } & \dots & 0
) .
\end{align}
When differentiating a Cartan-Polyakov loop, the result is a delta function. However, the matrix above contains non-Cartan-Polyakov loops with charges in the interval $(N_c+1)/2,\ldots, N_c-1$. For these cases one can calculate these derivatives using:
\begin{align}
\dv{\Plrep{\Fund}{a}}{ \Plrep{\Fund}{b} }
& =
\pdv{\Plrep{\Fund}{a}}{ \Plrep{\Fund}{b} }
+
\sum_{i=\frac{N_c+1}{2}}^{a-1}
\qty(
\pdv{\Plrep{\Fund}{a}}{ \Plrep{\Fund}{i} }
)
\qty(
\dv{\Plrep{\Fund}{i}}{ \Plrep{\Fund}{b} } 
) ,
\\
\dv{\Plrep{\Fund}{a}}{ \Plbarrep{\Fund}{b} }
& =
\pdv{\Plrep{\Fund}{a}}{ \Plbarrep{\Fund}{b} }
+
\sum_{i=\frac{N_c+1}{2}}^{a-1}
\qty(
\pdv{\Plrep{\Fund}{a}}{ \Plrep{\Fund}{i} }
)
\qty(
\dv{\Plrep{\Fund}{i}}{ \Plbarrep{\Fund}{b} }
) ,
\\
\dv{\Plbarrep{\Fund}{a}}{ \Plrep{\Fund}{b} }
& =
\pdv{\Plbarrep{\Fund}{a}}{ \Plrep{\Fund}{b} }
+
\sum_{i=\frac{N_c+1}{2}}^{a-1}
\qty(
\pdv{\Plbarrep{\Fund}{a}}{ \Plbarrep{\Fund}{i} }
)
\qty(
\dv{\Plbarrep{\Fund}{i}}{ \Plrep{\Fund}{b} } 
) ,
\\
\dv{\Plbarrep{\Fund}{a}}{ \Plbarrep{\Fund}{b} }
& =
\pdv{\Plbarrep{\Fund}{a}}{ \Plbarrep{\Fund}{b} }
+
\sum_{i=\frac{N_c+1}{2}}^{a-1}
\qty(
\pdv{\Plbarrep{\Fund}{a}}{ \Plbarrep{\Fund}{i} }
)
\qty(
\dv{\Plbarrep{\Fund}{i}}{ \Plbarrep{\Fund}{b} } 
) .
\end{align}
For more details see Appendix \ref{differentiatingNonCartanLoops} where all the partial derivatives that are necessary in the above equations are calculated, for any representation, $\Rep$. With this result one can write the gap equations of the PNJL model for any value of $N_c$.

This model can be used, as is, to study the phase diagram of the PNJL model for increasing values of $N_c$. Numerically one can define all the necessary quantities, impose the gap equations and calculate all thermodynamics functions of interest for a given value of $N_c$. Of course, apart from the increase in the number of degrees of freedom, increasing $N_c$ means that the largest matrix in $c_k ( -L_\Fund )$ or $c_k ( -L_\Fund^\dagger )$, has dimensions $(N_c-1)/2 \times (N_c-1)/2$. Thus, naturally, the complexity of the numerical calculation, using the thermodynamic potential of Eq. (\ref{thermodynamicPotentialArbitraryNc}), gets larger for larger values of $N_c$. Also, Eq. (\ref{thermodynamicPotentialArbitraryNc}) can be a good starting point to write down a thermodynamic potential for the large $N_c$ limit. Such potential could then be used to study the phase diagram of strongly interacting matter at large $N_c$, and the quarkyonic phase of matter. We leave both aforementioned studies, the numerical analysis and the large$-N_c$ model, as future research endeavors.

We would like to highlight that the formalism employed here, to define the PNJL for different values of $N_c$, can also be applied to other fermionic models as long as the thermodynamic potential can be written in a similar way to the one given in Eq. (\ref{thermoPotPNJLNc3}).

\subsection{Glue effective potential with an arbitrary $N_c$}

In this section we apply the concept of Cartan-Polyakov loops to extend the one-loop gluon effective potential to arbitrary values of $N_c$. The starting point is the usual one-loop contribution to an effective potential of gluons with a spatially uniform Polyakov loop \cite{Meisinger:2001fi,Meisinger:2001cq,Sasaki:2012bi,Megias:2013xaa}: 
\begin{align}
\Omega_g (N_c, T)
=
2T
\int_{\mathrm{reg}}
\frac{\dd[3]{p}}{ \qty(2 \pi)^3 }
\qty{
\beta N_A E_g
+
\ln 
\det
\qty[
1 - L_\Adj \e^{ -\beta E_g }
]
} .
\label{gluonEffectivePotential}
\end{align}
Here, $N_\Adj=N_c^2-1$, is the dimension of the adjoint representation, $E_g=\sqrt{p^2+M_g^2}$ ($M_g$ is a phenomenological gluon effective mass \cite{Sasaki:2012bi,Megias:2013xaa}) and, $L_A$ is the thermal Wilson line in the adjoint representation of the $\SU{N_c}$ group (see Eq. (\ref{def_thermalWilsonLine})). This contribution to the effective glue potential drives the system to a state of spontaneously broken $\Z{N_c}$ symmetry \cite{Lo:2021qkw}. In order to study the transition from a $\Z{N_c}$ symmetric state to the broken state, one can consider a confining contribution to the potential in the form of the Haar measure \cite{Lo:2021qkw} which, as previously discussed, can be written in terms of the Cartan-Polyakov loop set, $\mathcal{C}_{N_c}$. Hence, the total effective glue potential, $\Omega(N_c, T)$, is given by:
\begin{align}
\Omega (N_c, T)
=
k( N_c, T ) \ln \qty[ H( \mathcal{C}_{N_c} ) ]
+ \Omega_g (N_c, T)
.
\label{gluonEffectivePotentialComplete}
\end{align}
Here, $H( \mathcal{C}_{N_c} )$ is the Haar measure and $k(N_c,T)$ is a phenomenological parameter, similar to the one introduced for the PNJL model with an arbitrary $N_c$ (see the discussion in Section \ref{polyakovEffectivePotentialPNJLArbitraryNc}). The effective glue potential defined above was also derived, for the $N_c=3$ case, in Ref. \cite{Sasaki:2012bi} where the background field approximation was applied to the generating functional of the $\SU{3}$ Yang-Mills theory, in the presence of a uniform gluon field. This model was also considered in Ref. \cite{Lo:2021qkw}, in order to study fluctuations of the order parameter in an $\SU{N_c}$ effective model.

The strategy to extend this model will be the same as the one used for the PNJL model. The confining part of the potential, written in Eq. (\ref{gluonEffectivePotentialComplete}), was previously defined for an arbitrary number of colors (see Eqs. (\ref{logarithmicPolyakovPotentialNc}) and (\ref{haarMeasureCartanAndNonCartanLoops})). Indeed, in Section \ref{polyakovEffectivePotentialPNJLArbitraryNc} we discussed how to write the Haar measure using exclusively the elements which constitute the Cartan-Polyakov loop set. The final step is to evaluate the gluonic determinant, present in Eq. (\ref{gluonEffectivePotential}), for an arbitrary number of colors. The gluonic determinant is defined by:
\begin{align}
\det \qty[ 1 - L_\Adj \e^{ -\beta E_g } ] 
=
h_g^{ N_\Adj } \det \qty[ h_g^{-1} I - L_\Adj ]  . 
\label{determinantLA}
\end{align}
Like in the case of the fermionic determinant, in the above, we defined the non-zero quantity, $h_g$, as:
\begin{align}
h_g = \e^{ -\beta E_g } .
\label{h_g_definition}
\end{align}
As done for the fermionic determinant, the goal is to calculate the gluonic determinant for an arbitrary $N_c$ using only the corresponding Cartan-Polyakov loop set. To this end, consider the characteristic polynomial $p \qty( L_\Adj, h_g^{-1} ) $ (see Eq. (\ref{characteristic_polynomial_LRep})). We can write the gluonic determinant as:
\begin{align*}
h_g^{N_\Adj} \det \qty[ h_g^{-1} I - L_\Adj ] 
& =
1
+ 
\sum_{k=1}^{ N_c^2-2 }
c_k ( L_\Adj ) h_g^{k} 
+
\qty(-1)^{ (N_c^2-1) }
h_g^{ (N_c^2-1) } .
\numberthis
\label{determinantLA2}
\end{align*}
Where we used the fact that $\det \qty[L_\Adj ] = 1$. The coefficients, $c_k ( L_\Adj )$, which arise in the sum, are calculated using Eq. (\ref{ckLRep_def}) with the appropriate choice of representation. Observing the structure of the matrix $J_{\Adj,k}$, defined in Eq. (\ref{JRepk_matrix_def}), one can conclude that the $k$ coefficient will depend on the non-Cartan-Polyakov loops $\Plrep{\Adj}{k} $, with charges $1,\ldots,k$. Since we want to write the model in terms of Cartan-Polyakov loops, we must find a way to write the non-Cartan-Polyakov loop of charges $k$, in the adjoint representation, $\Plrep{\Adj}{k} $, in terms of Cartan-Polyakov loops. In order to accomplish this one can use some results from group theory.

%\footnote{
%Considering two representations of the group $G$, $\sigma$ and $\rho$, the following arithmetic properties hold:
%\begin{align*}
%\chi_{\rho \oplus \sigma} (g) & = 
%\chi_{\rho} (g)
%+
%\chi_{\sigma} (g) ,
%\\
%\chi_{\rho \otimes \sigma} (g) & = 
%\chi_{\rho} (g)
%\chi_{\sigma} (g) .
%\end{align*}
%Here, $\chi_{\rho} (g) = \tr \big[ g_\rho \big]$ and $g_\rho$ is the group element $g$ in the representation $\rho$.
%}

A known result from representation theory allows for the decomposition of the product between the fundamental and the conjugate representations of the $\SU{N_c}$ group ($\Fund$ and $\bar{\Fund}$, respectively) with its adjoint and the singlet representations \cite{Elvang:2003ue}:
\begin{align}
\Fund \otimes \bar{\Fund}
=
\Adj \oplus 1.
\label{irrepDecomp}
\end{align}

This result can be proven using Young tableaux or by sandwiching the fundamental and anti-fundamental thermal Wilson lines between two $\SU{N_c}$ generators in order to get an object with adjoint indexes \cite{Dumitru:2003hp}. Since powers of the thermal Wilson line, $L^k$, are elements of the $\SU{N_c}$ group, we can write the following identities using character theory \cite{Megias:2013xaa}:
\begin{align}
\tr \qty[ L_{\rho \oplus \sigma}^k ]
& =
\tr \qty[ L_{\rho}^k ]
+
\tr \qty[ L_{\sigma}^k ] ,
\\
\tr \qty[ L_{\rho \otimes \sigma}^k ]
& =
\tr \qty[ L_{\rho}^k ]
\tr \qty[ L_{\sigma}^k ] .
\end{align}
Where $\rho$ and $\sigma$ are different representations of the $\SU{N_c}$ group. Using these properties, alongside the decomposition made in Eq. (\ref{irrepDecomp}), one can write:
\begin{align*}
\tr \big[ L_{ \Fund \otimes \bar{\Fund}}^k \big]
& =
\tr \qty[ L_{ \Adj \oplus 1 }^k ] 
\Leftrightarrow
\\
\tr \big[ L_{\Fund}^k \big]
\tr \big[ {L_{\Fund}^\dagger}^k \big] 
& =
\tr \big[ L_{ \Adj }^k \big]
+
1 \Leftrightarrow
\\
N_{\Fund}^2 
\Plrep{\Fund}{k}  
\Plbarrep{\Fund}{k}  
& =
N_{\Adj} 
\Plrep{\Adj}{k}  
+
1 .
\numberthis
\end{align*}
Thus, one can write the Polyakov loop with charge $k$, in the adjoint representation, in terms of Polyakov loops with charge $k$, in the fundamental and antifundamental representations. Manipulating the equation above yields \cite{Dumitru:2003hp,Fukushima:2017csk}:
\begin{align}
\Plrep{\Adj}{k}  
& =
\frac{ 1 }{ 1 - 1/N_c^2  }
\Plrep{\Fund}{k}  
\Plbarrep{\Fund}{k} 
- 
\frac{1}{ N_c^2 - 1 } .
\label{polyakovLoopAdjointUsingFundament}
\end{align}
As mentioned for the non-Cartan loops in the fundamental representation, this relation can also be checked by hand, by considering a particular gauge choice, e.g., using the already mentioned Polyakov loop gauge \cite{Sasaki:2012bi}. 

With this result, one can readily write the gluonic determinant as a function of fundamental and anti-fundamental Polyakov loops of different charges. Moreover, recursively, one can use the non-Cartan-Polyakov loop decomposition formula of Eq. (\ref{lk_decomposing_formula_JRk}) for loops with charge $k<N_c$ and Eq. (\ref{cayleyHamilton3}) for loops with charge $k \geq N_c$, in order to write the gluonic determinant of Eq. (\ref{determinantLA2}) in terms of Cartan-Polyakov loops only. We point out that, just like for the derivation of the non-Cartan decomposition formulas, or the calculation of fermionic determinant, one can make use of the identity between the characteristic polynomial coefficients of unitary matrices, written in Eq. (\ref{relation_cj_and_cjdagger_in_paper}) and shown in Appendix \ref{characteristic_polynomial_appendix}. Making use of this identity implies that only $(N_c^2-1)/2$ different coefficients have to be calculated, with the remaining ones arising automatically.  For $N_c=3$, there are a total of seven coefficients, but only four have to be calculated. Explicitly, for $N_c=3$, one gets:
\begin{align*}
c_1 ( L_\Adj ) 
= 
c_7 ( L_\Adj ) 
& =
-8 \Plrep{\Adj}{1} 
\\
& =
1 - 9 \Plrep{\Fund}{1} \Plbarrep{\Fund}{1} ,
\numberthis
\\
c_2 ( L_\Adj ) 
=
c_6 ( L_\Adj ) 
& =
32 \Plrep{\Adj}{1}^2 - 4 \Plrep{\Adj}{2}
\\
& =
1 
- 27 ( \Plrep{\Fund}{1} \Plbarrep{\Fund}{1} -\Plrep{\Fund}{1}^3 - \Plbarrep{\Fund}{1}^3 ) ,
\numberthis
\\
c_3 ( L_\Adj ) 
=
c_5 ( L_\Adj ) 
& =
-\frac{8}{3} 
\qty(
32 \Plrep{\Adj}{1}^3 
- 12 \Plrep{\Adj}{1} \Plrep{\Adj}{2} + \Plrep{\Adj}{3}
)
\\
& = 
-81 \Plrep{\Fund}{1}^2 \Plbarrep{\Fund}{1}^2 + 27 \Plrep{\Fund}{1} \Plbarrep{\Fund}{1} - 2 ,
\numberthis
\\
c_4 ( L_\Adj ) 
& =
\frac{8}{3} 
\qty(64 \Plrep{\Adj}{1}^4 - 48 \Plrep{\Adj}{2} \Plrep{\Adj}{1}^2 + 8 \Plrep{\Adj}{1} \Plrep{\Adj}{3} + 3 \Plrep{\Adj}{2}^2 
)
- 2 \Plrep{\Adj}{4}
\\
& = 
162 \Plrep{\Fund}{1}^2 \Plbarrep{\Fund}{1}^2 + 18 \Plrep{\Fund}{1} \Plbarrep{\Fund}{1} - 54 \Plbarrep{\Fund}{1}^3 - 54 \Plrep{\Fund}{1}^3 - 2 .
\numberthis
\end{align*}
Here, we have written the coefficients first in terms of charged Polyakov loops in the adjoint representation and then used Eq. (\ref{polyakovLoopAdjointUsingFundament}) alongside the non-Cartan decomposition formulas, in order to write the coefficients in terms of Cartan-Polyakov loops. These results agree with the ones present in the literature, see Ref. \cite{Sasaki:2012bi}. For $N_c=5$, one has to calculate twelve coefficients, the first three are given by:
\begin{align*}
c_1 ( L_\Adj ) 
= 
c_{23} ( L_\Adj ) 
& =
-24 \Plrep{\Adj}{1}
\\
& =
1 - 25 \Plrep{\Fund}{1} \Plbarrep{\Fund}{1},
\numberthis
\\
c_2 ( L_\Adj ) 
=
c_{22} ( L_\Adj ) 
& =
288 \Plrep{\Adj}{1}^2 - 12 \Plrep{\Adj}{2}
\\
& = 
1 
+ \frac{625}{2} \Plrep{\Fund}{1}^2 \Plbarrep{\Fund}{1}^2 
- 25 \Plrep{\Fund}{1} \Plbarrep{\Fund}{1} 
- \frac{25}{2} \Plrep{\Fund}{2} \Plbarrep{\Fund}{2} ,
\numberthis
\\
c_3 ( L_\Adj ) 
=
c_{21} ( L_\Adj ) 
& =
-8 
\qty(
288 \Plrep{\Adj}{1}^3 - 36 \Plrep{\Adj}{1} \Plrep{\Adj}{2}  + \Plrep{\Adj}{3}
)
\\
& = 
1
- \frac{15625}{4} \Plrep{\Fund}{1}^3 \Plbarrep{\Fund}{1}^3
+ \frac{3125}{4} \Plrep{\Fund}{1}^3 \Plbarrep{\Fund}{1} \Plbarrep{\Fund}{2}
- \frac{625}{4} \Plrep{\Fund}{1}^2 \Plbarrep{\Fund}{1}^2
+ \frac{375}{4} \Plrep{\Fund}{1}^2 \Plbarrep{\Fund}{2}
\\
&
\quad + \frac{3125}{4} \Plrep{\Fund}{1} \Plrep{\Fund}{2} \Plbarrep{\Fund}{1}^3
- 25 \Plrep{\Fund}{1} \Plbarrep{\Fund}{1}
- \frac{625}{4} \Plrep{\Fund}{1} \Plrep{\Fund}{2} \Plbarrep{\Fund}{1} \Plbarrep{\Fund}{2}
+ \frac{375}{4} \Plrep{\Fund}{2} \Plbarrep{\Fund}{1}^2 
+ \frac{375}{4} \Plbarrep{\Fund}{1} \Plbarrep{\Fund}{2}^2
\\
&
\quad
+ \frac{3125}{4} \Plbarrep{\Fund}{1}^5
- 625 \Plbarrep{\Fund}{1}^3 \Plbarrep{\Fund}{2}
- \frac{125}{4} \Plrep{\Fund}{2} \Plbarrep{\Fund}{2}
+ \frac{3125}{4} \Plrep{\Fund}{1}^5
- 625 \Plrep{\Fund}{1}^3 \Plrep{\Fund}{2} 
+ \frac{375}{4} \Plrep{\Fund}{1} \Plrep{\Fund}{2}^2  .
\numberthis
\end{align*}
One can immediately observe that the complexity of the coefficients, as expressions written in terms of Cartan-Polyakov loops, rapidly increases for larger numbers of colors, as well as, for higher order coefficients for a fixed $N_c$. Indeed, the $c_{12} ( L_\Adj )$ coefficient, for example, has 104 different terms when written in terms of the Cartan-Polyakov loop set. In practice, for numerical calculations, the overall size of the analytical expressions for a given coefficient at fixed $N_c$ is not important. In such approach, using as variables the Cartan-Polyakov loops, one can build all non-Cartan-Polyakov loops using the decomposition formulas, make use of Eq. (\ref{polyakovLoopAdjointUsingFundament}) to directly obtain different charged Polyakov loops in the adjoint representation and then get the coefficients.

We point out that, since we are dealing with a gluonic theory, the coefficients only have terms which preserve $\Z{N_c}$ symmetry. Naturally, the opposite was found for the fermionic coefficients where the terms explicitly broke the $\Z{N_c}$ symmetry.

Finally, one can write the glue effective potential defined in Eq. (\ref{gluonEffectivePotential}), for any value of $N_c$, using only Cartan-Polyakov loops. It is explicitly given by:
\begin{align*}
\Omega (N_c, T)
& =
k( N_c, T ) \ln \qty[ H( \mathcal{C}_{N_c} ) ]
+
2 \qty(N_c^2-1)
\int_{\mathrm{reg}}
\frac{\dd[3]{p}}{ \qty(2 \pi)^3 }
E_g
\\
& +
2T
\int_{\mathrm{reg}}
\frac{\dd[3]{p}}{ \qty(2 \pi)^3 }
\ln 
\qty[
1
+ 
\sum_{k=1}^{ N_c^2-2 }
c_k ( L_\Adj ) 
\e^{ - k \beta E_g }
+
\qty(-1)^{ (N_c^2-1) }
\e^{ -(N_c^2-1) \beta E_g }
]
.
\numberthis
\label{glueThermodynamicPotentialArbitraryNc}
\end{align*}

It is known that the mean field approximation is not suitable to study gluon effective potentials written in terms of fundamental and anti-fundamental Polyakov loops \cite{Zhang:2010kn}. The reason for such can be understood in the basis of the relation between the adjoint and the fundamental loops, written in Eq. (\ref{polyakovLoopAdjointUsingFundament}). In the naive mean field approximation, considered in this work, one can see that in the confined phase, where the expectation values of charged Polyakov loops in the fundamental representation are zero, the expectation values of charged adjoint Polyakov loops are negative \cite{Zhang:2010kn}. As pointed out in Ref. \cite{Fukushima:2017csk}, this implies that charged Polyakov loops in the adjoint representation, $\Plrep{\Adj}{k}$, cannot be used as an order parameters for gluon confinement. Also, the thermodynamic potential obtained in Eq. (\ref{glueThermodynamicPotentialArbitraryNc}) leads to a negative pressure, negative entropy density and thus, to thermodynamic instability \cite{Zhang:2010kn,Sasaki:2012bi}. This was discussed in Ref. \cite{Sasaki:2012bi}, using the glue model of Eq. (\ref{gluonEffectivePotential}) for $N_c=3$, by considering the low temperature expansion of the effective action. In such limit, the last term inside the logarithm, $\qty(-1)^{ (N_c^2-1) }
\exp\qty[ -(N_c^2-1) \beta E_g ]$, prevails and the incorrect sign for Bose statistics in front of the exponential is obtained for $N_c=3$ \cite{Sasaki:2012bi}. To overcome this feature of the glue effective potential for $N_c=3$, an hybrid model of glueballs in the confined phase and Polyakov loops in the deconfined phase was suggested, with the transition between models occurring at the transition temperature. For further details see Ref. \cite{Sasaki:2012bi}. We point out an interesting feature of the arbitrary $N_c$ effective potential written in Eq. (\ref{glueThermodynamicPotentialArbitraryNc}): the incorrect Bose statistics sign is obtained for and odd number of colors while, for an even number of colors, the sign is the one expected for a system of bosons.

Although one can use the approach used here to extended glue effective potentials to arbitrary values of $N_c$, as discussed above, the simple mean field approach might not be sufficient and corrections coming from connected diagrams in Eq. (\ref{polyakovLoopAdjointUsingFundament}), might be necessary \cite{Fukushima:2017csk}. However, at large $N_c$,  these corrections become negligible since expectation values factorize \cite{Dumitru:2003hp}. Thus, in this limit, the relation between the adjoint loops of charge $k$ and the fundamental and anti-fundamental loops of charge $k$ become much simpler:
\begin{align}
\Plrep{\Adj}{k}  
&  \sim
\Plrep{\Fund}{k}  
\Plbarrep{\Fund}{k}
+
O\qty( \frac{1}{N_c} )
.
\end{align}
In fact, at large $N_c$, the expectation value of any Polyakov loop can be written as powers of the expectation value of the fundamental and anti-fundamental Polyakov loops \cite{Dumitru:2003hp}. Hence, at large $N_c$, this model is a proper tool to study the behavior of the deconfining transition of strongly interacting matter.

As a final note, we point out that this model can be used to extend the PNJL model with an arbitrary $N_c$ obtained earlier in this work. For instance, instead of using the Haar measure and the kinetic term proposed in Eq. (\ref{kineticPartPolyakovLoopEffectivePotential}) for the effective Polyakov loop potential in the PNJL model, one could use the above effective glue potential. Indeed, one can show that, for $N_c=3$, under certain assumptions, it is possible to recover from Eq. (\ref{glueThermodynamicPotentialArbitraryNc}) the $- \bar{\Phi} \Phi$ used in the usual $N_c=3$ PNJL model, which is essential in order to get a first-order phase transition \cite{Sasaki:2012bi}. Not only that, using this effective glue potential, for the gluonic sector of PNJL model, would remove the necessity of defining so many $N_c-$dependent phenomenological parameters present in the effective Polyakov loop potential (see Eq. (\ref{logarithmicPolyakovPotentialNc})). In such case, the only parameter coming from the gluonic sector would be the one in front of the Haar measure \cite{Sasaki:2012bi}. Of course, the problem of the negative contribution to the pressure in the confined phase, for $\Plrep{\Fund}{k}=0$, would have to be addressed \cite{Sasaki:2012bi,Lo:2021qkw}.  This issue is partially resolved in Ref. \cite{Lo:2021qkw} by incorporating, in the thermodynamic potential, a contribution coming from the ghost fields. Such contribution reinforces the confining part of the potential, opposing the gluonic (deconfining) part and it might be able to overturn the negative tendency of the gluonic pressure in the confined phase \cite{Lo:2021qkw}. In the same work, it is discussed a phenomenological solution to this problem for $N_c=3$: subtract, from the glue effective potential, its contribution at $\Plrep{\Fund}{1}=\Plbarrep{\Fund}{1}=0$. Employing this recipe does not change the results for the Polyakov loops, but requires one to correct the number of gluonic degrees of freedom at large temperatures. Naturally, for other values of $N_c$, this phenomenological approach would also work, albeit one would have to subtract the potential at $\Plrep{\Fund}{k}=\Plbarrep{\Fund}{k}=0$, with $k = 1, \ldots, (N_c-1)/2$, i.e., considering all elements of the Cartan-Polyakov loop set to vanish. To explore additional works that address and offer solutions to this problem, beyond the ones already mentioned, see Refs. \cite{Tsai:2008je,Ruggieri:2012ny,Alba:2014lda,Islam:2021qwh}.

\section{Conclusions}

In this work we defined Cartan-Polyakov loops, a special subset of Polyakov loops in the fundamental representation of the $\SU{N_c}$ group which, under $\Z{N_c}$ transformations, behave like fields with charges $k=1,\ldots,(N_c-1)/2$. This set can be used as independent degrees of freedom to parametrize effective models of quarks and gluons with any number of colors. When building such phenomenological models, non-Cartan-Polyakov loops, Polyakov loops with charges higher than $(N_c-1)/2$, naturally arise. In order to write these models as functions of the loops contained in the Cartan set, we derived a decomposition formula for non-Cartan-Polyakov loops. Such formula was derived using properties of the characteristic polynomial of unitary matrices and the Cayley-Hamilton theorem. Fixing some values of $N_c$, we consistently decomposed some non-Cartan loops until they were only functions of Cartan-Polyakov loops. Finally, we showed how to build two distinct effective models for any $N_c$, including a version of the PNJL model and an effective glue potential. These models were built using the naive mean field approximation commonly used when dealing with the PNJL model: the Cartan-Polyakov loops were directly substituted by their respective expectation values and the expectation values of products between them factorized. We highlight that the potentials presented in this work, were derived using a particular gauge choice and/or required some phenomenological parameters (see Eq. (\ref{thermodynamicPotentialArbitraryNc}) for the thermodynamic potential of the PNJL model and Eq. (\ref{glueThermodynamicPotentialArbitraryNc}) for the glue effective potential, both written for an arbitrary $N_c$). For instance, the number of colors and temperature dependent prefactors $B\qty( N_c, T ) $ and $k( N_c, T )$, in Eqs. (\ref{logarithmicPolyakovPotentialNc}) and (\ref{gluonEffectivePotentialComplete}) respectively, were included because of phenomenological considerations. They were not derived from pure gauge theory.

The formalism developed in this work, allows for a consistent study on the effect of increasing the number of colors in effective models of quarks and gluons. Numerically, one can build tools that, for a given $N_c$, automatically builds the Cartan-Polyakov loop set and calculate all the necessary non-Cartan-Polyakov loops from it. In the case of the PNJL model, for example, one can build numerical tools in order to solve the model, within the mean field approximation, for a fixed value of $N_c$. Having the solution of such model, one can analyze the phase diagram of the model for that particular value of $N_c$, especially the behavior of the critical endpoint and the interplay between the chiral transition and the deconfinement transition. One can then repeat the calculation to increasing values of $N_c$ and study the behavior of the critical endpoint for large values of $N_c$, as well as study the conjectured quarkyonic phase of matter at high chemical potentials and moderate temperatures. Indeed, in Ref. \cite{Kovacs:2022zcl}, the fate of the critical endpoint at large $N_c$ was studied using a Polyakov loop quark-meson model by varying the number of colors. Within this work it was also confirmed the existence of the quarkyonic phase, in which chiral symmetry is restored but quarks are still confined \cite{Kovacs:2022zcl}. It would be very interesting to see if similar features were obtained within the model proposed in this work. In the case of the glue effective model defined in this work, it can be used to study the fluctuations of the confined-deconfined order parameters, given by different susceptibilities of the Polyakov loops, and the curvature masses associated with the Cartan-Polyakov loop set, for any value of $N_c$. These observables can be extremely useful tools to study the deconfinement phase transition in the Yang-Mills theory \cite{Lo:2013etb}, providing useful information about the link between the structure of the vacuum and gluon properties \cite{Lo:2021qkw}. For example, in Ref. \cite{Lo:2013etb}, the susceptibilities of the modulus, the real part and the imaginary part of the Polyakov loop, in the fundamental representation, were calculated in $\SU{3}$ lattice gauge theory (for studies including quark fields see Refs. \cite{Lo:2013hla,Bazavov:2016uvm,Clarke:2019tzf}). In that work the authors were able to show that different ratios of these susceptibilities are independent of the renormalization of the Polyakov and that these ratios depend weakly on the system size. Additionally, from the point of view of the phase transition, in the same work, it was shown that these ratios are discontinuous at the phase transition point and that above and below this point they are almost temperature independent. For other studies involving the susceptibilities of the Polyakov loop, see Refs. \cite{Lo:2013hla,Lo:2014vba,Miura:2016kmd}. A similar study carried out for larger values of $N_c$, using $\SU{N_c}$ lattice gauge theory, might be extremely useful in order to verify the relevance of the Cartan-Polyakov loop set. In such study one can evaluate different observables in the form of different ratios of different elements of the Cartan-Polyakov loop set. For instance, for $N_c=5$, the Cartan-Polyakov set is  $\mathcal{C}_5=\qty{ \Plrep{\Fund}{1}, \Plrep{\Fund}{2},  \Plbarrep{\Fund}{1}, \Plbarrep{\Fund}{2} }$ and the ratio of the modulus, and real and imaginary parts of distinct elements of the set, as a function of temperature, can be analyzed. 

The Cartan-Polyakov loop set can also be relevant to study the so-called Casimir scaling hypothesis. For Polyakov loops, this hypothesis states that there is a scaling relation between the Polyakov loops in different representations of the $\SU{N_c}$ group \cite{Gupta:2007ax,Abuki:2009dt,Megias:2013xaa}. The Casimir hypothesis is not only applicable to Polyakov loops but also to other observables \cite{Mykkanen:2012ri}. For values of $N_c>3$, it can be insightful to understand if the hypothesis holds for all elements of the Cartan-Polyakov loop set, besides the $\Plrep{\Fund}{1}$ and $\Plbarrep{\Fund}{1}$. For some works where the Casimir scaling hypothesis is discussed and studied, see Refs. \cite{Bali:2000un,Gupta:2007ax,Abuki:2009dt,Megias:2013xaa,Petreczky:2015yta,Petreczky:2015yta,Mykkanen:2012ri}.

As future work, we plan to use the formalism developed in this paper to build numerical tools which are prepared to solve the PNJL model for any given $N_c$, with the goal of studying the  behavior of the critical endpoint with increasing $N_c$. As pointed out, this can be accomplished recursively without further approximations, except the ones introduced in this work. At the same time, we will build an effective PNJL model at large $N_c$. In this limit, we plan to study the phase diagram of the model and the properties of the expected quarkyonic phase at large baryon densities. Having a numerical calculation together with a large $N_c$ approximation will allow us to compare both approaches and better understand the behavior of strongly interacting matter at large $N_c$.

\section*{Acknowledgments}

The authors would like to thank João Moreira and Orlando Oliveira for useful discussions. The authors also thank Tomáš Brauner for elucidating how to express the $\SU{3}$ Haar measure in terms of the character of the fundamental representation, in an online document entitled ``Haar measure on the unitary groups'' \cite{haarMeasureBrauner}. This work was supported by project CERN/FIS-PAR/0040/2019.

\appendix

\section{The characteristic polynomial}
\label{characteristic_polynomial_appendix}

The characteristic polynomial of a square $N \times N$ matrix $A$ is a polynomial, $p\qty(A,\lambda)$, in the variable $\lambda$, whose roots are the eigenvalues of the matrix $A$. This polynomial is defined by \cite{brown_1992,prasolov1996problems,Curtright:2020cta}:
\begin{align}
p \qty( A,\lambda )
\equiv
\det \qty[ A - \lambda I ] ,
\label{characteristic_equation_def}
\end{align}
where $I$ is the identity matrix. It can be shown that the characteristic polynomial can be written as:
\begin{align}
p \qty( A,\lambda ) 
= 
\qty(-1)^N
\qty[
\lambda^N + 
c_1 (A) \lambda^{N-1} + 
c_2 (A) \lambda^{N-2} + 
\dots + 
c_{N-1} (A) \lambda +
c_{N} (A) 
] .
\label{characteristic_polynomial_def}
\end{align}
By considering the special case of $\lambda = 0$, the $c_N$ coefficient is readily obtained. It is given by:
\begin{align}
c_N (A) = \qty(-1)^N  \det \qty[ A ] .
\label{cN_coefficient_def}
\end{align}
The coefficients $c_k (A)$ can be calculated using \cite{brown_1992,prasolov1996problems,Curtright:2020cta}:
\begin{align}
c_k (A) = 
\frac{ (-1)^k }{ k! }
\mqty|
t_{1}   & 1       & 0       & 0       & 0      & 0   \\ 
t_{2}   & t_{1}   & 2       & 0       & 0      & 0   \\ 
t_{3}   & t_{2}   & t_1     & 3       & 0      & 0   \\ 
\vdots  & \vdots  & \vdots  & \vdots  & \ddots & 0   \\ 
t_{k-1} & t_{k-2} & t_{k-3} & t_{k-4} & \dots  & k-1 \\ 
t_{k}   & t_{k-1} & t_{k-2} & t_{k-3} & \dots  & t_1 
| .
\label{c_k_definition}
\end{align}
Here, $t_k = \tr \qty[ A^k ]$. One can also formally define a zeroth order coefficient, $c_0=1$.

\subsection{ Relation between coefficients for $\U{N}$ matrices }

Consider an unitary matrix $M$ with dimension $N \times N$. Using Eq. (\ref{characteristic_polynomial_def}), its characteristic polynomial, $p \qty( M,\lambda )$, is:
\begin{align*}
p \qty( M,\lambda ) 
& = 
\det \qty[ M - \lambda I ] 
\\
& =
\qty(-1)^N
\qty(
\lambda^N
+ 
\sum_{i=1}^{N-1}
c_i (M) \lambda^{N-i} 
+
\qty(-1)^N  \det \qty[ M ]
) .
\numberthis
\label{characteristic_polynomial_M}
\end{align*}
Using this definition and the properties of the unitary matrix $M$, one can write the polynomial $p \qty( M,\lambda )$ as follows:
\begin{align*}
p \qty( M,\lambda ) 
& = 
\det \qty[ M - \lambda M M^\dagger ]
\\
& = 
\det 
\qty[ -\lambda M \qty( M^\dagger -\bar{\lambda} I  ) ]
\\
& = 
\det \qty[ -\lambda M ] 
\det
\qty[ M^\dagger -\bar{\lambda} I  ]
\\
& = 
\qty(-1)^N
\lambda^N
\det \qty[ M ] 
\det
\qty[ M^\dagger -\bar{\lambda} I  ] .
\numberthis
\label{comparing_det_M_and_Mdagger}
\end{align*}
Here, we assumed that $\lambda$ is non-zero and defined $\bar{\lambda}=\lambda^{-1}$. The determinant on the right hand side of this equation, $\det \qty[ M^\dagger -\bar{\lambda} I  ]$, can also be written in terms of a characteristic polynomial, $p \qty( M^\dagger, \bar{\lambda} ) $. It yields:
\begin{align*}
p \qty( M,\lambda ) 
& =
\qty(-1)^N
\lambda^N
\det \qty[ M ] 
\,
p \qty( M^\dagger, \bar{\lambda} )
\\
& =
\qty(-1)^N
\lambda^N
\det \qty[ M ] 
\qty(
\qty(-1)^N
\bar{\lambda}^N
+ 
\qty(-1)^N
\sum_{i=1}^{N-1}
c_i (M^\dagger) \bar{\lambda}^{N-i} 
+
\det \qty[ M^\dagger ]
) 
\\
& =
\qty(-1)^N
\qty(
\qty(-1)^N
\det \qty[ M ] 
+ 
\qty(-1)^N
\det \qty[ M ] 
\sum_{i=1}^{N-1}
c_i (M^\dagger) \lambda^{i} 
+
\lambda^N
\det \qty[ M M^\dagger ] 
) 
\\
& =
\qty(-1)^N
\qty(
\lambda^N
+ 
\sum_{i=1}^{N-1}
\qty(-1)^N
\det \qty[ M ] 
c_{N-i} (M^\dagger) 
\lambda^{N-i} 
+
\qty(-1)^N
\det \qty[ M ] 
) .
\numberthis
\end{align*}
Comparing this polynomial, with the polynomial in Eq. (\ref{characteristic_polynomial_M}), one can establish the following identities:
\begin{align*}
c_1 (M) & = \qty(-1)^N \det \qty[ M ] c_{N-1} (M^\dagger),
\\
c_2 (M) & = \qty(-1)^N \det \qty[ M ] c_{N-2} (M^\dagger),
\\
& \;\; \vdots \numberthis
\\
c_{N-1} (M) & = \qty(-1)^N \det \qty[ M ] c_{1} (M^\dagger).
\end{align*}
For a general coefficient with index $j$, one can write:
\begin{align}
c_j (M) & = \qty(-1)^N \det \qty[ M ] c_{N-j} (M^\dagger),
\label{relation_cj_and_cjdagger}
\end{align}
with $j=0,1,2,3,\dots,N-1,N$. We extended the relation for the trivial $0-$th and $N-$th order coefficients, $c_0(M)$ and $c_N(M)$. This extension can be checked explicitly (by definition, $c_0=1$):
\begin{align*}
c_0 (M) 
& = 
\qty(-1)^N \det \qty[ M ] c_{N} (M^\dagger)
= 
\qty(-1)^{2N} \det \qty[ M M^\dagger ] 
 = 
1 .
\end{align*}
In the above we made use of Eq. (\ref{cN_coefficient_def}). We highlight that $N$ is the dimension of the unitary matrix $M$ and $c_j(M)$ is calculated from Eq. (\ref{c_k_definition}).

\section{Differentiating non-Cartan-Polyakov loops}
\label{differentiatingNonCartanLoops}

For a given $N_c$, one can write a non-Cartan-Polyakov loop of charge $k$ in the $\Rep$ representation of the $\SU{N_c}$ group, as a function of the Cartan-Polyakov loop set for that specific $N_c$, using the decomposition formulas introduced in Eqs.(\ref{lk_decomposing_formula_JRk}), (\ref{lk_decomposing_formula_RRk}) and (\ref{cayleyHamilton3}). For more details, see Section \ref{cartanPolyakovLoops}. Hence, one can write a formula for the total derivative of a non-Cartan-Polyakov loop with respect to some element of the Cartan set. This result is very important in the evaluation of the gap equations of the PNJL model, arising when one is calculating derivatives of the Haar measure with respect to Cartan-Polyakov loops.

%\begin{align}
%\Plrep{\Rep}{q} 
%= 
%\Plrep{\Rep}{q}
%\qty( 
%\mathcal{C}_{N_c} ,
%\Plrep{\Rep}{q-1}
%\qty(\mathcal{C}_{N_c},\Plrep{\Rep}{q-2}(\mathcal{C}_{N_c},\ldots),\ldots)
%,
%\Plrep{\Rep}{q-2}
%\qty(\mathcal{C}_{N_c},\Plrep{\Rep}{q-3}(\mathcal{C}_{N_c},\ldots),\ldots)
%).
%\end{align}

Consider a non-Cartan-Polyakov loop of charge $a$, in some representation of the $\SU{N_c}$ group, $\Plrep{\Rep}{a}$. As discussed in this work, one can always build these quantities recursively using Cartan loops. In practice, this non-Cartan loop can be considered as function of the elements in the Cartan set and as a function of all non-Cartan loops of smaller charge, which, in turn, are functions of the Cartan set and so on. Thus, we can write:
\begin{align}
\Plrep{\Rep}{a} 
= 
\Plrep{\Rep}{a}
\qty( 
\mathcal{C}_{N_c} ,
\Plrep{\Rep}{a-1}
,
\Plrep{\Rep}{a-2}
,\ldots
).
\end{align}
Thus, the total derivative of this non-Cartan loop, $\Plrep{\Rep}{a} $, with respect to a Cartan-Polyakov loop $\Plrep{\Rep}{b}$, is given by the following expression:
\begin{align}
\dv{\Plrep{\Rep}{a}}{ \Plrep{\Rep}{b} }
& =
\pdv{\Plrep{\Rep}{a}}{ \Plrep{\Rep}{b} }
+
\sum_{i=\frac{N_c+1}{2}}^{a-1}
\qty(
\pdv{\Plrep{\Rep}{a}}{ \Plrep{\Rep}{i} }
)
\qty(
\dv{\Plrep{\Rep}{i}}{ \Plrep{\Rep}{b} } 
) .
\end{align}
This formula also works recursively: in order to get $\dv*{\Plrep{\Rep}{a}}{ \Plrep{\Rep}{b} }$ one needs to first calculate all the total derivatives of non-Cartan loops with smaller charges, $\dv*{\Plrep{\Rep}{i}}{ \Plrep{\Rep}{b} }$, with $i \in \qty{ \nicefrac{(N_c+1)}{2}, \ldots, a-1 }$. Likewise, one can write the total derivative of a non-Cartan loop with respect to some adjoint Cartan loop, $\dv*{\Plrep{\Rep}{a}}{ \Plbarrep{\Rep}{b} }$ or $\dv*{\Plbarrep{\Rep}{a}}{ \Plrep{\Rep}{b} }$ and $\dv*{\Plbarrep{\Rep}{a}}{ \Plbarrep{\Rep}{b} }$. They are given by:
\begin{align}
\dv{\Plrep{\Rep}{a}}{ \Plbarrep{\Rep}{b} }
& =
\pdv{\Plrep{\Rep}{a}}{ \Plbarrep{\Rep}{b} }
+
\sum_{i=\frac{N_c+1}{2}}^{a-1}
\qty(
\pdv{\Plrep{\Rep}{a}}{ \Plrep{\Rep}{i} }
)
\qty(
\dv{\Plrep{\Rep}{i}}{ \Plbarrep{\Rep}{b} }
) ,
\\
\dv{\Plbarrep{\Rep}{a}}{ \Plrep{\Rep}{b} }
& =
\pdv{\Plbarrep{\Rep}{a}}{ \Plrep{\Rep}{b} }
+
\sum_{i=\frac{N_c+1}{2}}^{a-1}
\qty(
\pdv{\Plbarrep{\Rep}{a}}{ \Plbarrep{\Rep}{i} }
)
\qty(
\dv{\Plbarrep{\Rep}{i}}{ \Plrep{\Rep}{b} } 
) ,
\\
\dv{\Plbarrep{\Rep}{a}}{ \Plbarrep{\Rep}{b} }
& =
\pdv{\Plbarrep{\Rep}{a}}{ \Plbarrep{\Rep}{b} }
+
\sum_{i=\frac{N_c+1}{2}}^{a-1}
\qty(
\pdv{\Plbarrep{\Rep}{a}}{ \Plbarrep{\Rep}{i} }
)
\qty(
\dv{\Plbarrep{\Rep}{i}}{ \Plbarrep{\Rep}{b} } 
) .
\end{align}

The different partial derivatives, necessary to evaluate the total derivatives, can be obtained by partially differentiating the decomposition formulas derived in this work. If the charge $a$ is smaller than $N_\Rep$, the partial derivative can be calculated using the decomposition formula of Eq. (\ref{lk_decomposing_formula_JRk}) (a similar calculation can be done using Eq. (\ref{lk_decomposing_formula_RRk}) as a starting point) while, for charges $a \geq N_\Rep$, one can calculate the partial derivative using Eq. (\ref{cayleyHamilton3}). Lets first deal with the case $a < N_\Rep$. Calculating the partial derivative of Eq. (\ref{lk_decomposing_formula_JRk}), with respect to the Cartan-Polyakov loop $\Plrep{\Rep}{b}$, yields:
\begin{align*}
\left.
\pdv{\Plrep{\Rep}{a}}{ \Plrep{\Rep}{b} }
\right|_{a < N_\Rep}
& = 
(-1)^{a+1}
\frac{ a }{ N_\Rep \qty(N_\Rep-a)! }
\pdv{}{ \Plrep{\Rep}{b} }
\qty(
\det \qty[ \bar{J}_{\Rep,N_\Rep-a} ]
)
-
\frac{ a }{ a! }
\sum_{i=1}^{a-1}  
\qty(-1)^{i-a} 
\pdv{}{ \Plrep{\Rep}{b} }
\qty(
\Plrep{\Rep}{i} \det \qty[J_{\Rep,a}^{(i,1)}] 
)
\\
& = 
-
\frac{ a }{ a! }
\qty(
\qty(-1)^{b-a} 
\det \qty[J_{\Rep,a}^{(b,1)}] 
+
\sum_{i=1}^{a-1}  
\qty(-1)^{i-a} 
\Plrep{\Rep}{i} 
\det \qty[J_{\Rep,a}^{(i,1)}] 
\tr\qty[ 
\qty[ J_{\Rep,a}^{(i,1)} ]^{-1}
\pdv{ J_{\Rep,a}^{(i,1)} }{ \Plrep{\Rep}{b} }
]
) .
\numberthis
\label{partialDerivativeChargeSmallerThanNrep}
\end{align*}
Here, we used Jacobi's formula in order to express the derivative of the determinant,
$\dv{}{t} \det [A(t)] =$ $ \det[A(t)] \times $ $\tr [ A(t)^{-1} \dv{A(t)}{t} ]$. We also used the fact that $\pdv{}{ \Plrep{\Rep}{b} }
\qty(
\det \qty[ \bar{J}_{\Rep,N_\Rep-a} ]
)=0$, since the matrix $\bar{J}_{\Rep,N_\Rep-a}$ only depends on adjoint Cartan-Polyakov loops. We recall that in the notation used throughout this work, the subscript $(i,1)$ in the matrix $J_{\Rep,a}^{(i,1)}$ is equivalent to the matrix $J_{\Rep,a}$ with the $i-$th line and first column removed. 

The matrix $\pdv{ J_{\Rep,a} }{ \Plrep{\Rep}{b} }$ can be found very easily by taking into account its definition given in Eq. (\ref{JRepk_matrix_def}). It is explicitly given by:
\begin{align*}
\pdv{ J_{\Rep,a} }{ \Plrep{\Rep}{b} }
& = 
N_\Rep
\mqty(
\delta_{b,1} & 0 & 0 & 0 & 0 & 0          \\ 
\delta_{b,2} &  \delta_{b,1} & 0 & 0 & 0 & 0   \\ 
\delta_{b,3} & \delta_{b,2}  & \delta_{b,1} & 0 & 0 & 0 \\ 
\vdots & \vdots & \vdots & \vdots & \ddots & 0 \\ 
\delta_{b,a-1} & \delta_{b,a-2} & \delta_{b,a-3} & \delta_{b,a-4} & \dots  & 0        \\ 
\delta_{b,a}     & \delta_{b,a-1} & \delta_{b,a-2} & \delta_{b,a-3} & \dots  & \delta_{b,1}
)  .
\numberthis
\label{dJRepkdelleb_matrix}
\end{align*}
If the non-Cartan loop has charge $a \geq N_\Rep$, one must calculate the partial derivative of Eq. (\ref{lk_decomposing_formula_JRk}). It yields:
\begin{align*}
\left.
\pdv{\Plrep{\Rep}{a}}{ \Plrep{\Rep}{b} }
\right|_{a \geq N_\Rep}
& = 
\qty(-1)^{N_\Rep+1}
\pdv{  }{ \Plrep{\Rep}{b} }
\Plrep{\Rep}{a-N_\Rep} 
-
\sum_{i=1}^{{N_\Rep}-1}
\pdv{  }{ \Plrep{\Rep}{b} }
\qty(
c_i ( L_\Rep ) 
\Plrep{\Rep}{a-i} 
)
\\
& = 
\qty(-1)^{N_\Rep+1}
\delta_{ b,a-N_\Rep }
-
\sum_{i=1}^{{N_\Rep}-1}
\delta_{ b,a-i }
c_i ( L_\Rep ) 
-
\sum_{i=1}^{{N_\Rep}-1}
c_i ( L_\Rep ) 
\Plrep{\Rep}{a-i} 
\tr\qty[ 
\qty[ J_{\Rep,a} ]^{-1}
\pdv{ J_{\Rep,a} }{ \Plrep{\Rep}{b} }
].
\numberthis
\label{partialDerivativeChargeGreaterEqualThanNrep}
\end{align*}
Likewise, one can calculate the other partial derivatives. Summarizing, one gets:
\begin{align*}
\left.
\pdv{\Plbarrep{\Rep}{a}}{ \Plbarrep{\Rep}{b} }
\right|_{a < N_\Rep}
& = 
-
\frac{ a }{ a! }
\qty(
\qty(-1)^{b-a} 
\det \qty[ \bar{J}_{\Rep,a}^{(b,1)}] 
+
\sum_{i=1}^{a-1}  
\qty(-1)^{i-a} 
\Plbarrep{\Rep}{i} 
\det \qty[ \bar{J}_{\Rep,a}^{(i,1)}] 
\tr\qty[ 
\qty[ \bar{J}_{\Rep,a}^{(i,1)} ]^{-1}
\pdv{ \bar{J}_{\Rep,a}^{(i,1)} }{ \Plbarrep{\Rep}{b} }
]
) ,
\numberthis
\label{ellebarEllebarPartialDerivativeChargeSmallerThanNrep}
\\
\left.
\pdv{\Plbarrep{\Rep}{a}}{ \Plbarrep{\Rep}{b} }
\right|_{a \geq N_\Rep}
& = 
\qty(-1)^{N_\Rep+1}
\delta_{ b,a-N_\Rep }
-
\sum_{i=1}^{{N_\Rep}-1}
\delta_{ b,a-i }
c_i ( L_\Rep^\dagger ) 
-
\sum_{i=1}^{{N_\Rep}-1}
c_i ( L_\Rep^\dagger ) 
\Plbarrep{\Rep}{a-i} 
\tr\qty[ 
\qty[ \bar{J}_{\Rep,a} ]^{-1}
\pdv{ \bar{J}_{\Rep,a} }{ \Plbarrep{\Rep}{b} }
] ,
\numberthis
\label{ellebarEllebarPartialDerivativeChargeGreaterEqualThanNrep}
\\
\left.
\pdv{\Plbarrep{\Rep}{a}}{ \Plrep{\Rep}{b} }
\right|_{a < N_\Rep}
& = 
(-1)^{a+1}
\frac{ a }{ N_\Rep \qty(N_\Rep-a)! }
\det \qty[ J_{\Rep,N_\Rep-a} ]
\tr
\qty[
\qty[ J_{\Rep,N_\Rep-a} ]^{-1}
\pdv{ J_{\Rep,N_\Rep-a} }{ \Plrep{\Rep}{b} }
] ,
\label{elleBarEllePartialDerivativeChargeSmallerThanNrep}
\numberthis
\\
\left.
\pdv{\Plbarrep{\Rep}{a}}{ \Plrep{\Rep}{b} }
\right|_{a \geq N_\Rep}
& = 
0 ,
\numberthis
\label{elleBarEllePartialDerivativeChargeGreaterEqualThanNrep}
\\
\left.
\pdv{\Plrep{\Rep}{a}}{ \Plbarrep{\Rep}{b} }
\right|_{a < N_\Rep}
& = 
(-1)^{a+1}
\frac{ a }{ N_\Rep \qty(N_\Rep-a)! }
\det \qty[ \bar{J}_{\Rep,N_\Rep-a} ]
\tr
\qty[
\qty[ \bar{J}_{\Rep,N_\Rep-a} ]^{-1}
\pdv{ \bar{J}_{\Rep,N_\Rep-a} }{ \Plbarrep{\Rep}{b} }
] ,
\label{elleEllebarPartialDerivativeChargeSmallerThanNrep}
\numberthis
\\
\left.
\pdv{\Plrep{\Rep}{a}}{ \Plbarrep{\Rep}{b} }
\right|_{a \geq N_\Rep}
& = 
0 .
\numberthis
\label{elleEllebarPartialDerivativeChargeGreaterEqualThanNrep}
\end{align*}

With these tools one is able to evaluate the total derivatives of quantities which depend on non-Cartan-Polyakov loops, as is the case with the Haar measure (see Eq. (\ref{haarMeasureCartanAndNonCartanLoops})). So, one can readily evaluate the derivatives of the Haar measure with respect to the elements of the Cartan-Polyakov loop set and obtain the gap equations of the PNJL model for any value of $N_c$.

\bibliography{article}

\begin{thebibliography}{99}%
\makeatletter
\providecommand \@ifxundefined [1]{%
 \@ifx{#1\undefined}
}%
\providecommand \@ifnum [1]{%
 \ifnum #1\expandafter \@firstoftwo
 \else \expandafter \@secondoftwo
 \fi
}%
\providecommand \@ifx [1]{%
 \ifx #1\expandafter \@firstoftwo
 \else \expandafter \@secondoftwo
 \fi
}%
\providecommand \natexlab [1]{#1}%
\providecommand \enquote  [1]{``#1''}%
\providecommand \bibnamefont  [1]{#1}%
\providecommand \bibfnamefont [1]{#1}%
\providecommand \citenamefont [1]{#1}%
\providecommand \href@noop [0]{\@secondoftwo}%
\providecommand \href [0]{\begingroup \@sanitize@url \@href}%
\providecommand \@href[1]{\@@startlink{#1}\@@href}%
\providecommand \@@href[1]{\endgroup#1\@@endlink}%
\providecommand \@sanitize@url [0]{\catcode `\\12\catcode `\$12\catcode
  `\&12\catcode `\#12\catcode `\^12\catcode `\_12\catcode `\%12\relax}%
\providecommand \@@startlink[1]{}%
\providecommand \@@endlink[0]{}%
\providecommand \url  [0]{\begingroup\@sanitize@url \@url }%
\providecommand \@url [1]{\endgroup\@href {#1}{\urlprefix }}%
\providecommand \urlprefix  [0]{URL }%
\providecommand \Eprint [0]{\href }%
\providecommand \doibase [0]{https://doi.org/}%
\providecommand \selectlanguage [0]{\@gobble}%
\providecommand \bibinfo  [0]{\@secondoftwo}%
\providecommand \bibfield  [0]{\@secondoftwo}%
\providecommand \translation [1]{[#1]}%
\providecommand \BibitemOpen [0]{}%
\providecommand \bibitemStop [0]{}%
\providecommand \bibitemNoStop [0]{.\EOS\space}%
\providecommand \EOS [0]{\spacefactor3000\relax}%
\providecommand \BibitemShut  [1]{\csname bibitem#1\endcsname}%
\let\auto@bib@innerbib\@empty
%</preamble>
\bibitem [{\citenamefont {Skands}(2013)}]{Skands:2012ts}%
  \BibitemOpen
  \bibfield  {author} {\bibinfo {author} {\bibfnamefont {P.}~\bibnamefont
  {Skands}},\ }\bibfield  {title} {\bibinfo {title} {{Introduction to QCD}},\
  }in\ \href {https://doi.org/10.1142/9789814525220_0008} {\emph {\bibinfo
  {booktitle} {{Searching for New Physics at Small and Large Scales}}}}\
  (\bibinfo {year} {2013})\ pp.\ \bibinfo {pages} {341--420},\ \Eprint
  {https://arxiv.org/abs/1207.2389} {arXiv:1207.2389 [hep-ph]} \BibitemShut
  {NoStop}%
\bibitem [{\citenamefont {Halasz}\ \emph {et~al.}(1998)\citenamefont {Halasz},
  \citenamefont {Jackson}, \citenamefont {Shrock}, \citenamefont {Stephanov},\
  and\ \citenamefont {Verbaarschot}}]{Halasz:1998qr}%
  \BibitemOpen
  \bibfield  {author} {\bibinfo {author} {\bibfnamefont {A.~M.}\ \bibnamefont
  {Halasz}}, \bibinfo {author} {\bibfnamefont {A.~D.}\ \bibnamefont {Jackson}},
  \bibinfo {author} {\bibfnamefont {R.~E.}\ \bibnamefont {Shrock}}, \bibinfo
  {author} {\bibfnamefont {M.~A.}\ \bibnamefont {Stephanov}},\ and\ \bibinfo
  {author} {\bibfnamefont {J.~J.~M.}\ \bibnamefont {Verbaarschot}},\ }\bibfield
   {title} {\bibinfo {title} {{On the phase diagram of QCD}},\ }\href
  {https://doi.org/10.1103/PhysRevD.58.096007} {\bibfield  {journal} {\bibinfo
  {journal} {Phys. Rev. D}\ }\textbf {\bibinfo {volume} {58}},\ \bibinfo
  {pages} {096007} (\bibinfo {year} {1998})},\ \Eprint
  {https://arxiv.org/abs/hep-ph/9804290} {arXiv:hep-ph/9804290} \BibitemShut
  {NoStop}%
\bibitem [{\citenamefont {Hansen}\ \emph {et~al.}(2020)\citenamefont {Hansen},
  \citenamefont {Stiele},\ and\ \citenamefont {Costa}}]{Hansen:2019lnf}%
  \BibitemOpen
  \bibfield  {author} {\bibinfo {author} {\bibfnamefont {H.}~\bibnamefont
  {Hansen}}, \bibinfo {author} {\bibfnamefont {R.}~\bibnamefont {Stiele}},\
  and\ \bibinfo {author} {\bibfnamefont {P.}~\bibnamefont {Costa}},\ }\bibfield
   {title} {\bibinfo {title} {{Quark and Polyakov-loop correlations in
  effective models at zero and nonvanishing density}},\ }\href
  {https://doi.org/10.1103/PhysRevD.101.094001} {\bibfield  {journal} {\bibinfo
   {journal} {Phys. Rev. D}\ }\textbf {\bibinfo {volume} {101}},\ \bibinfo
  {pages} {094001} (\bibinfo {year} {2020})},\ \Eprint
  {https://arxiv.org/abs/1904.08965} {arXiv:1904.08965 [hep-ph]} \BibitemShut
  {NoStop}%
\bibitem [{\citenamefont {Schmidt}\ and\ \citenamefont
  {Sharma}(2017)}]{Schmidt:2017bjt}%
  \BibitemOpen
  \bibfield  {author} {\bibinfo {author} {\bibfnamefont {C.}~\bibnamefont
  {Schmidt}}\ and\ \bibinfo {author} {\bibfnamefont {S.}~\bibnamefont
  {Sharma}},\ }\bibfield  {title} {\bibinfo {title} {{The phase structure of
  QCD}},\ }\href {https://doi.org/10.1088/1361-6471/aa824a} {\bibfield
  {journal} {\bibinfo  {journal} {J. Phys. G}\ }\textbf {\bibinfo {volume}
  {44}},\ \bibinfo {pages} {104002} (\bibinfo {year} {2017})},\ \Eprint
  {https://arxiv.org/abs/1701.04707} {arXiv:1701.04707 [hep-lat]} \BibitemShut
  {NoStop}%
\bibitem [{\citenamefont {'t~Hooft}(1974)}]{tHooft:1973alw}%
  \BibitemOpen
  \bibfield  {author} {\bibinfo {author} {\bibfnamefont {G.}~\bibnamefont
  {'t~Hooft}},\ }\bibfield  {title} {\bibinfo {title} {{A Planar Diagram Theory
  for Strong Interactions}},\ }\href
  {https://doi.org/10.1016/0550-3213(74)90154-0} {\bibfield  {journal}
  {\bibinfo  {journal} {Nucl. Phys. B}\ }\textbf {\bibinfo {volume} {72}},\
  \bibinfo {pages} {461} (\bibinfo {year} {1974})}\BibitemShut {NoStop}%
\bibitem [{\citenamefont {Witten}(1979)}]{Witten:1979kh}%
  \BibitemOpen
  \bibfield  {author} {\bibinfo {author} {\bibfnamefont {E.}~\bibnamefont
  {Witten}},\ }\bibfield  {title} {\bibinfo {title} {{Baryons in the 1/n
  Expansion}},\ }\href {https://doi.org/10.1016/0550-3213(79)90232-3}
  {\bibfield  {journal} {\bibinfo  {journal} {Nucl. Phys. B}\ }\textbf
  {\bibinfo {volume} {160}},\ \bibinfo {pages} {57} (\bibinfo {year}
  {1979})}\BibitemShut {NoStop}%
\bibitem [{\citenamefont {Jenkins}(1998)}]{Jenkins:1998wy}%
  \BibitemOpen
  \bibfield  {author} {\bibinfo {author} {\bibfnamefont {E.~E.}\ \bibnamefont
  {Jenkins}},\ }\bibfield  {title} {\bibinfo {title} {{Large N(c) baryons}},\
  }\href {https://doi.org/10.1146/annurev.nucl.48.1.81} {\bibfield  {journal}
  {\bibinfo  {journal} {Ann. Rev. Nucl. Part. Sci.}\ }\textbf {\bibinfo
  {volume} {48}},\ \bibinfo {pages} {81} (\bibinfo {year} {1998})},\ \Eprint
  {https://arxiv.org/abs/hep-ph/9803349} {arXiv:hep-ph/9803349} \BibitemShut
  {NoStop}%
\bibitem [{\citenamefont {Bonanno}\ and\ \citenamefont
  {Giacosa}(2011)}]{Bonanno:2011yr}%
  \BibitemOpen
  \bibfield  {author} {\bibinfo {author} {\bibfnamefont {L.}~\bibnamefont
  {Bonanno}}\ and\ \bibinfo {author} {\bibfnamefont {F.}~\bibnamefont
  {Giacosa}},\ }\bibfield  {title} {\bibinfo {title} {{Does nuclear matter bind
  at large $N_c$?}},\ }\href {https://doi.org/10.1016/j.nuclphysa.2011.04.012}
  {\bibfield  {journal} {\bibinfo  {journal} {Nucl. Phys. A}\ }\textbf
  {\bibinfo {volume} {859}},\ \bibinfo {pages} {49} (\bibinfo {year} {2011})},\
  \Eprint {https://arxiv.org/abs/1102.3367} {arXiv:1102.3367 [hep-ph]}
  \BibitemShut {NoStop}%
\bibitem [{\citenamefont {Lucini}\ and\ \citenamefont
  {Panero}(2014)}]{Lucini:2013qja}%
  \BibitemOpen
  \bibfield  {author} {\bibinfo {author} {\bibfnamefont {B.}~\bibnamefont
  {Lucini}}\ and\ \bibinfo {author} {\bibfnamefont {M.}~\bibnamefont
  {Panero}},\ }\bibfield  {title} {\bibinfo {title} {{Introductory lectures to
  large-$\scriptsize{N}$ QCD phenomenology and lattice results}},\ }\href
  {https://doi.org/10.1016/j.ppnp.2014.01.001} {\bibfield  {journal} {\bibinfo
  {journal} {Prog. Part. Nucl. Phys.}\ }\textbf {\bibinfo {volume} {75}},\
  \bibinfo {pages} {1} (\bibinfo {year} {2014})},\ \Eprint
  {https://arxiv.org/abs/1309.3638} {arXiv:1309.3638 [hep-th]} \BibitemShut
  {NoStop}%
\bibitem [{\citenamefont {Richardson}\ \emph {et~al.}(2021)\citenamefont
  {Richardson}, \citenamefont {Schindler}, \citenamefont {Pastore},\ and\
  \citenamefont {Springer}}]{Richardson:2021xiu}%
  \BibitemOpen
  \bibfield  {author} {\bibinfo {author} {\bibfnamefont {T.~R.}\ \bibnamefont
  {Richardson}}, \bibinfo {author} {\bibfnamefont {M.~R.}\ \bibnamefont
  {Schindler}}, \bibinfo {author} {\bibfnamefont {S.}~\bibnamefont {Pastore}},\
  and\ \bibinfo {author} {\bibfnamefont {R.~P.}\ \bibnamefont {Springer}},\
  }\bibfield  {title} {\bibinfo {title} {{Large-$N_c$ analysis of two-nucleon
  neutrinoless double-$\beta$ decay and charge-independence-breaking contact
  terms}},\ }\href {https://doi.org/10.1103/PhysRevC.103.055501} {\bibfield
  {journal} {\bibinfo  {journal} {Phys. Rev. C}\ }\textbf {\bibinfo {volume}
  {103}},\ \bibinfo {pages} {055501} (\bibinfo {year} {2021})},\ \Eprint
  {https://arxiv.org/abs/2102.02184} {arXiv:2102.02184 [nucl-th]} \BibitemShut
  {NoStop}%
\bibitem [{\citenamefont {Lucha}\ \emph {et~al.}(2021)\citenamefont {Lucha},
  \citenamefont {Melikhov},\ and\ \citenamefont {Sazdjian}}]{Lucha:2021mwx}%
  \BibitemOpen
  \bibfield  {author} {\bibinfo {author} {\bibfnamefont {W.}~\bibnamefont
  {Lucha}}, \bibinfo {author} {\bibfnamefont {D.}~\bibnamefont {Melikhov}},\
  and\ \bibinfo {author} {\bibfnamefont {H.}~\bibnamefont {Sazdjian}},\
  }\bibfield  {title} {\bibinfo {title} {{Tetraquarks in large-Nc QCD}},\
  }\href {https://doi.org/10.1016/j.ppnp.2021.103867} {\bibfield  {journal}
  {\bibinfo  {journal} {Prog. Part. Nucl. Phys.}\ }\textbf {\bibinfo {volume}
  {120}},\ \bibinfo {pages} {103867} (\bibinfo {year} {2021})},\ \Eprint
  {https://arxiv.org/abs/2102.02542} {arXiv:2102.02542 [hep-ph]} \BibitemShut
  {NoStop}%
\bibitem [{\citenamefont {Flores-Mendieta}\ \emph {et~al.}(2021)\citenamefont
  {Flores-Mendieta}, \citenamefont {Garcia},\ and\ \citenamefont
  {Hernandez}}]{Flores-Mendieta:2021wzh}%
  \BibitemOpen
  \bibfield  {author} {\bibinfo {author} {\bibfnamefont {R.}~\bibnamefont
  {Flores-Mendieta}}, \bibinfo {author} {\bibfnamefont {C.~I.}\ \bibnamefont
  {Garcia}},\ and\ \bibinfo {author} {\bibfnamefont {J.}~\bibnamefont
  {Hernandez}},\ }\bibfield  {title} {\bibinfo {title} {{Baryon axial vector
  current in large-$N_c$ chiral perturbation theory: Complete analysis for
  $N_c=3$}},\ }\href {https://doi.org/10.1103/PhysRevD.103.094032} {\bibfield
  {journal} {\bibinfo  {journal} {Phys. Rev. D}\ }\textbf {\bibinfo {volume}
  {103}},\ \bibinfo {pages} {094032} (\bibinfo {year} {2021})},\ \Eprint
  {https://arxiv.org/abs/2102.06100} {arXiv:2102.06100 [hep-ph]} \BibitemShut
  {NoStop}%
\bibitem [{\citenamefont {Hern\'andez}\ and\ \citenamefont
  {Romero-L\'opez}(2021)}]{Hernandez:2020tbc}%
  \BibitemOpen
  \bibfield  {author} {\bibinfo {author} {\bibfnamefont {P.}~\bibnamefont
  {Hern\'andez}}\ and\ \bibinfo {author} {\bibfnamefont {F.}~\bibnamefont
  {Romero-L\'opez}},\ }\bibfield  {title} {\bibinfo {title} {{The large $N_{c}$
  limit of QCD on the lattice}},\ }\href
  {https://doi.org/10.1140/epja/s10050-021-00374-2} {\bibfield  {journal}
  {\bibinfo  {journal} {Eur. Phys. J. A}\ }\textbf {\bibinfo {volume} {57}},\
  \bibinfo {pages} {52} (\bibinfo {year} {2021})},\ \Eprint
  {https://arxiv.org/abs/2012.03331} {arXiv:2012.03331 [hep-lat]} \BibitemShut
  {NoStop}%
\bibitem [{\citenamefont {Karthik}\ and\ \citenamefont
  {Narayanan}(2022)}]{Karthik:2022fdb}%
  \BibitemOpen
  \bibfield  {author} {\bibinfo {author} {\bibfnamefont {N.}~\bibnamefont
  {Karthik}}\ and\ \bibinfo {author} {\bibfnamefont {R.}~\bibnamefont
  {Narayanan}},\ }\bibfield  {title} {\bibinfo {title} {{Parton physics of the
  large-Nc mesons}},\ }\href {https://doi.org/10.1103/PhysRevD.106.014503}
  {\bibfield  {journal} {\bibinfo  {journal} {Phys. Rev. D}\ }\textbf {\bibinfo
  {volume} {106}},\ \bibinfo {pages} {014503} (\bibinfo {year} {2022})},\
  \Eprint {https://arxiv.org/abs/2205.02252} {arXiv:2205.02252 [hep-lat]}
  \BibitemShut {NoStop}%
\bibitem [{\citenamefont {Teper}(1997)}]{Teper:1997tq}%
  \BibitemOpen
  \bibfield  {author} {\bibinfo {author} {\bibfnamefont {M.}~\bibnamefont
  {Teper}},\ }\bibfield  {title} {\bibinfo {title} {{SU(N(c)) gauge theories
  for all N(c) in three-dimensions and four-dimensions}},\ }\href
  {https://doi.org/10.1016/S0370-2693(97)00181-0} {\bibfield  {journal}
  {\bibinfo  {journal} {Phys. Lett. B}\ }\textbf {\bibinfo {volume} {397}},\
  \bibinfo {pages} {223} (\bibinfo {year} {1997})},\ \Eprint
  {https://arxiv.org/abs/hep-lat/9701003} {arXiv:hep-lat/9701003} \BibitemShut
  {NoStop}%
\bibitem [{\citenamefont {Teper}(1999)}]{Teper:1998te}%
  \BibitemOpen
  \bibfield  {author} {\bibinfo {author} {\bibfnamefont {M.~J.}\ \bibnamefont
  {Teper}},\ }\bibfield  {title} {\bibinfo {title} {{SU(N) gauge theories in
  (2+1)-dimensions}},\ }\href {https://doi.org/10.1103/PhysRevD.59.014512}
  {\bibfield  {journal} {\bibinfo  {journal} {Phys. Rev. D}\ }\textbf {\bibinfo
  {volume} {59}},\ \bibinfo {pages} {014512} (\bibinfo {year} {1999})},\
  \Eprint {https://arxiv.org/abs/hep-lat/9804008} {arXiv:hep-lat/9804008}
  \BibitemShut {NoStop}%
\bibitem [{\citenamefont {Di~Vecchia}(1999)}]{DiVecchia:1999yr}%
  \BibitemOpen
  \bibfield  {author} {\bibinfo {author} {\bibfnamefont {P.}~\bibnamefont
  {Di~Vecchia}},\ }\bibfield  {title} {\bibinfo {title} {{Large N gauge
  theories and AdS / CFT correspondence}},\ }in\ \href@noop {} {\emph {\bibinfo
  {booktitle} {{ICTP Trieste Spring Workshop on Superstrings and Related
  Matters}}}}\ (\bibinfo {year} {1999})\ pp.\ \bibinfo {pages} {1--70},\
  \Eprint {https://arxiv.org/abs/hep-th/9908148} {arXiv:hep-th/9908148}
  \BibitemShut {NoStop}%
\bibitem [{\citenamefont {DeGrand}\ and\ \citenamefont
  {Liu}(2016)}]{DeGrand:2016pur}%
  \BibitemOpen
  \bibfield  {author} {\bibinfo {author} {\bibfnamefont {T.}~\bibnamefont
  {DeGrand}}\ and\ \bibinfo {author} {\bibfnamefont {Y.}~\bibnamefont {Liu}},\
  }\bibfield  {title} {\bibinfo {title} {{Lattice study of large $N_c$ QCD}},\
  }\href {https://doi.org/10.1103/PhysRevD.94.034506} {\bibfield  {journal}
  {\bibinfo  {journal} {Phys. Rev. D}\ }\textbf {\bibinfo {volume} {94}},\
  \bibinfo {pages} {034506} (\bibinfo {year} {2016})},\ \bibinfo {note}
  {[Erratum: Phys.Rev.D 95, 019902 (2017)]},\ \Eprint
  {https://arxiv.org/abs/1606.01277} {arXiv:1606.01277 [hep-lat]} \BibitemShut
  {NoStop}%
\bibitem [{\citenamefont {Garc\'\i{}a~P\'erez}(2020)}]{GarciaPerez:2020gnf}%
  \BibitemOpen
  \bibfield  {author} {\bibinfo {author} {\bibfnamefont {M.}~\bibnamefont
  {Garc\'\i{}a~P\'erez}},\ }\bibfield  {title} {\bibinfo {title} {{Prospects
  for large N gauge theories on the lattice}},\ }\href
  {https://doi.org/10.22323/1.363.0276} {\bibfield  {journal} {\bibinfo
  {journal} {PoS}\ }\textbf {\bibinfo {volume} {LATTICE2019}},\ \bibinfo
  {pages} {276} (\bibinfo {year} {2020})},\ \Eprint
  {https://arxiv.org/abs/2001.10859} {arXiv:2001.10859 [hep-lat]} \BibitemShut
  {NoStop}%
\bibitem [{\citenamefont {Polchinski}(1992)}]{Polchinski:1991tw}%
  \BibitemOpen
  \bibfield  {author} {\bibinfo {author} {\bibfnamefont {J.}~\bibnamefont
  {Polchinski}},\ }\bibfield  {title} {\bibinfo {title} {{High temperature
  limit of the confining phase}},\ }\href
  {https://doi.org/10.1103/PhysRevLett.68.1267} {\bibfield  {journal} {\bibinfo
   {journal} {Phys. Rev. Lett.}\ }\textbf {\bibinfo {volume} {68}},\ \bibinfo
  {pages} {1267} (\bibinfo {year} {1992})},\ \Eprint
  {https://arxiv.org/abs/hep-th/9109007} {arXiv:hep-th/9109007} \BibitemShut
  {NoStop}%
\bibitem [{\citenamefont {Makeenko}(2010)}]{Makeenko:2009dw}%
  \BibitemOpen
  \bibfield  {author} {\bibinfo {author} {\bibfnamefont {Y.}~\bibnamefont
  {Makeenko}},\ }\bibfield  {title} {\bibinfo {title} {{A Brief Introduction to
  Wilson Loops and Large N}},\ }\href
  {https://doi.org/10.1134/S106377881005011X} {\bibfield  {journal} {\bibinfo
  {journal} {Phys. Atom. Nucl.}\ }\textbf {\bibinfo {volume} {73}},\ \bibinfo
  {pages} {878} (\bibinfo {year} {2010})},\ \Eprint
  {https://arxiv.org/abs/0906.4487} {arXiv:0906.4487 [hep-th]} \BibitemShut
  {NoStop}%
\bibitem [{\citenamefont {Maldacena}(1998)}]{Maldacena:1997re}%
  \BibitemOpen
  \bibfield  {author} {\bibinfo {author} {\bibfnamefont {J.~M.}\ \bibnamefont
  {Maldacena}},\ }\bibfield  {title} {\bibinfo {title} {{The Large N limit of
  superconformal field theories and supergravity}},\ }\href
  {https://doi.org/10.1023/A:1026654312961} {\bibfield  {journal} {\bibinfo
  {journal} {Adv. Theor. Math. Phys.}\ }\textbf {\bibinfo {volume} {2}},\
  \bibinfo {pages} {231} (\bibinfo {year} {1998})},\ \Eprint
  {https://arxiv.org/abs/hep-th/9711200} {arXiv:hep-th/9711200} \BibitemShut
  {NoStop}%
\bibitem [{\citenamefont {Aharony}\ \emph {et~al.}(2000)\citenamefont
  {Aharony}, \citenamefont {Gubser}, \citenamefont {Maldacena}, \citenamefont
  {Ooguri},\ and\ \citenamefont {Oz}}]{Aharony:1999ti}%
  \BibitemOpen
  \bibfield  {author} {\bibinfo {author} {\bibfnamefont {O.}~\bibnamefont
  {Aharony}}, \bibinfo {author} {\bibfnamefont {S.~S.}\ \bibnamefont {Gubser}},
  \bibinfo {author} {\bibfnamefont {J.~M.}\ \bibnamefont {Maldacena}}, \bibinfo
  {author} {\bibfnamefont {H.}~\bibnamefont {Ooguri}},\ and\ \bibinfo {author}
  {\bibfnamefont {Y.}~\bibnamefont {Oz}},\ }\bibfield  {title} {\bibinfo
  {title} {{Large N field theories, string theory and gravity}},\ }\href
  {https://doi.org/10.1016/S0370-1573(99)00083-6} {\bibfield  {journal}
  {\bibinfo  {journal} {Phys. Rept.}\ }\textbf {\bibinfo {volume} {323}},\
  \bibinfo {pages} {183} (\bibinfo {year} {2000})},\ \Eprint
  {https://arxiv.org/abs/hep-th/9905111} {arXiv:hep-th/9905111} \BibitemShut
  {NoStop}%
\bibitem [{\citenamefont {Yang}\ and\ \citenamefont
  {Yuan}(2022)}]{Yang:2020hun}%
  \BibitemOpen
  \bibfield  {author} {\bibinfo {author} {\bibfnamefont {Y.}~\bibnamefont
  {Yang}}\ and\ \bibinfo {author} {\bibfnamefont {P.-H.}\ \bibnamefont
  {Yuan}},\ }\bibfield  {title} {\bibinfo {title} {{QCD phase diagram by
  holography}},\ }\href {https://doi.org/10.1016/j.physletb.2022.137212}
  {\bibfield  {journal} {\bibinfo  {journal} {Phys. Lett. B}\ }\textbf
  {\bibinfo {volume} {832}},\ \bibinfo {pages} {137212} (\bibinfo {year}
  {2022})},\ \Eprint {https://arxiv.org/abs/2011.11941} {arXiv:2011.11941
  [hep-th]} \BibitemShut {NoStop}%
\bibitem [{\citenamefont {McLerran}\ and\ \citenamefont
  {Pisarski}(2007)}]{McLerran:2007qj}%
  \BibitemOpen
  \bibfield  {author} {\bibinfo {author} {\bibfnamefont {L.}~\bibnamefont
  {McLerran}}\ and\ \bibinfo {author} {\bibfnamefont {R.~D.}\ \bibnamefont
  {Pisarski}},\ }\bibfield  {title} {\bibinfo {title} {{Phases of cold, dense
  quarks at large N(c)}},\ }\href
  {https://doi.org/10.1016/j.nuclphysa.2007.08.013} {\bibfield  {journal}
  {\bibinfo  {journal} {Nucl. Phys. A}\ }\textbf {\bibinfo {volume} {796}},\
  \bibinfo {pages} {83} (\bibinfo {year} {2007})},\ \Eprint
  {https://arxiv.org/abs/0706.2191} {arXiv:0706.2191 [hep-ph]} \BibitemShut
  {NoStop}%
\bibitem [{\citenamefont {McLerran}\ \emph {et~al.}(2009)\citenamefont
  {McLerran}, \citenamefont {Redlich},\ and\ \citenamefont
  {Sasaki}}]{McLerran:2008ua}%
  \BibitemOpen
  \bibfield  {author} {\bibinfo {author} {\bibfnamefont {L.}~\bibnamefont
  {McLerran}}, \bibinfo {author} {\bibfnamefont {K.}~\bibnamefont {Redlich}},\
  and\ \bibinfo {author} {\bibfnamefont {C.}~\bibnamefont {Sasaki}},\
  }\bibfield  {title} {\bibinfo {title} {{Quarkyonic Matter and Chiral Symmetry
  Breaking}},\ }\href {https://doi.org/10.1016/j.nuclphysa.2009.04.001}
  {\bibfield  {journal} {\bibinfo  {journal} {Nucl. Phys. A}\ }\textbf
  {\bibinfo {volume} {824}},\ \bibinfo {pages} {86} (\bibinfo {year} {2009})},\
  \Eprint {https://arxiv.org/abs/0812.3585} {arXiv:0812.3585 [hep-ph]}
  \BibitemShut {NoStop}%
\bibitem [{\citenamefont {Torrieri}\ and\ \citenamefont
  {Mishustin}(2010)}]{Torrieri:2010gz}%
  \BibitemOpen
  \bibfield  {author} {\bibinfo {author} {\bibfnamefont {G.}~\bibnamefont
  {Torrieri}}\ and\ \bibinfo {author} {\bibfnamefont {I.}~\bibnamefont
  {Mishustin}},\ }\bibfield  {title} {\bibinfo {title} {{The nuclear liquid-gas
  phase transition at large $N_c$ in the Van der Waals approximation}},\ }\href
  {https://doi.org/10.1103/PhysRevC.82.055202} {\bibfield  {journal} {\bibinfo
  {journal} {Phys. Rev. C}\ }\textbf {\bibinfo {volume} {82}},\ \bibinfo
  {pages} {055202} (\bibinfo {year} {2010})},\ \Eprint
  {https://arxiv.org/abs/1006.2471} {arXiv:1006.2471 [nucl-th]} \BibitemShut
  {NoStop}%
\bibitem [{\citenamefont {Buisseret}\ and\ \citenamefont
  {Lacroix}(2012)}]{Buisseret:2011ms}%
  \BibitemOpen
  \bibfield  {author} {\bibinfo {author} {\bibfnamefont {F.}~\bibnamefont
  {Buisseret}}\ and\ \bibinfo {author} {\bibfnamefont {G.}~\bibnamefont
  {Lacroix}},\ }\bibfield  {title} {\bibinfo {title} {{A large-$N_c$ PNJL model
  with explicit Z$_{N_c}$ symmetry}},\ }\href
  {https://doi.org/10.1103/PhysRevD.85.016009} {\bibfield  {journal} {\bibinfo
  {journal} {Phys. Rev. D}\ }\textbf {\bibinfo {volume} {85}},\ \bibinfo
  {pages} {016009} (\bibinfo {year} {2012})},\ \Eprint
  {https://arxiv.org/abs/1107.4672} {arXiv:1107.4672 [hep-ph]} \BibitemShut
  {NoStop}%
\bibitem [{\citenamefont {Kov\'acs}\ \emph {et~al.}(2022)\citenamefont
  {Kov\'acs}, \citenamefont {Kov\'acs},\ and\ \citenamefont
  {Giacosa}}]{Kovacs:2022zcl}%
  \BibitemOpen
  \bibfield  {author} {\bibinfo {author} {\bibfnamefont {P.}~\bibnamefont
  {Kov\'acs}}, \bibinfo {author} {\bibfnamefont {G.}~\bibnamefont {Kov\'acs}},\
  and\ \bibinfo {author} {\bibfnamefont {F.}~\bibnamefont {Giacosa}},\
  }\bibfield  {title} {\bibinfo {title} {{Fate of the critical endpoint at
  large Nc}},\ }\href {https://doi.org/10.1103/PhysRevD.106.116016} {\bibfield
  {journal} {\bibinfo  {journal} {Phys. Rev. D}\ }\textbf {\bibinfo {volume}
  {106}},\ \bibinfo {pages} {116016} (\bibinfo {year} {2022})},\ \Eprint
  {https://arxiv.org/abs/2209.09568} {arXiv:2209.09568 [hep-ph]} \BibitemShut
  {NoStop}%
\bibitem [{\citenamefont {Giacosa}\ and\ \citenamefont
  {Pagliara}(2017)}]{Giacosa:2017mis}%
  \BibitemOpen
  \bibfield  {author} {\bibinfo {author} {\bibfnamefont {F.}~\bibnamefont
  {Giacosa}}\ and\ \bibinfo {author} {\bibfnamefont {G.}~\bibnamefont
  {Pagliara}},\ }\bibfield  {title} {\bibinfo {title} {{Neutron stars in the
  large-$N_{c}$ limit}},\ }\href
  {https://doi.org/10.1016/j.nuclphysa.2017.08.006} {\bibfield  {journal}
  {\bibinfo  {journal} {Nucl. Phys. A}\ }\textbf {\bibinfo {volume} {968}},\
  \bibinfo {pages} {366} (\bibinfo {year} {2017})},\ \Eprint
  {https://arxiv.org/abs/1707.02644} {arXiv:1707.02644 [nucl-th]} \BibitemShut
  {NoStop}%
\bibitem [{\citenamefont {Margueron}\ \emph {et~al.}(2021)\citenamefont
  {Margueron}, \citenamefont {Hansen}, \citenamefont {Proust},\ and\
  \citenamefont {Chanfray}}]{Margueron:2021dtx}%
  \BibitemOpen
  \bibfield  {author} {\bibinfo {author} {\bibfnamefont {J.}~\bibnamefont
  {Margueron}}, \bibinfo {author} {\bibfnamefont {H.}~\bibnamefont {Hansen}},
  \bibinfo {author} {\bibfnamefont {P.}~\bibnamefont {Proust}},\ and\ \bibinfo
  {author} {\bibfnamefont {G.}~\bibnamefont {Chanfray}},\ }\bibfield  {title}
  {\bibinfo {title} {{Quarkyonic stars with isospin-flavor asymmetry}},\ }\href
  {https://doi.org/10.1103/PhysRevC.104.055803} {\bibfield  {journal} {\bibinfo
   {journal} {Phys. Rev. C}\ }\textbf {\bibinfo {volume} {104}},\ \bibinfo
  {pages} {055803} (\bibinfo {year} {2021})},\ \Eprint
  {https://arxiv.org/abs/2103.10209} {arXiv:2103.10209 [nucl-th]} \BibitemShut
  {NoStop}%
\bibitem [{\citenamefont {Pisarski}(2002)}]{Pisarski:2002ji}%
  \BibitemOpen
  \bibfield  {author} {\bibinfo {author} {\bibfnamefont {R.~D.}\ \bibnamefont
  {Pisarski}},\ }\bibfield  {title} {\bibinfo {title} {{Notes on the
  deconfining phase transition}},\ }in\ \href@noop {} {\emph {\bibinfo
  {booktitle} {{Cargese Summer School on QCD Perspectives on Hot and Dense
  Matter}}}}\ (\bibinfo {year} {2002})\ pp.\ \bibinfo {pages} {353--384},\
  \Eprint {https://arxiv.org/abs/hep-ph/0203271} {arXiv:hep-ph/0203271}
  \BibitemShut {NoStop}%
\bibitem [{\citenamefont {Fukushima}(2004)}]{Fukushima:2003fw}%
  \BibitemOpen
  \bibfield  {author} {\bibinfo {author} {\bibfnamefont {K.}~\bibnamefont
  {Fukushima}},\ }\bibfield  {title} {\bibinfo {title} {{Chiral effective model
  with the Polyakov loop}},\ }\href
  {https://doi.org/10.1016/j.physletb.2004.04.027} {\bibfield  {journal}
  {\bibinfo  {journal} {Phys. Lett.}\ }\textbf {\bibinfo {volume} {B591}},\
  \bibinfo {pages} {277} (\bibinfo {year} {2004})},\ \Eprint
  {https://arxiv.org/abs/hep-ph/0310121} {arXiv:hep-ph/0310121 [hep-ph]}
  \BibitemShut {NoStop}%
%%CITATION = HEP-PH/0310121;%%
\bibitem [{\citenamefont {Fukushima}\ and\ \citenamefont
  {Skokov}(2017)}]{Fukushima:2017csk}%
  \BibitemOpen
  \bibfield  {author} {\bibinfo {author} {\bibfnamefont {K.}~\bibnamefont
  {Fukushima}}\ and\ \bibinfo {author} {\bibfnamefont {V.}~\bibnamefont
  {Skokov}},\ }\bibfield  {title} {\bibinfo {title} {{Polyakov loop modeling
  for hot QCD}},\ }\href {https://doi.org/10.1016/j.ppnp.2017.05.002}
  {\bibfield  {journal} {\bibinfo  {journal} {Prog. Part. Nucl. Phys.}\
  }\textbf {\bibinfo {volume} {96}},\ \bibinfo {pages} {154} (\bibinfo {year}
  {2017})},\ \Eprint {https://arxiv.org/abs/1705.00718} {arXiv:1705.00718
  [hep-ph]} \BibitemShut {NoStop}%
\bibitem [{\citenamefont {McLerran}\ and\ \citenamefont
  {Svetitsky}(1981)}]{McLerran:1981pb}%
  \BibitemOpen
  \bibfield  {author} {\bibinfo {author} {\bibfnamefont {L.~D.}\ \bibnamefont
  {McLerran}}\ and\ \bibinfo {author} {\bibfnamefont {B.}~\bibnamefont
  {Svetitsky}},\ }\bibfield  {title} {\bibinfo {title} {{Quark Liberation at
  High Temperature: A Monte Carlo Study of SU(2) Gauge Theory}},\ }\href
  {https://doi.org/10.1103/PhysRevD.24.450} {\bibfield  {journal} {\bibinfo
  {journal} {Phys. Rev. D}\ }\textbf {\bibinfo {volume} {24}},\ \bibinfo
  {pages} {450} (\bibinfo {year} {1981})}\BibitemShut {NoStop}%
\bibitem [{\citenamefont {Ratti}()}]{QuarkGluonPlasma}%
  \BibitemOpen
  \bibfield  {author} {\bibinfo {author} {\bibfnamefont {C.}~\bibnamefont
  {Ratti}},\ }\bibfield  {title} {\bibinfo {title} {Thermodynamics of the
  quark-gluon plasma},\ }in\ \href@noop {} {\emph {\bibinfo {booktitle}
  {International School on Quark-Gluon Plasma and Heavy Ion Collisions : past,
  present, future}}}\BibitemShut {NoStop}%
\bibitem [{\citenamefont {Zee}(2016)}]{Zee:2016fuk}%
  \BibitemOpen
  \bibfield  {author} {\bibinfo {author} {\bibfnamefont {A.}~\bibnamefont
  {Zee}},\ }\href@noop {} {\emph {\bibinfo {title} {{Group Theory in a Nutshell
  for Physicists}}}}\ (\bibinfo  {publisher} {Princeton University Press},\
  \bibinfo {address} {USA},\ \bibinfo {year} {2016})\BibitemShut {NoStop}%
\bibitem [{\citenamefont {Hidaka}\ and\ \citenamefont
  {Pisarski}(2010)}]{Hidaka:2009ma}%
  \BibitemOpen
  \bibfield  {author} {\bibinfo {author} {\bibfnamefont {Y.}~\bibnamefont
  {Hidaka}}\ and\ \bibinfo {author} {\bibfnamefont {R.~D.}\ \bibnamefont
  {Pisarski}},\ }\bibfield  {title} {\bibinfo {title} {{Small shear viscosity
  in the semi quark gluon plasma}},\ }\href
  {https://doi.org/10.1103/PhysRevD.81.076002} {\bibfield  {journal} {\bibinfo
  {journal} {Phys. Rev. D}\ }\textbf {\bibinfo {volume} {81}},\ \bibinfo
  {pages} {076002} (\bibinfo {year} {2010})},\ \Eprint
  {https://arxiv.org/abs/0912.0940} {arXiv:0912.0940 [hep-ph]} \BibitemShut
  {NoStop}%
\bibitem [{\citenamefont {Dumitru}\ \emph {et~al.}(2004)\citenamefont
  {Dumitru}, \citenamefont {Hatta}, \citenamefont {Lenaghan}, \citenamefont
  {Orginos},\ and\ \citenamefont {Pisarski}}]{Dumitru:2003hp}%
  \BibitemOpen
  \bibfield  {author} {\bibinfo {author} {\bibfnamefont {A.}~\bibnamefont
  {Dumitru}}, \bibinfo {author} {\bibfnamefont {Y.}~\bibnamefont {Hatta}},
  \bibinfo {author} {\bibfnamefont {J.}~\bibnamefont {Lenaghan}}, \bibinfo
  {author} {\bibfnamefont {K.}~\bibnamefont {Orginos}},\ and\ \bibinfo {author}
  {\bibfnamefont {R.~D.}\ \bibnamefont {Pisarski}},\ }\bibfield  {title}
  {\bibinfo {title} {{Deconfining phase transition as a matrix model of
  renormalized Polyakov loops}},\ }\href
  {https://doi.org/10.1103/PhysRevD.70.034511} {\bibfield  {journal} {\bibinfo
  {journal} {Phys. Rev. D}\ }\textbf {\bibinfo {volume} {70}},\ \bibinfo
  {pages} {034511} (\bibinfo {year} {2004})},\ \Eprint
  {https://arxiv.org/abs/hep-th/0311223} {arXiv:hep-th/0311223} \BibitemShut
  {NoStop}%
\bibitem [{\citenamefont {Gupta}\ \emph {et~al.}(2008)\citenamefont {Gupta},
  \citenamefont {Huebner},\ and\ \citenamefont {Kaczmarek}}]{Gupta:2007ax}%
  \BibitemOpen
  \bibfield  {author} {\bibinfo {author} {\bibfnamefont {S.}~\bibnamefont
  {Gupta}}, \bibinfo {author} {\bibfnamefont {K.}~\bibnamefont {Huebner}},\
  and\ \bibinfo {author} {\bibfnamefont {O.}~\bibnamefont {Kaczmarek}},\
  }\bibfield  {title} {\bibinfo {title} {{Renormalized Polyakov loops in many
  representations}},\ }\href {https://doi.org/10.1103/PhysRevD.77.034503}
  {\bibfield  {journal} {\bibinfo  {journal} {Phys. Rev. D}\ }\textbf {\bibinfo
  {volume} {77}},\ \bibinfo {pages} {034503} (\bibinfo {year} {2008})},\
  \Eprint {https://arxiv.org/abs/0711.2251} {arXiv:0711.2251 [hep-lat]}
  \BibitemShut {NoStop}%
\bibitem [{\citenamefont {Abuki}\ and\ \citenamefont
  {Fukushima}(2009)}]{Abuki:2009dt}%
  \BibitemOpen
  \bibfield  {author} {\bibinfo {author} {\bibfnamefont {H.}~\bibnamefont
  {Abuki}}\ and\ \bibinfo {author} {\bibfnamefont {K.}~\bibnamefont
  {Fukushima}},\ }\bibfield  {title} {\bibinfo {title} {{Gauge dynamics in the
  PNJL model: Color neutrality and Casimir scaling}},\ }\href
  {https://doi.org/10.1016/j.physletb.2009.04.078} {\bibfield  {journal}
  {\bibinfo  {journal} {Phys. Lett. B}\ }\textbf {\bibinfo {volume} {676}},\
  \bibinfo {pages} {57} (\bibinfo {year} {2009})},\ \Eprint
  {https://arxiv.org/abs/0901.4821} {arXiv:0901.4821 [hep-ph]} \BibitemShut
  {NoStop}%
\bibitem [{\citenamefont {Ayyar}\ \emph {et~al.}(2018)\citenamefont {Ayyar},
  \citenamefont {DeGrand}, \citenamefont {Hackett}, \citenamefont {Jay},
  \citenamefont {Neil}, \citenamefont {Shamir},\ and\ \citenamefont
  {Svetitsky}}]{Ayyar:2018ppa}%
  \BibitemOpen
  \bibfield  {author} {\bibinfo {author} {\bibfnamefont {V.}~\bibnamefont
  {Ayyar}}, \bibinfo {author} {\bibfnamefont {T.}~\bibnamefont {DeGrand}},
  \bibinfo {author} {\bibfnamefont {D.~C.}\ \bibnamefont {Hackett}}, \bibinfo
  {author} {\bibfnamefont {W.~I.}\ \bibnamefont {Jay}}, \bibinfo {author}
  {\bibfnamefont {E.~T.}\ \bibnamefont {Neil}}, \bibinfo {author}
  {\bibfnamefont {Y.}~\bibnamefont {Shamir}},\ and\ \bibinfo {author}
  {\bibfnamefont {B.}~\bibnamefont {Svetitsky}},\ }\bibfield  {title} {\bibinfo
  {title} {{Finite-temperature phase structure of SU(4) gauge theory with
  multiple fermion representations}},\ }\href
  {https://doi.org/10.1103/PhysRevD.97.114502} {\bibfield  {journal} {\bibinfo
  {journal} {Phys. Rev. D}\ }\textbf {\bibinfo {volume} {97}},\ \bibinfo
  {pages} {114502} (\bibinfo {year} {2018})},\ \Eprint
  {https://arxiv.org/abs/1802.09644} {arXiv:1802.09644 [hep-lat]} \BibitemShut
  {NoStop}%
\bibitem [{\citenamefont {Megias}\ \emph {et~al.}(2006)\citenamefont {Megias},
  \citenamefont {Ruiz~Arriola},\ and\ \citenamefont {Salcedo}}]{Megias:2004hj}%
  \BibitemOpen
  \bibfield  {author} {\bibinfo {author} {\bibfnamefont {E.}~\bibnamefont
  {Megias}}, \bibinfo {author} {\bibfnamefont {E.}~\bibnamefont
  {Ruiz~Arriola}},\ and\ \bibinfo {author} {\bibfnamefont {L.~L.}\ \bibnamefont
  {Salcedo}},\ }\bibfield  {title} {\bibinfo {title} {{Polyakov loop in chiral
  quark models at finite temperature}},\ }\href
  {https://doi.org/10.1103/PhysRevD.74.065005} {\bibfield  {journal} {\bibinfo
  {journal} {Phys. Rev. D}\ }\textbf {\bibinfo {volume} {74}},\ \bibinfo
  {pages} {065005} (\bibinfo {year} {2006})},\ \Eprint
  {https://arxiv.org/abs/hep-ph/0412308} {arXiv:hep-ph/0412308} \BibitemShut
  {NoStop}%
\bibitem [{\citenamefont {Lo}\ \emph {et~al.}(2021)\citenamefont {Lo},
  \citenamefont {Redlich},\ and\ \citenamefont {Sasaki}}]{Lo:2021qkw}%
  \BibitemOpen
  \bibfield  {author} {\bibinfo {author} {\bibfnamefont {P.~M.}\ \bibnamefont
  {Lo}}, \bibinfo {author} {\bibfnamefont {K.}~\bibnamefont {Redlich}},\ and\
  \bibinfo {author} {\bibfnamefont {C.}~\bibnamefont {Sasaki}},\ }\bibfield
  {title} {\bibinfo {title} {{Fluctuations of the order parameter in an
  $SU(N_c)$ effective model}},\ }\href
  {https://doi.org/10.1103/PhysRevD.103.074026} {\bibfield  {journal} {\bibinfo
   {journal} {Phys. Rev. D}\ }\textbf {\bibinfo {volume} {103}},\ \bibinfo
  {pages} {074026} (\bibinfo {year} {2021})},\ \Eprint
  {https://arxiv.org/abs/2101.12663} {arXiv:2101.12663 [hep-ph]} \BibitemShut
  {NoStop}%
\bibitem [{\citenamefont {Mocsy}\ \emph {et~al.}(2004)\citenamefont {Mocsy},
  \citenamefont {Sannino},\ and\ \citenamefont {Tuominen}}]{Mocsy:2003qw}%
  \BibitemOpen
  \bibfield  {author} {\bibinfo {author} {\bibfnamefont {A.}~\bibnamefont
  {Mocsy}}, \bibinfo {author} {\bibfnamefont {F.}~\bibnamefont {Sannino}},\
  and\ \bibinfo {author} {\bibfnamefont {K.}~\bibnamefont {Tuominen}},\
  }\bibfield  {title} {\bibinfo {title} {{Confinement versus chiral
  symmetry}},\ }\href {https://doi.org/10.1103/PhysRevLett.92.182302}
  {\bibfield  {journal} {\bibinfo  {journal} {Phys. Rev. Lett.}\ }\textbf
  {\bibinfo {volume} {92}},\ \bibinfo {pages} {182302} (\bibinfo {year}
  {2004})},\ \Eprint {https://arxiv.org/abs/hep-ph/0308135}
  {arXiv:hep-ph/0308135} \BibitemShut {NoStop}%
\bibitem [{\citenamefont {Ratti}\ \emph {et~al.}(2006)\citenamefont {Ratti},
  \citenamefont {Thaler},\ and\ \citenamefont {Weise}}]{Ratti:2005jh}%
  \BibitemOpen
  \bibfield  {author} {\bibinfo {author} {\bibfnamefont {C.}~\bibnamefont
  {Ratti}}, \bibinfo {author} {\bibfnamefont {M.~A.}\ \bibnamefont {Thaler}},\
  and\ \bibinfo {author} {\bibfnamefont {W.}~\bibnamefont {Weise}},\ }\bibfield
   {title} {\bibinfo {title} {{Phases of QCD: Lattice thermodynamics and a
  field theoretical model}},\ }\href
  {https://doi.org/10.1103/PhysRevD.73.014019} {\bibfield  {journal} {\bibinfo
  {journal} {Phys. Rev. D}\ }\textbf {\bibinfo {volume} {73}},\ \bibinfo
  {pages} {014019} (\bibinfo {year} {2006})},\ \Eprint
  {https://arxiv.org/abs/hep-ph/0506234} {arXiv:hep-ph/0506234} \BibitemShut
  {NoStop}%
\bibitem [{\citenamefont {Hansen}\ \emph {et~al.}(2007)\citenamefont {Hansen},
  \citenamefont {Alberico}, \citenamefont {Beraudo}, \citenamefont {Molinari},
  \citenamefont {Nardi},\ and\ \citenamefont {Ratti}}]{Hansen:2006ee}%
  \BibitemOpen
  \bibfield  {author} {\bibinfo {author} {\bibfnamefont {H.}~\bibnamefont
  {Hansen}}, \bibinfo {author} {\bibfnamefont {W.~M.}\ \bibnamefont
  {Alberico}}, \bibinfo {author} {\bibfnamefont {A.}~\bibnamefont {Beraudo}},
  \bibinfo {author} {\bibfnamefont {A.}~\bibnamefont {Molinari}}, \bibinfo
  {author} {\bibfnamefont {M.}~\bibnamefont {Nardi}},\ and\ \bibinfo {author}
  {\bibfnamefont {C.}~\bibnamefont {Ratti}},\ }\bibfield  {title} {\bibinfo
  {title} {{Mesonic correlation functions at finite temperature and density in
  the Nambu-Jona-Lasinio model with a Polyakov loop}},\ }\href
  {https://doi.org/10.1103/PhysRevD.75.065004} {\bibfield  {journal} {\bibinfo
  {journal} {Phys. Rev. D}\ }\textbf {\bibinfo {volume} {75}},\ \bibinfo
  {pages} {065004} (\bibinfo {year} {2007})},\ \Eprint
  {https://arxiv.org/abs/hep-ph/0609116} {arXiv:hep-ph/0609116} \BibitemShut
  {NoStop}%
\bibitem [{\citenamefont {Moreira}\ \emph {et~al.}(2018)\citenamefont
  {Moreira}, \citenamefont {Morais}, \citenamefont {Hiller}, \citenamefont
  {Osipov},\ and\ \citenamefont {Blin}}]{Moreira:2018xsp}%
  \BibitemOpen
  \bibfield  {author} {\bibinfo {author} {\bibfnamefont {J.}~\bibnamefont
  {Moreira}}, \bibinfo {author} {\bibfnamefont {J.}~\bibnamefont {Morais}},
  \bibinfo {author} {\bibfnamefont {B.}~\bibnamefont {Hiller}}, \bibinfo
  {author} {\bibfnamefont {A.~A.}\ \bibnamefont {Osipov}},\ and\ \bibinfo
  {author} {\bibfnamefont {A.~H.}\ \bibnamefont {Blin}},\ }\bibfield  {title}
  {\bibinfo {title} {{Thermodynamical properties of strongly interacting matter
  in a model with explicit chiral symmetry breaking interactions}},\ }\href
  {https://doi.org/10.1103/PhysRevD.98.074010} {\bibfield  {journal} {\bibinfo
  {journal} {Phys. Rev. D}\ }\textbf {\bibinfo {volume} {98}},\ \bibinfo
  {pages} {074010} (\bibinfo {year} {2018})},\ \Eprint
  {https://arxiv.org/abs/1806.00327} {arXiv:1806.00327 [hep-ph]} \BibitemShut
  {NoStop}%
\bibitem [{\citenamefont {C\^amara~Pereira}\ \emph {et~al.}(2020)\citenamefont
  {C\^amara~Pereira}, \citenamefont {Moreira},\ and\ \citenamefont
  {Costa}}]{CamaraPereira:2020rtu}%
  \BibitemOpen
  \bibfield  {author} {\bibinfo {author} {\bibfnamefont {R.}~\bibnamefont
  {C\^amara~Pereira}}, \bibinfo {author} {\bibfnamefont {J.~a.}\ \bibnamefont
  {Moreira}},\ and\ \bibinfo {author} {\bibfnamefont {P.}~\bibnamefont
  {Costa}},\ }\bibfield  {title} {\bibinfo {title} {{The strange critical
  endpoint and isentropic trajectories in an extended PNJL model with eight
  Quark interactions}},\ }\href
  {https://doi.org/10.1140/epja/s10050-020-00223-8} {\bibfield  {journal}
  {\bibinfo  {journal} {Eur. Phys. J. A}\ }\textbf {\bibinfo {volume} {56}},\
  \bibinfo {pages} {214} (\bibinfo {year} {2020})},\ \Eprint
  {https://arxiv.org/abs/2006.02385} {arXiv:2006.02385 [hep-ph]} \BibitemShut
  {NoStop}%
\bibitem [{\citenamefont {Pereira}(2021)}]{Pereira:2021xxv}%
  \BibitemOpen
  \bibfield  {author} {\bibinfo {author} {\bibfnamefont {R.~C.}\ \bibnamefont
  {Pereira}},\ }\emph {\bibinfo {title} {{Quantum Chromodynamics Phase Diagram
  Under Extreme Conditions}}},\ \href@noop {} {Ph.D. thesis},\ \bibinfo
  {school} {Coimbra U.} (\bibinfo {year} {2021})\BibitemShut {NoStop}%
\bibitem [{\citenamefont {Ferreira}\ \emph {et~al.}(2021)\citenamefont
  {Ferreira}, \citenamefont {C\^amara~Pereira},\ and\ \citenamefont
  {Provid\^encia}}]{Ferreira:2021osk}%
  \BibitemOpen
  \bibfield  {author} {\bibinfo {author} {\bibfnamefont {M.}~\bibnamefont
  {Ferreira}}, \bibinfo {author} {\bibfnamefont {R.}~\bibnamefont
  {C\^amara~Pereira}},\ and\ \bibinfo {author} {\bibfnamefont {C.}~\bibnamefont
  {Provid\^encia}},\ }\bibfield  {title} {\bibinfo {title} {{Hybrid stars with
  large strange quark cores constrained by GW170817}},\ }\href
  {https://doi.org/10.1103/PhysRevD.103.123020} {\bibfield  {journal} {\bibinfo
   {journal} {Phys. Rev. D}\ }\textbf {\bibinfo {volume} {103}},\ \bibinfo
  {pages} {123020} (\bibinfo {year} {2021})},\ \Eprint
  {https://arxiv.org/abs/2105.06239} {arXiv:2105.06239 [nucl-th]} \BibitemShut
  {NoStop}%
\bibitem [{\citenamefont {Haas}\ \emph {et~al.}(2013)\citenamefont {Haas},
  \citenamefont {Stiele}, \citenamefont {Braun}, \citenamefont {Pawlowski},\
  and\ \citenamefont {Schaffner-Bielich}}]{Haas:2013qwp}%
  \BibitemOpen
  \bibfield  {author} {\bibinfo {author} {\bibfnamefont {L.~M.}\ \bibnamefont
  {Haas}}, \bibinfo {author} {\bibfnamefont {R.}~\bibnamefont {Stiele}},
  \bibinfo {author} {\bibfnamefont {J.}~\bibnamefont {Braun}}, \bibinfo
  {author} {\bibfnamefont {J.~M.}\ \bibnamefont {Pawlowski}},\ and\ \bibinfo
  {author} {\bibfnamefont {J.}~\bibnamefont {Schaffner-Bielich}},\ }\bibfield
  {title} {\bibinfo {title} {{Improved Polyakov-loop potential for effective
  models from functional calculations}},\ }\href
  {https://doi.org/10.1103/PhysRevD.87.076004} {\bibfield  {journal} {\bibinfo
  {journal} {Phys. Rev. D}\ }\textbf {\bibinfo {volume} {87}},\ \bibinfo
  {pages} {076004} (\bibinfo {year} {2013})},\ \Eprint
  {https://arxiv.org/abs/1302.1993} {arXiv:1302.1993 [hep-ph]} \BibitemShut
  {NoStop}%
\bibitem [{\citenamefont {Lo}\ \emph {et~al.}(2018)\citenamefont {Lo},
  \citenamefont {Szyma\'nski}, \citenamefont {Redlich},\ and\ \citenamefont
  {Sasaki}}]{Lo:2018wdo}%
  \BibitemOpen
  \bibfield  {author} {\bibinfo {author} {\bibfnamefont {P.~M.}\ \bibnamefont
  {Lo}}, \bibinfo {author} {\bibfnamefont {M.}~\bibnamefont {Szyma\'nski}},
  \bibinfo {author} {\bibfnamefont {K.}~\bibnamefont {Redlich}},\ and\ \bibinfo
  {author} {\bibfnamefont {C.}~\bibnamefont {Sasaki}},\ }\bibfield  {title}
  {\bibinfo {title} {{Polyakov loop fluctuations in the presence of external
  fields}},\ }\href {https://doi.org/10.1103/PhysRevD.97.114006} {\bibfield
  {journal} {\bibinfo  {journal} {Phys. Rev. D}\ }\textbf {\bibinfo {volume}
  {97}},\ \bibinfo {pages} {114006} (\bibinfo {year} {2018})},\ \Eprint
  {https://arxiv.org/abs/1801.08040} {arXiv:1801.08040 [hep-ph]} \BibitemShut
  {NoStop}%
\bibitem [{\citenamefont {R{\"o}{\ss}ner}(2006)}]{Rossnerthesis}%
  \BibitemOpen
  \bibfield  {author} {\bibinfo {author} {\bibfnamefont {S.}~\bibnamefont
  {R{\"o}{\ss}ner}},\ }\emph {\bibinfo {title} {Field theoretical modelling of
  the QCD phase diagram}},\ \href@noop {} {Ph.D. thesis},\ \bibinfo  {school}
  {Technische Universit{\"a}t M{\"u}nchen} (\bibinfo {year} {2006})\BibitemShut
  {NoStop}%
\bibitem [{\citenamefont {Hell}\ \emph {et~al.}(2010)\citenamefont {Hell},
  \citenamefont {Rossner}, \citenamefont {Cristoforetti},\ and\ \citenamefont
  {Weise}}]{Hell:2009by}%
  \BibitemOpen
  \bibfield  {author} {\bibinfo {author} {\bibfnamefont {T.}~\bibnamefont
  {Hell}}, \bibinfo {author} {\bibfnamefont {S.}~\bibnamefont {Rossner}},
  \bibinfo {author} {\bibfnamefont {M.}~\bibnamefont {Cristoforetti}},\ and\
  \bibinfo {author} {\bibfnamefont {W.}~\bibnamefont {Weise}},\ }\bibfield
  {title} {\bibinfo {title} {{Thermodynamics of a three-flavor nonlocal
  Polyakov-Nambu-Jona-Lasinio model}},\ }\href
  {https://doi.org/10.1103/PhysRevD.81.074034} {\bibfield  {journal} {\bibinfo
  {journal} {Phys. Rev. D}\ }\textbf {\bibinfo {volume} {81}},\ \bibinfo
  {pages} {074034} (\bibinfo {year} {2010})},\ \Eprint
  {https://arxiv.org/abs/0911.3510} {arXiv:0911.3510 [hep-ph]} \BibitemShut
  {NoStop}%
\bibitem [{\citenamefont {Hell}(2010)}]{Hellthesis}%
  \BibitemOpen
  \bibfield  {author} {\bibinfo {author} {\bibfnamefont {T.}~\bibnamefont
  {Hell}},\ }\emph {\bibinfo {title} {Modeling the Thermodynamics of QCD}},\
  \href@noop {} {Ph.D. thesis},\ \bibinfo  {school} {Technische Universit{\"a}t
  M{\"u}nchen} (\bibinfo {year} {2010})\BibitemShut {NoStop}%
\bibitem [{\citenamefont {Sasaki}\ and\ \citenamefont
  {Redlich}(2012)}]{Sasaki:2012bi}%
  \BibitemOpen
  \bibfield  {author} {\bibinfo {author} {\bibfnamefont {C.}~\bibnamefont
  {Sasaki}}\ and\ \bibinfo {author} {\bibfnamefont {K.}~\bibnamefont
  {Redlich}},\ }\bibfield  {title} {\bibinfo {title} {{An Effective gluon
  potential and hybrid approach to Yang-Mills thermodynamics}},\ }\href
  {https://doi.org/10.1103/PhysRevD.86.014007} {\bibfield  {journal} {\bibinfo
  {journal} {Phys. Rev. D}\ }\textbf {\bibinfo {volume} {86}},\ \bibinfo
  {pages} {014007} (\bibinfo {year} {2012})},\ \Eprint
  {https://arxiv.org/abs/1204.4330} {arXiv:1204.4330 [hep-ph]} \BibitemShut
  {NoStop}%
\bibitem [{\citenamefont {Hell}\ \emph {et~al.}(2009)\citenamefont {Hell},
  \citenamefont {Roessner}, \citenamefont {Cristoforetti},\ and\ \citenamefont
  {Weise}}]{Hell:2008cc}%
  \BibitemOpen
  \bibfield  {author} {\bibinfo {author} {\bibfnamefont {T.}~\bibnamefont
  {Hell}}, \bibinfo {author} {\bibfnamefont {S.}~\bibnamefont {Roessner}},
  \bibinfo {author} {\bibfnamefont {M.}~\bibnamefont {Cristoforetti}},\ and\
  \bibinfo {author} {\bibfnamefont {W.}~\bibnamefont {Weise}},\ }\bibfield
  {title} {\bibinfo {title} {{Dynamics and thermodynamics of a non-local PNJL
  model with running coupling}},\ }\href
  {https://doi.org/10.1103/PhysRevD.79.014022} {\bibfield  {journal} {\bibinfo
  {journal} {Phys. Rev. D}\ }\textbf {\bibinfo {volume} {79}},\ \bibinfo
  {pages} {014022} (\bibinfo {year} {2009})},\ \Eprint
  {https://arxiv.org/abs/0810.1099} {arXiv:0810.1099 [hep-ph]} \BibitemShut
  {NoStop}%
\bibitem [{\citenamefont {Drouffe}\ and\ \citenamefont
  {Zuber}(1983)}]{DROUFFE19831}%
  \BibitemOpen
  \bibfield  {author} {\bibinfo {author} {\bibfnamefont {J.-M.}\ \bibnamefont
  {Drouffe}}\ and\ \bibinfo {author} {\bibfnamefont {J.-B.}\ \bibnamefont
  {Zuber}},\ }\bibfield  {title} {\bibinfo {title} {Strong coupling and mean
  field methods in lattice gauge theories},\ }\href
  {https://doi.org/https://doi.org/10.1016/0370-1573(83)90034-0} {\bibfield
  {journal} {\bibinfo  {journal} {Physics Reports}\ }\textbf {\bibinfo {volume}
  {102}},\ \bibinfo {pages} {1} (\bibinfo {year} {1983})}\BibitemShut {NoStop}%
\bibitem [{\citenamefont {Roessner}\ \emph {et~al.}(2007)\citenamefont
  {Roessner}, \citenamefont {Ratti},\ and\ \citenamefont
  {Weise}}]{Roessner:2006xn}%
  \BibitemOpen
  \bibfield  {author} {\bibinfo {author} {\bibfnamefont {S.}~\bibnamefont
  {Roessner}}, \bibinfo {author} {\bibfnamefont {C.}~\bibnamefont {Ratti}},\
  and\ \bibinfo {author} {\bibfnamefont {W.}~\bibnamefont {Weise}},\ }\bibfield
   {title} {\bibinfo {title} {{Polyakov loop, diquarks and the two-flavour
  phase diagram}},\ }\href {https://doi.org/10.1103/PhysRevD.75.034007}
  {\bibfield  {journal} {\bibinfo  {journal} {Phys. Rev. D}\ }\textbf {\bibinfo
  {volume} {75}},\ \bibinfo {pages} {034007} (\bibinfo {year} {2007})},\
  \Eprint {https://arxiv.org/abs/hep-ph/0609281} {arXiv:hep-ph/0609281}
  \BibitemShut {NoStop}%
\bibitem [{\citenamefont {Haag}(1958)}]{Haag:1958vt}%
  \BibitemOpen
  \bibfield  {author} {\bibinfo {author} {\bibfnamefont {R.}~\bibnamefont
  {Haag}},\ }\bibfield  {title} {\bibinfo {title} {{Quantum field theories with
  composite particles and asymptotic conditions}},\ }\href
  {https://doi.org/10.1103/PhysRev.112.669} {\bibfield  {journal} {\bibinfo
  {journal} {Phys. Rev.}\ }\textbf {\bibinfo {volume} {112}},\ \bibinfo {pages}
  {669} (\bibinfo {year} {1958})}\BibitemShut {NoStop}%
\bibitem [{\citenamefont {Chisholm}(1961)}]{CHISHOLM1961469}%
  \BibitemOpen
  \bibfield  {author} {\bibinfo {author} {\bibfnamefont {J.}~\bibnamefont
  {Chisholm}},\ }\bibfield  {title} {\bibinfo {title} {Change of variables in
  quantum field theories},\ }\href
  {https://doi.org/https://doi.org/10.1016/0029-5582(61)90106-7} {\bibfield
  {journal} {\bibinfo  {journal} {Nuclear Physics}\ }\textbf {\bibinfo {volume}
  {26}},\ \bibinfo {pages} {469} (\bibinfo {year} {1961})}\BibitemShut
  {NoStop}%
\bibitem [{\citenamefont {Kamefuchi}\ \emph {et~al.}(1961)\citenamefont
  {Kamefuchi}, \citenamefont {O'Raifeartaigh},\ and\ \citenamefont
  {Salam}}]{KAMEFUCHI1961529}%
  \BibitemOpen
  \bibfield  {author} {\bibinfo {author} {\bibfnamefont {S.}~\bibnamefont
  {Kamefuchi}}, \bibinfo {author} {\bibfnamefont {L.}~\bibnamefont
  {O'Raifeartaigh}},\ and\ \bibinfo {author} {\bibfnamefont {A.}~\bibnamefont
  {Salam}},\ }\bibfield  {title} {\bibinfo {title} {Change of variables and
  equivalence theorems in quantum field theories},\ }\href
  {https://doi.org/https://doi.org/10.1016/0029-5582(61)90056-6} {\bibfield
  {journal} {\bibinfo  {journal} {Nuclear Physics}\ }\textbf {\bibinfo {volume}
  {28}},\ \bibinfo {pages} {529} (\bibinfo {year} {1961})}\BibitemShut
  {NoStop}%
\bibitem [{\citenamefont {Buballa}(2005)}]{Buballa:2003qv}%
  \BibitemOpen
  \bibfield  {author} {\bibinfo {author} {\bibfnamefont {M.}~\bibnamefont
  {Buballa}},\ }\bibfield  {title} {\bibinfo {title} {{NJL model analysis of
  quark matter at large density}},\ }\href
  {https://doi.org/10.1016/j.physrep.2004.11.004} {\bibfield  {journal}
  {\bibinfo  {journal} {Phys. Rept.}\ }\textbf {\bibinfo {volume} {407}},\
  \bibinfo {pages} {205} (\bibinfo {year} {2005})},\ \Eprint
  {https://arxiv.org/abs/hep-ph/0402234} {arXiv:hep-ph/0402234} \BibitemShut
  {NoStop}%
\bibitem [{\citenamefont {Stiele}(2014)}]{Rainerthesis}%
  \BibitemOpen
  \bibfield  {author} {\bibinfo {author} {\bibfnamefont {R.}~\bibnamefont
  {Stiele}},\ }\emph {\bibinfo {title} {On the Thermodynamics and Phase
  Structure of Strongly-Interacting Matter in a Polyakov-loop–extended
  Constituent-Quark Model}},\ \href@noop {} {Ph.D. thesis},\ \bibinfo  {school}
  {Ruperto-Carola University of Heidelberg} (\bibinfo {year}
  {2014})\BibitemShut {NoStop}%
\bibitem [{\citenamefont {Costa}\ \emph {et~al.}(2010)\citenamefont {Costa},
  \citenamefont {Ruivo}, \citenamefont {de~Sousa},\ and\ \citenamefont
  {Hansen}}]{Costa:2010zw}%
  \BibitemOpen
  \bibfield  {author} {\bibinfo {author} {\bibfnamefont {P.}~\bibnamefont
  {Costa}}, \bibinfo {author} {\bibfnamefont {M.~C.}\ \bibnamefont {Ruivo}},
  \bibinfo {author} {\bibfnamefont {C.~A.}\ \bibnamefont {de~Sousa}},\ and\
  \bibinfo {author} {\bibfnamefont {H.}~\bibnamefont {Hansen}},\ }\bibfield
  {title} {\bibinfo {title} {{Phase diagram and critical properties within an
  effective model of QCD: the Nambu-Jona-Lasinio model coupled to the Polyakov
  loop}},\ }\href {https://doi.org/10.3390/sym2031338} {\bibfield  {journal}
  {\bibinfo  {journal} {Symmetry}\ }\textbf {\bibinfo {volume} {2}},\ \bibinfo
  {pages} {1338} (\bibinfo {year} {2010})},\ \Eprint
  {https://arxiv.org/abs/1007.1380} {arXiv:1007.1380 [hep-ph]} \BibitemShut
  {NoStop}%
\bibitem [{\citenamefont {Boyd}\ \emph {et~al.}(1996)\citenamefont {Boyd},
  \citenamefont {Engels}, \citenamefont {Karsch}, \citenamefont {Laermann},
  \citenamefont {Legeland}, \citenamefont {Lutgemeier},\ and\ \citenamefont
  {Petersson}}]{Boyd:1996bx}%
  \BibitemOpen
  \bibfield  {author} {\bibinfo {author} {\bibfnamefont {G.}~\bibnamefont
  {Boyd}}, \bibinfo {author} {\bibfnamefont {J.}~\bibnamefont {Engels}},
  \bibinfo {author} {\bibfnamefont {F.}~\bibnamefont {Karsch}}, \bibinfo
  {author} {\bibfnamefont {E.}~\bibnamefont {Laermann}}, \bibinfo {author}
  {\bibfnamefont {C.}~\bibnamefont {Legeland}}, \bibinfo {author}
  {\bibfnamefont {M.}~\bibnamefont {Lutgemeier}},\ and\ \bibinfo {author}
  {\bibfnamefont {B.}~\bibnamefont {Petersson}},\ }\bibfield  {title} {\bibinfo
  {title} {{Thermodynamics of SU(3) lattice gauge theory}},\ }\href
  {https://doi.org/10.1016/0550-3213(96)00170-8} {\bibfield  {journal}
  {\bibinfo  {journal} {Nucl. Phys. B}\ }\textbf {\bibinfo {volume} {469}},\
  \bibinfo {pages} {419} (\bibinfo {year} {1996})},\ \Eprint
  {https://arxiv.org/abs/hep-lat/9602007} {arXiv:hep-lat/9602007} \BibitemShut
  {NoStop}%
\bibitem [{\citenamefont {Kaczmarek}\ \emph {et~al.}(2002)\citenamefont
  {Kaczmarek}, \citenamefont {Karsch}, \citenamefont {Petreczky},\ and\
  \citenamefont {Zantow}}]{Kaczmarek:2002mc}%
  \BibitemOpen
  \bibfield  {author} {\bibinfo {author} {\bibfnamefont {O.}~\bibnamefont
  {Kaczmarek}}, \bibinfo {author} {\bibfnamefont {F.}~\bibnamefont {Karsch}},
  \bibinfo {author} {\bibfnamefont {P.}~\bibnamefont {Petreczky}},\ and\
  \bibinfo {author} {\bibfnamefont {F.}~\bibnamefont {Zantow}},\ }\bibfield
  {title} {\bibinfo {title} {{Heavy quark anti-quark free energy and the
  renormalized Polyakov loop}},\ }\href
  {https://doi.org/10.1016/S0370-2693(02)02415-2} {\bibfield  {journal}
  {\bibinfo  {journal} {Phys. Lett. B}\ }\textbf {\bibinfo {volume} {543}},\
  \bibinfo {pages} {41} (\bibinfo {year} {2002})},\ \Eprint
  {https://arxiv.org/abs/hep-lat/0207002} {arXiv:hep-lat/0207002} \BibitemShut
  {NoStop}%
\bibitem [{\citenamefont {Karsch}(2002)}]{Karsch:2001cy}%
  \BibitemOpen
  \bibfield  {author} {\bibinfo {author} {\bibfnamefont {F.}~\bibnamefont
  {Karsch}},\ }\bibfield  {title} {\bibinfo {title} {{Lattice QCD at high
  temperature and density}},\ }\href {https://doi.org/10.1007/3-540-45792-5_6}
  {\bibfield  {journal} {\bibinfo  {journal} {Lect. Notes Phys.}\ }\textbf
  {\bibinfo {volume} {583}},\ \bibinfo {pages} {209} (\bibinfo {year}
  {2002})},\ \Eprint {https://arxiv.org/abs/hep-lat/0106019}
  {arXiv:hep-lat/0106019} \BibitemShut {NoStop}%
\bibitem [{\citenamefont {Borsanyi}\ \emph {et~al.}(2012)\citenamefont
  {Borsanyi}, \citenamefont {Endrodi}, \citenamefont {Fodor}, \citenamefont
  {Katz},\ and\ \citenamefont {Szabo}}]{Borsanyi:2012ve}%
  \BibitemOpen
  \bibfield  {author} {\bibinfo {author} {\bibfnamefont {S.}~\bibnamefont
  {Borsanyi}}, \bibinfo {author} {\bibfnamefont {G.}~\bibnamefont {Endrodi}},
  \bibinfo {author} {\bibfnamefont {Z.}~\bibnamefont {Fodor}}, \bibinfo
  {author} {\bibfnamefont {S.~D.}\ \bibnamefont {Katz}},\ and\ \bibinfo
  {author} {\bibfnamefont {K.~K.}\ \bibnamefont {Szabo}},\ }\bibfield  {title}
  {\bibinfo {title} {{Precision SU(3) lattice thermodynamics for a large
  temperature range}},\ }\href {https://doi.org/10.1007/JHEP07(2012)056}
  {\bibfield  {journal} {\bibinfo  {journal} {JHEP}\ }\textbf {\bibinfo
  {volume} {07}},\ \bibinfo {pages} {056}},\ \Eprint
  {https://arxiv.org/abs/1204.6184} {arXiv:1204.6184 [hep-lat]} \BibitemShut
  {NoStop}%
\bibitem [{\citenamefont {Schaefer}\ \emph {et~al.}(2007)\citenamefont
  {Schaefer}, \citenamefont {Pawlowski},\ and\ \citenamefont
  {Wambach}}]{Schaefer:2007pw}%
  \BibitemOpen
  \bibfield  {author} {\bibinfo {author} {\bibfnamefont {B.-J.}\ \bibnamefont
  {Schaefer}}, \bibinfo {author} {\bibfnamefont {J.~M.}\ \bibnamefont
  {Pawlowski}},\ and\ \bibinfo {author} {\bibfnamefont {J.}~\bibnamefont
  {Wambach}},\ }\bibfield  {title} {\bibinfo {title} {{The Phase Structure of
  the Polyakov--Quark-Meson Model}},\ }\href
  {https://doi.org/10.1103/PhysRevD.76.074023} {\bibfield  {journal} {\bibinfo
  {journal} {Phys. Rev. D}\ }\textbf {\bibinfo {volume} {76}},\ \bibinfo
  {pages} {074023} (\bibinfo {year} {2007})},\ \Eprint
  {https://arxiv.org/abs/0704.3234} {arXiv:0704.3234 [hep-ph]} \BibitemShut
  {NoStop}%
\bibitem [{\citenamefont {Pisarski}(2000)}]{Pisarski:2000eq}%
  \BibitemOpen
  \bibfield  {author} {\bibinfo {author} {\bibfnamefont {R.~D.}\ \bibnamefont
  {Pisarski}},\ }\bibfield  {title} {\bibinfo {title} {{Quark gluon plasma as a
  condensate of SU(3) Wilson lines}},\ }\href
  {https://doi.org/10.1103/PhysRevD.62.111501} {\bibfield  {journal} {\bibinfo
  {journal} {Phys. Rev. D}\ }\textbf {\bibinfo {volume} {62}},\ \bibinfo
  {pages} {111501} (\bibinfo {year} {2000})},\ \Eprint
  {https://arxiv.org/abs/hep-ph/0006205} {arXiv:hep-ph/0006205} \BibitemShut
  {NoStop}%
\bibitem [{\citenamefont {Lo}\ \emph {et~al.}(2013{\natexlab{a}})\citenamefont
  {Lo}, \citenamefont {Friman}, \citenamefont {Kaczmarek}, \citenamefont
  {Redlich},\ and\ \citenamefont {Sasaki}}]{Lo:2013hla}%
  \BibitemOpen
  \bibfield  {author} {\bibinfo {author} {\bibfnamefont {P.~M.}\ \bibnamefont
  {Lo}}, \bibinfo {author} {\bibfnamefont {B.}~\bibnamefont {Friman}}, \bibinfo
  {author} {\bibfnamefont {O.}~\bibnamefont {Kaczmarek}}, \bibinfo {author}
  {\bibfnamefont {K.}~\bibnamefont {Redlich}},\ and\ \bibinfo {author}
  {\bibfnamefont {C.}~\bibnamefont {Sasaki}},\ }\bibfield  {title} {\bibinfo
  {title} {{Polyakov loop fluctuations in SU(3) lattice gauge theory and an
  effective gluon potential}},\ }\href
  {https://doi.org/10.1103/PhysRevD.88.074502} {\bibfield  {journal} {\bibinfo
  {journal} {Phys. Rev. D}\ }\textbf {\bibinfo {volume} {88}},\ \bibinfo
  {pages} {074502} (\bibinfo {year} {2013}{\natexlab{a}})},\ \Eprint
  {https://arxiv.org/abs/1307.5958} {arXiv:1307.5958 [hep-lat]} \BibitemShut
  {NoStop}%
\bibitem [{\citenamefont {Lucini}\ \emph {et~al.}(2002)\citenamefont {Lucini},
  \citenamefont {Teper},\ and\ \citenamefont {Wenger}}]{Lucini:2002ku}%
  \BibitemOpen
  \bibfield  {author} {\bibinfo {author} {\bibfnamefont {B.}~\bibnamefont
  {Lucini}}, \bibinfo {author} {\bibfnamefont {M.}~\bibnamefont {Teper}},\ and\
  \bibinfo {author} {\bibfnamefont {U.}~\bibnamefont {Wenger}},\ }\bibfield
  {title} {\bibinfo {title} {{The Deconfinement transition in SU(N) gauge
  theories}},\ }\href {https://doi.org/10.1016/S0370-2693(02)02556-X}
  {\bibfield  {journal} {\bibinfo  {journal} {Phys. Lett. B}\ }\textbf
  {\bibinfo {volume} {545}},\ \bibinfo {pages} {197} (\bibinfo {year}
  {2002})},\ \Eprint {https://arxiv.org/abs/hep-lat/0206029}
  {arXiv:hep-lat/0206029} \BibitemShut {NoStop}%
\bibitem [{\citenamefont {Lucini}\ \emph {et~al.}(2004)\citenamefont {Lucini},
  \citenamefont {Teper},\ and\ \citenamefont {Wenger}}]{Lucini:2003jp}%
  \BibitemOpen
  \bibfield  {author} {\bibinfo {author} {\bibfnamefont {B.}~\bibnamefont
  {Lucini}}, \bibinfo {author} {\bibfnamefont {M.}~\bibnamefont {Teper}},\ and\
  \bibinfo {author} {\bibfnamefont {U.}~\bibnamefont {Wenger}},\ }\bibfield
  {title} {\bibinfo {title} {{SU(N) gauge theories near T(c)}},\ }\href
  {https://doi.org/10.1016/S0920-5632(03)02644-6} {\bibfield  {journal}
  {\bibinfo  {journal} {Nucl. Phys. B Proc. Suppl.}\ }\textbf {\bibinfo
  {volume} {129}},\ \bibinfo {pages} {569} (\bibinfo {year} {2004})},\ \Eprint
  {https://arxiv.org/abs/hep-lat/0309009} {arXiv:hep-lat/0309009} \BibitemShut
  {NoStop}%
\bibitem [{\citenamefont {Liddle}\ and\ \citenamefont
  {Teper}(2008)}]{Liddle:2008kk}%
  \BibitemOpen
  \bibfield  {author} {\bibinfo {author} {\bibfnamefont {J.}~\bibnamefont
  {Liddle}}\ and\ \bibinfo {author} {\bibfnamefont {M.}~\bibnamefont {Teper}},\
  }\bibfield  {title} {\bibinfo {title} {{The Deconfining phase transition in
  D=2+1 SU(N) gauge theories}}\ }(\bibinfo {year} {2008})\ \Eprint
  {https://arxiv.org/abs/0803.2128} {arXiv:0803.2128 [hep-lat]} \BibitemShut
  {NoStop}%
\bibitem [{\citenamefont {Lucini}\ \emph {et~al.}(2012)\citenamefont {Lucini},
  \citenamefont {Rago},\ and\ \citenamefont {Rinaldi}}]{Lucini:2012wq}%
  \BibitemOpen
  \bibfield  {author} {\bibinfo {author} {\bibfnamefont {B.}~\bibnamefont
  {Lucini}}, \bibinfo {author} {\bibfnamefont {A.}~\bibnamefont {Rago}},\ and\
  \bibinfo {author} {\bibfnamefont {E.}~\bibnamefont {Rinaldi}},\ }\bibfield
  {title} {\bibinfo {title} {{SU($N_c$) gauge theories at deconfinement}},\
  }\href {https://doi.org/10.1016/j.physletb.2012.04.070} {\bibfield  {journal}
  {\bibinfo  {journal} {Phys. Lett. B}\ }\textbf {\bibinfo {volume} {712}},\
  \bibinfo {pages} {279} (\bibinfo {year} {2012})},\ \Eprint
  {https://arxiv.org/abs/1202.6684} {arXiv:1202.6684 [hep-lat]} \BibitemShut
  {NoStop}%
\bibitem [{\citenamefont {Zhang}\ \emph {et~al.}(2010)\citenamefont {Zhang},
  \citenamefont {Brauner},\ and\ \citenamefont {Rischke}}]{Zhang:2010kn}%
  \BibitemOpen
  \bibfield  {author} {\bibinfo {author} {\bibfnamefont {T.}~\bibnamefont
  {Zhang}}, \bibinfo {author} {\bibfnamefont {T.}~\bibnamefont {Brauner}},\
  and\ \bibinfo {author} {\bibfnamefont {D.~H.}\ \bibnamefont {Rischke}},\
  }\bibfield  {title} {\bibinfo {title} {{QCD-like theories at nonzero
  temperature and density}},\ }\href {https://doi.org/10.1007/JHEP06(2010)064}
  {\bibfield  {journal} {\bibinfo  {journal} {JHEP}\ }\textbf {\bibinfo
  {volume} {06}},\ \bibinfo {pages} {064}},\ \Eprint
  {https://arxiv.org/abs/1005.2928} {arXiv:1005.2928 [hep-ph]} \BibitemShut
  {NoStop}%
\bibitem [{\citenamefont {Brauner}()}]{haarMeasureBrauner}%
  \BibitemOpen
  \bibfield  {author} {\bibinfo {author} {\bibfnamefont {T.}~\bibnamefont
  {Brauner}},\ }\href
  {https://drive.google.com/file/d/1rL0PqJLIPaY_BTyr7RFaoOZ2fmlcy9Gr/view}
  {\bibinfo {title} {Haar measure on the unitary groups}},\ \bibinfo {note}
  {[Online; last downloaded at 15-February-2023]}\BibitemShut {NoStop}%
\bibitem [{\citenamefont {Ripka}(1997)}]{Ripka:1997zb}%
  \BibitemOpen
  \bibfield  {author} {\bibinfo {author} {\bibfnamefont {G.}~\bibnamefont
  {Ripka}},\ }\href@noop {} {\emph {\bibinfo {title} {{Quarks bound by chiral
  fields: The quark-structure of the vacuum and of light mesons and
  baryons}}}}\ (\bibinfo {year} {1997})\BibitemShut {NoStop}%
\bibitem [{\citenamefont {Meisinger}\ and\ \citenamefont
  {Ogilvie}(2002)}]{Meisinger:2001fi}%
  \BibitemOpen
  \bibfield  {author} {\bibinfo {author} {\bibfnamefont {P.~N.}\ \bibnamefont
  {Meisinger}}\ and\ \bibinfo {author} {\bibfnamefont {M.~C.}\ \bibnamefont
  {Ogilvie}},\ }\bibfield  {title} {\bibinfo {title} {{Complete high
  temperature expansions for one loop finite temperature effects}},\ }\href
  {https://doi.org/10.1103/PhysRevD.65.056013} {\bibfield  {journal} {\bibinfo
  {journal} {Phys. Rev. D}\ }\textbf {\bibinfo {volume} {65}},\ \bibinfo
  {pages} {056013} (\bibinfo {year} {2002})},\ \Eprint
  {https://arxiv.org/abs/hep-ph/0108026} {arXiv:hep-ph/0108026} \BibitemShut
  {NoStop}%
\bibitem [{\citenamefont {Meisinger}\ \emph {et~al.}(2002)\citenamefont
  {Meisinger}, \citenamefont {Miller},\ and\ \citenamefont
  {Ogilvie}}]{Meisinger:2001cq}%
  \BibitemOpen
  \bibfield  {author} {\bibinfo {author} {\bibfnamefont {P.~N.}\ \bibnamefont
  {Meisinger}}, \bibinfo {author} {\bibfnamefont {T.~R.}\ \bibnamefont
  {Miller}},\ and\ \bibinfo {author} {\bibfnamefont {M.~C.}\ \bibnamefont
  {Ogilvie}},\ }\bibfield  {title} {\bibinfo {title} {{Phenomenological
  equations of state for the quark gluon plasma}},\ }\href
  {https://doi.org/10.1103/PhysRevD.65.034009} {\bibfield  {journal} {\bibinfo
  {journal} {Phys. Rev. D}\ }\textbf {\bibinfo {volume} {65}},\ \bibinfo
  {pages} {034009} (\bibinfo {year} {2002})},\ \Eprint
  {https://arxiv.org/abs/hep-ph/0108009} {arXiv:hep-ph/0108009} \BibitemShut
  {NoStop}%
\bibitem [{\citenamefont {Megias}\ \emph {et~al.}(2014)\citenamefont {Megias},
  \citenamefont {Ruiz~Arriola},\ and\ \citenamefont
  {Salcedo}}]{Megias:2013xaa}%
  \BibitemOpen
  \bibfield  {author} {\bibinfo {author} {\bibfnamefont {E.}~\bibnamefont
  {Megias}}, \bibinfo {author} {\bibfnamefont {E.}~\bibnamefont
  {Ruiz~Arriola}},\ and\ \bibinfo {author} {\bibfnamefont {L.~L.}\ \bibnamefont
  {Salcedo}},\ }\bibfield  {title} {\bibinfo {title} {{Polyakov loop in various
  representations in the confined phase of QCD}},\ }\href
  {https://doi.org/10.1103/PhysRevD.89.076006} {\bibfield  {journal} {\bibinfo
  {journal} {Phys. Rev. D}\ }\textbf {\bibinfo {volume} {89}},\ \bibinfo
  {pages} {076006} (\bibinfo {year} {2014})},\ \Eprint
  {https://arxiv.org/abs/1311.2814} {arXiv:1311.2814 [hep-ph]} \BibitemShut
  {NoStop}%
\bibitem [{\citenamefont {Elvang}\ \emph {et~al.}(2003)\citenamefont {Elvang},
  \citenamefont {Cvitanovic},\ and\ \citenamefont {Kennedy}}]{Elvang:2003ue}%
  \BibitemOpen
  \bibfield  {author} {\bibinfo {author} {\bibfnamefont {H.}~\bibnamefont
  {Elvang}}, \bibinfo {author} {\bibfnamefont {P.}~\bibnamefont {Cvitanovic}},\
  and\ \bibinfo {author} {\bibfnamefont {A.~D.}\ \bibnamefont {Kennedy}},\
  }\bibfield  {title} {\bibinfo {title} {{Diagrammatic young projection
  operators for U(n)}}\ }(\bibinfo {year} {2003})\ \Eprint
  {https://arxiv.org/abs/hep-th/0307186} {arXiv:hep-th/0307186} \BibitemShut
  {NoStop}%
\bibitem [{\citenamefont {Tsai}\ and\ \citenamefont
  {Muller}(2009)}]{Tsai:2008je}%
  \BibitemOpen
  \bibfield  {author} {\bibinfo {author} {\bibfnamefont {H.-M.}\ \bibnamefont
  {Tsai}}\ and\ \bibinfo {author} {\bibfnamefont {B.}~\bibnamefont {Muller}},\
  }\bibfield  {title} {\bibinfo {title} {{Phenomenology of the three-flavour
  PNJL model and thermal strange quark production}},\ }\href
  {https://doi.org/10.1088/0954-3899/36/7/075101} {\bibfield  {journal}
  {\bibinfo  {journal} {J. Phys. G}\ }\textbf {\bibinfo {volume} {36}},\
  \bibinfo {pages} {075101} (\bibinfo {year} {2009})},\ \Eprint
  {https://arxiv.org/abs/0811.2216} {arXiv:0811.2216 [hep-ph]} \BibitemShut
  {NoStop}%
\bibitem [{\citenamefont {Ruggieri}\ \emph {et~al.}(2012)\citenamefont
  {Ruggieri}, \citenamefont {Alba}, \citenamefont {Castorina}, \citenamefont
  {Plumari}, \citenamefont {Ratti},\ and\ \citenamefont
  {Greco}}]{Ruggieri:2012ny}%
  \BibitemOpen
  \bibfield  {author} {\bibinfo {author} {\bibfnamefont {M.}~\bibnamefont
  {Ruggieri}}, \bibinfo {author} {\bibfnamefont {P.}~\bibnamefont {Alba}},
  \bibinfo {author} {\bibfnamefont {P.}~\bibnamefont {Castorina}}, \bibinfo
  {author} {\bibfnamefont {S.}~\bibnamefont {Plumari}}, \bibinfo {author}
  {\bibfnamefont {C.}~\bibnamefont {Ratti}},\ and\ \bibinfo {author}
  {\bibfnamefont {V.}~\bibnamefont {Greco}},\ }\bibfield  {title} {\bibinfo
  {title} {{Polyakov Loop and Gluon Quasiparticles in Yang-Mills
  Thermodynamics}},\ }\href {https://doi.org/10.1103/PhysRevD.86.054007}
  {\bibfield  {journal} {\bibinfo  {journal} {Phys. Rev. D}\ }\textbf {\bibinfo
  {volume} {86}},\ \bibinfo {pages} {054007} (\bibinfo {year} {2012})},\
  \Eprint {https://arxiv.org/abs/1204.5995} {arXiv:1204.5995 [hep-ph]}
  \BibitemShut {NoStop}%
\bibitem [{\citenamefont {Alba}\ \emph {et~al.}(2014)\citenamefont {Alba},
  \citenamefont {Alberico}, \citenamefont {Bluhm}, \citenamefont {Greco},
  \citenamefont {Ratti},\ and\ \citenamefont {Ruggieri}}]{Alba:2014lda}%
  \BibitemOpen
  \bibfield  {author} {\bibinfo {author} {\bibfnamefont {P.}~\bibnamefont
  {Alba}}, \bibinfo {author} {\bibfnamefont {W.}~\bibnamefont {Alberico}},
  \bibinfo {author} {\bibfnamefont {M.}~\bibnamefont {Bluhm}}, \bibinfo
  {author} {\bibfnamefont {V.}~\bibnamefont {Greco}}, \bibinfo {author}
  {\bibfnamefont {C.}~\bibnamefont {Ratti}},\ and\ \bibinfo {author}
  {\bibfnamefont {M.}~\bibnamefont {Ruggieri}},\ }\bibfield  {title} {\bibinfo
  {title} {{Polyakov loop and gluon quasiparticles: A self-consistent approach
  to Yang\textendash{}Mills thermodynamics}},\ }\href
  {https://doi.org/10.1016/j.nuclphysa.2014.11.011} {\bibfield  {journal}
  {\bibinfo  {journal} {Nucl. Phys. A}\ }\textbf {\bibinfo {volume} {934}},\
  \bibinfo {pages} {41} (\bibinfo {year} {2014})},\ \Eprint
  {https://arxiv.org/abs/1402.6213} {arXiv:1402.6213 [hep-ph]} \BibitemShut
  {NoStop}%
\bibitem [{\citenamefont {Islam}\ \emph {et~al.}(2022)\citenamefont {Islam},
  \citenamefont {Mustafa}, \citenamefont {Ray},\ and\ \citenamefont
  {Singha}}]{Islam:2021qwh}%
  \BibitemOpen
  \bibfield  {author} {\bibinfo {author} {\bibfnamefont {C.~A.}\ \bibnamefont
  {Islam}}, \bibinfo {author} {\bibfnamefont {M.~G.}\ \bibnamefont {Mustafa}},
  \bibinfo {author} {\bibfnamefont {R.}~\bibnamefont {Ray}},\ and\ \bibinfo
  {author} {\bibfnamefont {P.}~\bibnamefont {Singha}},\ }\bibfield  {title}
  {\bibinfo {title} {{Consistent approach to study gluon quasiparticles}},\
  }\href {https://doi.org/10.1103/PhysRevD.106.054002} {\bibfield  {journal}
  {\bibinfo  {journal} {Phys. Rev. D}\ }\textbf {\bibinfo {volume} {106}},\
  \bibinfo {pages} {054002} (\bibinfo {year} {2022})},\ \Eprint
  {https://arxiv.org/abs/2109.13321} {arXiv:2109.13321 [hep-ph]} \BibitemShut
  {NoStop}%
\bibitem [{\citenamefont {Lo}\ \emph {et~al.}(2013{\natexlab{b}})\citenamefont
  {Lo}, \citenamefont {Friman}, \citenamefont {Kaczmarek}, \citenamefont
  {Redlich},\ and\ \citenamefont {Sasaki}}]{Lo:2013etb}%
  \BibitemOpen
  \bibfield  {author} {\bibinfo {author} {\bibfnamefont {P.~M.}\ \bibnamefont
  {Lo}}, \bibinfo {author} {\bibfnamefont {B.}~\bibnamefont {Friman}}, \bibinfo
  {author} {\bibfnamefont {O.}~\bibnamefont {Kaczmarek}}, \bibinfo {author}
  {\bibfnamefont {K.}~\bibnamefont {Redlich}},\ and\ \bibinfo {author}
  {\bibfnamefont {C.}~\bibnamefont {Sasaki}},\ }\bibfield  {title} {\bibinfo
  {title} {{Probing Deconfinement with Polyakov Loop Susceptibilities}},\
  }\href {https://doi.org/10.1103/PhysRevD.88.014506} {\bibfield  {journal}
  {\bibinfo  {journal} {Phys. Rev. D}\ }\textbf {\bibinfo {volume} {88}},\
  \bibinfo {pages} {014506} (\bibinfo {year} {2013}{\natexlab{b}})},\ \Eprint
  {https://arxiv.org/abs/1306.5094} {arXiv:1306.5094 [hep-lat]} \BibitemShut
  {NoStop}%
\bibitem [{\citenamefont {Bazavov}\ \emph {et~al.}(2016)\citenamefont
  {Bazavov}, \citenamefont {Brambilla}, \citenamefont {Ding}, \citenamefont
  {Petreczky}, \citenamefont {Schadler}, \citenamefont {Vairo},\ and\
  \citenamefont {Weber}}]{Bazavov:2016uvm}%
  \BibitemOpen
  \bibfield  {author} {\bibinfo {author} {\bibfnamefont {A.}~\bibnamefont
  {Bazavov}}, \bibinfo {author} {\bibfnamefont {N.}~\bibnamefont {Brambilla}},
  \bibinfo {author} {\bibfnamefont {H.~T.}\ \bibnamefont {Ding}}, \bibinfo
  {author} {\bibfnamefont {P.}~\bibnamefont {Petreczky}}, \bibinfo {author}
  {\bibfnamefont {H.~P.}\ \bibnamefont {Schadler}}, \bibinfo {author}
  {\bibfnamefont {A.}~\bibnamefont {Vairo}},\ and\ \bibinfo {author}
  {\bibfnamefont {J.~H.}\ \bibnamefont {Weber}},\ }\bibfield  {title} {\bibinfo
  {title} {{Polyakov loop in 2+1 flavor QCD from low to high temperatures}},\
  }\href {https://doi.org/10.1103/PhysRevD.93.114502} {\bibfield  {journal}
  {\bibinfo  {journal} {Phys. Rev. D}\ }\textbf {\bibinfo {volume} {93}},\
  \bibinfo {pages} {114502} (\bibinfo {year} {2016})},\ \Eprint
  {https://arxiv.org/abs/1603.06637} {arXiv:1603.06637 [hep-lat]} \BibitemShut
  {NoStop}%
\bibitem [{\citenamefont {Clarke}\ \emph {et~al.}(2020)\citenamefont {Clarke},
  \citenamefont {Kaczmarek}, \citenamefont {Karsch},\ and\ \citenamefont
  {Lahiri}}]{Clarke:2019tzf}%
  \BibitemOpen
  \bibfield  {author} {\bibinfo {author} {\bibfnamefont {D.~A.}\ \bibnamefont
  {Clarke}}, \bibinfo {author} {\bibfnamefont {O.}~\bibnamefont {Kaczmarek}},
  \bibinfo {author} {\bibfnamefont {F.}~\bibnamefont {Karsch}},\ and\ \bibinfo
  {author} {\bibfnamefont {A.}~\bibnamefont {Lahiri}},\ }\bibfield  {title}
  {\bibinfo {title} {{Polyakov Loop Susceptibility and Correlators in the
  Chiral Limit}},\ }\href {https://doi.org/10.22323/1.363.0194} {\bibfield
  {journal} {\bibinfo  {journal} {PoS}\ }\textbf {\bibinfo {volume}
  {LATTICE2019}},\ \bibinfo {pages} {194} (\bibinfo {year} {2020})},\ \Eprint
  {https://arxiv.org/abs/1911.07668} {arXiv:1911.07668 [hep-lat]} \BibitemShut
  {NoStop}%
\bibitem [{\citenamefont {Lo}\ \emph {et~al.}(2014)\citenamefont {Lo},
  \citenamefont {Friman},\ and\ \citenamefont {Redlich}}]{Lo:2014vba}%
  \BibitemOpen
  \bibfield  {author} {\bibinfo {author} {\bibfnamefont {P.~M.}\ \bibnamefont
  {Lo}}, \bibinfo {author} {\bibfnamefont {B.}~\bibnamefont {Friman}},\ and\
  \bibinfo {author} {\bibfnamefont {K.}~\bibnamefont {Redlich}},\ }\bibfield
  {title} {\bibinfo {title} {{Polyakov loop fluctuations and deconfinement in
  the limit of heavy quarks}},\ }\href
  {https://doi.org/10.1103/PhysRevD.90.074035} {\bibfield  {journal} {\bibinfo
  {journal} {Phys. Rev. D}\ }\textbf {\bibinfo {volume} {90}},\ \bibinfo
  {pages} {074035} (\bibinfo {year} {2014})},\ \Eprint
  {https://arxiv.org/abs/1406.4050} {arXiv:1406.4050 [hep-ph]} \BibitemShut
  {NoStop}%
\bibitem [{\citenamefont {Miura}\ \emph {et~al.}(2017)\citenamefont {Miura},
  \citenamefont {Kawamoto}, \citenamefont {Nakano},\ and\ \citenamefont
  {Ohnishi}}]{Miura:2016kmd}%
  \BibitemOpen
  \bibfield  {author} {\bibinfo {author} {\bibfnamefont {K.}~\bibnamefont
  {Miura}}, \bibinfo {author} {\bibfnamefont {N.}~\bibnamefont {Kawamoto}},
  \bibinfo {author} {\bibfnamefont {T.~Z.}\ \bibnamefont {Nakano}},\ and\
  \bibinfo {author} {\bibfnamefont {A.}~\bibnamefont {Ohnishi}},\ }\bibfield
  {title} {\bibinfo {title} {{Polyakov loop effects on the phase diagram in
  strong-coupling lattice QCD}},\ }\href
  {https://doi.org/10.1103/PhysRevD.95.114505} {\bibfield  {journal} {\bibinfo
  {journal} {Phys. Rev. D}\ }\textbf {\bibinfo {volume} {95}},\ \bibinfo
  {pages} {114505} (\bibinfo {year} {2017})},\ \Eprint
  {https://arxiv.org/abs/1610.09288} {arXiv:1610.09288 [hep-lat]} \BibitemShut
  {NoStop}%
\bibitem [{\citenamefont {Mykkanen}\ \emph {et~al.}(2012)\citenamefont
  {Mykkanen}, \citenamefont {Panero},\ and\ \citenamefont
  {Rummukainen}}]{Mykkanen:2012ri}%
  \BibitemOpen
  \bibfield  {author} {\bibinfo {author} {\bibfnamefont {A.}~\bibnamefont
  {Mykkanen}}, \bibinfo {author} {\bibfnamefont {M.}~\bibnamefont {Panero}},\
  and\ \bibinfo {author} {\bibfnamefont {K.}~\bibnamefont {Rummukainen}},\
  }\bibfield  {title} {\bibinfo {title} {{Casimir scaling and renormalization
  of Polyakov loops in large-N gauge theories}},\ }\href
  {https://doi.org/10.1007/JHEP05(2012)069} {\bibfield  {journal} {\bibinfo
  {journal} {JHEP}\ }\textbf {\bibinfo {volume} {05}},\ \bibinfo {pages}
  {069}},\ \Eprint {https://arxiv.org/abs/1202.2762} {arXiv:1202.2762
  [hep-lat]} \BibitemShut {NoStop}%
\bibitem [{\citenamefont {Bali}(2000)}]{Bali:2000un}%
  \BibitemOpen
  \bibfield  {author} {\bibinfo {author} {\bibfnamefont {G.~S.}\ \bibnamefont
  {Bali}},\ }\bibfield  {title} {\bibinfo {title} {{Casimir scaling of SU(3)
  static potentials}},\ }\href {https://doi.org/10.1103/PhysRevD.62.114503}
  {\bibfield  {journal} {\bibinfo  {journal} {Phys. Rev. D}\ }\textbf {\bibinfo
  {volume} {62}},\ \bibinfo {pages} {114503} (\bibinfo {year} {2000})},\
  \Eprint {https://arxiv.org/abs/hep-lat/0006022} {arXiv:hep-lat/0006022}
  \BibitemShut {NoStop}%
\bibitem [{\citenamefont {Petreczky}\ and\ \citenamefont
  {Schadler}(2015)}]{Petreczky:2015yta}%
  \BibitemOpen
  \bibfield  {author} {\bibinfo {author} {\bibfnamefont {P.}~\bibnamefont
  {Petreczky}}\ and\ \bibinfo {author} {\bibfnamefont {H.~P.}\ \bibnamefont
  {Schadler}},\ }\bibfield  {title} {\bibinfo {title} {{Renormalization of the
  Polyakov loop with gradient flow}},\ }\href
  {https://doi.org/10.1103/PhysRevD.92.094517} {\bibfield  {journal} {\bibinfo
  {journal} {Phys. Rev. D}\ }\textbf {\bibinfo {volume} {92}},\ \bibinfo
  {pages} {094517} (\bibinfo {year} {2015})},\ \Eprint
  {https://arxiv.org/abs/1509.07874} {arXiv:1509.07874 [hep-lat]} \BibitemShut
  {NoStop}%
\bibitem [{\citenamefont {Brown}(1992)}]{brown_1992}%
  \BibitemOpen
  \bibfield  {author} {\bibinfo {author} {\bibfnamefont {L.~S.}\ \bibnamefont
  {Brown}},\ }\href {https://doi.org/10.1017/CBO9780511622649} {\emph {\bibinfo
  {title} {Quantum Field Theory}}}\ (\bibinfo  {publisher} {Cambridge
  University Press},\ \bibinfo {year} {1992})\BibitemShut {NoStop}%
\bibitem [{\citenamefont {Prasolov}(1996)}]{prasolov1996problems}%
  \BibitemOpen
  \bibfield  {author} {\bibinfo {author} {\bibfnamefont {V.}~\bibnamefont
  {Prasolov}},\ }\href {https://books.google.pt/books?id=W9qP4ahS0YEC} {\emph
  {\bibinfo {title} {Problems and Theorems in Linear Algebra}}},\ Translations
  of mathematical monographs\ (\bibinfo  {publisher} {American Mathematical
  Society},\ \bibinfo {year} {1996})\BibitemShut {NoStop}%
\bibitem [{\citenamefont {Curtright}\ \emph {et~al.}(2020)\citenamefont
  {Curtright}, \citenamefont {Fairlie},\ and\ \citenamefont
  {Alshal}}]{Curtright:2020cta}%
  \BibitemOpen
  \bibfield  {author} {\bibinfo {author} {\bibfnamefont {T.~L.}\ \bibnamefont
  {Curtright}}, \bibinfo {author} {\bibfnamefont {D.~B.}\ \bibnamefont
  {Fairlie}},\ and\ \bibinfo {author} {\bibfnamefont {H.}~\bibnamefont
  {Alshal}},\ }\bibfield  {title} {\bibinfo {title} {{A Galileon Primer}}\
  }(\bibinfo {year} {2020})\ \Eprint {https://arxiv.org/abs/1212.6972}
  {arXiv:1212.6972 [hep-th]} \BibitemShut {NoStop}%
\end{thebibliography}%

\end{document}